\documentclass[onecolumn,nofootinbib,reprint]{revtex4-1}
\usepackage[T1]{fontenc}
\usepackage[latin9]{inputenc}
\usepackage{verbatim}
\usepackage{amstext}
\usepackage{color}
\usepackage{graphicx,psfrag}
\usepackage{amssymb}

\def\eg{\emph{e.g.~}}
\def\ie{\emph{i.e.~}}
\def\mup{{\mu'}}
\def\nup{{\nu'}}
\def\aka{\emph{a.k.a.~}}
\def\beq{\begin{equation}}
\def\eeq{\end{equation}}
\def\bea{\begin{eqnarray}}
\def\eea{\end{eqnarray}}
\def\beg{\begin{lyxgreyedout}}
\def\eeg{\end{lyxgreyedout}}
\def\nn{\nonumber}
\def\eps{{\epsilon}}
\begin{document}

\title{Unification of Relativistic and Quantum Mechanics from Elementary Cycles Theory}

\author{Donatello Dolce}
\affiliation{University of Camerino, Piazza Cavour 19F, 62032 Camerino, Italy.}
%\date{\today}

\begin{abstract}
In Elementary Cycles theory elementary \emph{quantum} particles are consistently described as the manifestation of ultra-fast relativistic spacetime cyclic dynamics, \emph{classical} in the essence. The peculiar relativistic geometrodynamics of Elementary Cycles theory yields \emph{de facto} a unification of ordinary relativistic and quantum physics. In particular its classical-relativistic cyclic dynamics reproduce exactly from classical physics first principles all the fundamental aspects of Quantum Mechanics, such as all its axioms, the Feynman path integral, the Dirac quantisation prescription (second quantisation), quantum dynamics of statistical systems, non-relativistic quantum mechanics, atomic physics, superconductivity, graphene physics and so on. Furthermore the theory allows for the explicit derivation of gauge interactions, without postulating gauge invariance, directly from relativistic geometrodynamical transformations, in close analogy with the description of gravitational interaction in general relativity.  In this paper we summarise some of the major achievements, rigorously proven also in several recent peer-reviewed papers, of this innovative formulation of quantum particle physics.   
\end{abstract}

\maketitle

\newpage

\setcounter{tocdepth}{4}
\tableofcontents

\newpage

\section{Introduction} 

Quantum Mechanics (QM) is one of the pillars of physics. Its modern formulation, based on mere mathematical axioms, has been successfully tested by some of the most accurate predictions in physics. It is an undeniable scientific fact that the present mathematical formulation of QM is absolutely correct: \emph{this point is absolutely not questioned in this paper}. Our remarkable result is that all the axioms of QM as well as all the other fundamental aspects of QM are exactly derivable from the simple physical principle of elementary relativistic cyclic dynamics, in a unified description with relativistic mechanics. That is, elementary particles are intrinsically cyclic phenomena (as also implicitly suggested by wave-particle duality) and, as we will prove in this paper, the manifestation of these cyclic behaviours is QM in all its fundamental aspects. 

The elusive physical origin of QM has left a consistent number of unsolved riddles. Today specialists on the field still do not agree on its physical interpretation \cite{2013arXiv1301.1069S}. It is not surprising that some of the founding  fathers of modern physics (e.g. Einstein, Feynman, de Broglie, or, more recently, 't Hooft \cite{Hooft:2014kka}, Wilczek \cite{Wilczek:2012jt}, Weinberg \cite{Weinberg:2014ewa}, etc) have expressed  the necessity of a deeper understanding of its origin by means of simple first physical principles.   The discovery of possible physics beyond QM potentially represents the tipping point for the solution of long-standing problems of modern physics --- including particle physics and cosmology --- and, eventually, to formulate a possible ``theory of everything'' \cite{'tHooft:2005ym}.  

We rigorously prove, see also \cite{Dolce:cycles,Dolce:tune,Dolce:ADSCFT,Dolce:2009ce,Dolce:licata,Dolce:SuperC,Dolce:EPJP,Dolce:Dice2014,Dolce:TM2012,Dolce:Dice2012,Dolce:cycle,Dolce:ICHEP2012,Dolce:FQXi,Dolce:Dice,Dolce:2010ij,Dolce:2010zz,Dolce:2009cev4}, that simple elementary classical-relativistic cyclic systems representing elementary particles directly implies QM in all its fundamental aspects such as: all the axioms of QM; the Feynman Path Integral; the commutation relations, and thus second quantisation and Dirac quantisation rule;  quantum electrodynamics (QED); the QM of statistical systems (Matsubara theory); non-relativistic QM (Bohr-Sommerfeld quantisation, WKB method), atomic physics, and so on. We will refer to such fundamental elementary systems of nature as spacetime Elementary Cycles (ECs), defined in sec.(\ref{Relativistic:EC:model}).  
 
 Each classical EC exactly describes the quantum behaviour of a corresponding elementary particle, sec.(\ref{QM:free}-\ref{INT:QM}).  The postulate of ECs theory, from which the exact unified description of physics is uniquely derived, can be thus stated in the following way: 
 \begin{quote}
 \emph{A free elementary quantum particle of (persistent) energy $\omega$, observed from an inertial reference frame, is an elementary  relativistic cyclic system, classical in the essence, of persistent time periodicity $T = 2 \pi / \omega$}.  
 \end{quote}
 Equivalently, this can be stated in the following form:
 \begin{quote}
 \emph{Every free elementary particle is an elementary relativistic reference clock}.
 \end{quote}
  The postulate of ECs theory and its possible equivalent enunciations, is discussed in detail in, for instance, \cite{Dolce:licata}. Technically, this postulate is realised in terms of covariant compact spacetime dimensions, i.e. with a formalism typical of extra-dimensional theories or string theory. 
 
 ECs can be regarded as one dimensional classical-relativistic strings vibrating in spacetime \footnote{There are fundamental analogies with ordinary string theory such as assumption of a compact world-line parameter in the theory (which in EC theory is the proper time of a particle), but there are also fundamental differences: due to the peculiar assumption of compact world-line parameter EC theory does not requires extra dimensions to be self-consistent. EC theory is the full relativistic generalisation of the theory of sound, where sound sources (i.e. particles) can vibrate in time and not only in space. Nevertheless EC theory fully confirms originals proposals of ordinary String Theory and it can be actually regarded as a String Theory, see sec.({\ref{ONTICSPACETIME}}).} ---  more exactly the theory describes elementary particles as vibrations of spacetime itself. Their characteristic fundamental periods and topology determine the kinematical states of the elementary particles and the other quantum numbers (\text{e.g.} the spin), respectively.  
  
  The most fascinating aspect of EC theory is probably that it proofs the existence of an enriched, non-trivial nature of spacetime never considered before which, on one hand, is fully compatible with ordinary special and general relativistic physics and, on the other hand, implies QM directly from relativity dynamics, sec.(\ref{QM:free}-\ref{INT:QM}). The only quantisation condition from which the whole QM follows is the constraint of intrinsic periodicity of elementary particles. Such a cyclic nature is only referred to free elementary constituents of nature, \ie isolated elementary particles. Due to interactions composite systems are not periodic, in such a way that relativistic causality is fully preserved by the theory (not to be confused with Closed Timelike Cycles). 
 In other words, the clear outcome of our rigorous mathematical demonstrations is the QM mechanics in all its aspects is the manifestation of the intrinsically cyclic nature of spacetime: elementary particles are perfectly cyclic phenomena (\eg consider the wave-particle duality), elementary particles are the basic constituents of our universe, hence physics (in particular quantum physics) can be reduced to the composition of elementary cyclic systems.  
  
  Remarkably EC physics also provides \emph{de facto} a unified geometrodynamical description of gauge and gravitational mechanics, sec.(\ref{INT}) \cite{Dolce:tune,Dolce:cycles}.   Undulatory mechanics is encoded directly into spacetime geometrodynamics of the theory. EC theory proofs that \emph{the price to pay for a unified description of relativistic and quantum dynamics, as well as of gauge and gravitational interaction, is to give up with the emphatically  non-compact formulation of spacetime typical of ordinary particle physics}, sec.(\ref{God:dice}, \ref{ONTICSPACETIME}). In \cite{Dolce:licata,Dolce:tune,Dolce:ADSCFT,Dolce:2009ce,Dolce:cycles} we have discussed in details a similar approach was after all  suggested by Einstein in his attempt of derivation of QM from \emph{constrained} relativistic dynamics \cite{Einstein:1923,Pais:einstein}. The unified description of physics stemming from ECs is so straightforward and exhaustive that it deservers the most careful attention: we recall that ``\textit{in questions of science, the authority of a thousand
is not worth the humble reasoning of a single individual}'' as long as the claims are supported by undeniable evidences (mathematical demonstrations for what which concerns theoretical physics); this is the essence of science according to G. Galilei (see quotation an the end of the paper).

 %Indeed, similarly to 't Hooft proposal in CA models, the theory leads an extremely novel formulation of QM. 
 ECs theory  must not be confused with existing interpretations of QM (de Broglie-Bohm, many-words, stochastic, etc).
  In particular our approach --- being based on covariant BCs imposed to relativistic dynamics --- does not involve hidden-variables of any sort, so it evades Bell's or similar no-go theorems, allowing for a deterministic interpretation of QM. Finally, an EC turns out to be regarded as the covariant effective formulation of an \emph{ultra-fast} continuous periodic Cellular Automata (CA) proposed by 't Hooft, \emph{a.k.a.} cogwheels model.

%We will conclude that the difficulties of canonical physics in interpreting QM is related to a misleading use of the role of relativistic spacetime in the description of particles dynamics. 

\subsection{Basic ideas about Elementary Cycles in time}\label{CA:EC}

An EC can be heuristically introduced as a ``particle moving on a circle'' of time period $T$. This way to introduces EC shows the analogy with 't Hooft Cellular Automata model, which is
 a fascinating deterministic model of QM proposed by G. 't Hooft and reviewed in \cite{Hooft:2014kka}, see also \cite{hooft-2009-0,'tHooft:2007xi,'tHooft:2006sy,'tHooft:2001ct,'tHooft:2001fb,'tHooft:2001ar,'tHooft:1998fa}. In particular an EC share analogies with a ``continuous periodic CA'' \footnote{A periodic CA describes permutations among a generic number $N$ of neighbour ``ontic'' sites on a circle. The time step of the permutations  $\delta t_P$  is of the order of the Planck time. It can be assumed infinitesimal for the scope of this paper.  That is, in a continuous periodic CA (\aka  as continuous cogwheel model),  the ``ontic'' sites can be approximated to a continuum ``ontic'' time $t$ of periodicity $T$, \cite{Hooft:2014kka}. We only retain the cyclic aspects (``CA fractional variables'') of CA and approximate to a continuum the discrete ones (``CA discrete variables''). Our analysis clearly indicates that the correspondence with QM comes essentially from the former aspects whereas the latter could be relevant for physics at the Planck scale, but not for ordinary QM.}. 
 Though ECs and CA share some phenomenology, they are independent theories as they are based on different hypothesis about spacetime.

The continuous periodic variable $t$ of period $T$ parametrizing such a cyclic motion of an ECs can be simply addressed as the relativistic \emph{time} coordinate. It will playing the role of the relativistic time of the theory; in 't Hooft terminology it is named \emph{``ontic'' time}.  In natural units $\hbar = c = 1$, it is natural to assume that to such a cyclic system of period $T$  a fundamental energy $\omega = 2 \pi / T $ is associated, according to the de Broglie phase harmony condition in the ``ontic'' time: $\omega T = 2 \pi $.

We shall discuss in details that an EC can be actually regarded as the effective description of a classical ``particle moving [very fast] on a circle'', similarly to CA. Such a ``particle on a circle'' can be assumed to be massless (e.g. a photon on a circle) in ECs physics (the length of the circle is $\lambda = T $, in natural units). Actually, the effective mass associated to an EC will result from the EC intrinsic rest periodicity $T_C$ according to the Compton relation $T_C = 2 \pi / m$. In other words ECs will encode the so called ``internal clocks'' of elementary particles \cite{Broglie:1924,2008FoPh...38..659C,Lan01022013}. 

More generally, we can define an EC as an elementary cyclic system characterised by intrinsic periodicity in time (and space); to each EC of period $T$ is associated a fundamental energy $\omega = 2 \pi / T $  according to the phase harmony relation  \cite{Dolce:cycles,Dolce:tune,Dolce:ADSCFT,Dolce:2009ce,Dolce:licata,Dolce:SuperC,Dolce:EPJP,Dolce:Dice2014,Dolce:TM2012,Dolce:Dice2012,Dolce:cycle,Dolce:ICHEP2012,Dolce:FQXi,Dolce:Dice,Dolce:2010ij,Dolce:2010zz,Dolce:2009cev4}. We must bear in mind that the fundamental topology of an EC is that of the circle, \ie $\mathbb S^1$. 

An EC can be represented --- at a statistical level --- by the so-called physical state $\Phi(t)$. The periodic temporal dynamics characterising an EC implies that EC physical state is therefore characterised by Periodic Boundary Conditions (PBCs) on the relativistic time $\Phi(t) = \Phi(t + T)$. Such cyclic dynamics of an EC can be equivalently described by  the infinitesimal evolution law $ t  \rightarrow  t + d t + \text{mod}~T $, in analogy with CA description, where  $d t$ is an infinitesimal (``ontic'') time interval. 

It is straightforward to see that such an temporal EC has much in common with the time evolution of a normally ordered Quantum Harmonic Oscillator (QHO) --- as also noticed by 't Hooft in the context of CA.  
To introduce this correspondence we may notice that the intrinsic periodicity (\ie PBCs of ECs theory) in the relativistic time of an EC determines (e.g. through discrete Fourier transform) a ``quantum'' number $n$ labelling energy eigenstates $\phi_n(t)= e^{-i \omega_n t}/\sqrt{2 \pi}$. It in turn form a complete, orthogonal set with harmonic energy spectrum $\omega_n = n \omega = n \frac{2 \pi}{T}$ ($n \in \mathbb Z$). Hence, every EC naturally defines an Hilbert space of basis $| n \rangle$ such that $\langle t | n \rangle\,  \dot = \, \phi_n(t)$  and induced inner product $\langle n | n' \rangle = \delta_{n,n'}$. Similarly to a vibrating string, an EC turns out to be represented  by  the superposition of its energy eigenstates (vibrational modes) $\Phi(t) \dot = \sum_{n \in \mathbb Z} \alpha_n \phi_n(t)$, where $\alpha_n$ are Fourier coefficients whose physical meaning will be interpreted later. A first consequence of this analysis is that it is possible to describe an EC, which is a classical system, in a corresponding Hilbert space representation. 

An EC, with its spectral composition, is uniquely associated to a point in the corresponding Hilbert space $|\Phi \rangle = \sum_n \alpha_n |n \rangle$.  In this formalism an EC also naturally defines a  Hermitian operator $\mathcal H$ as the operator in the Hilbert space such that $\mathcal H | n \rangle\, \dot = \, \omega_n | n \rangle$, which is manifestly Hermitian operator due to the PBCs of the theory. Since the temporal evolution of every EC eigenmode fulfils  $i \partial_t \phi_n(t) = \omega_n \phi_n(t)$, in this Hilbert space formalism follows that the EC time evolution is given by the Schr\"odinger equation $i \partial_t  |\Phi(t) \rangle = \mathcal H  |\Phi(t) \rangle$.  Hence,  the unitary Hilbert operator $\mathcal U(dt)\, \dot = \, e^{-i \mathcal H dt}$ describes the EC time evolutions.  

Notice the analogy between the ECs classical dynamics in time mentioned above and the time evolution of a normally ordered QHO of period $T$, \ie of quanta energy $\omega = \frac{2\pi}{T}$. As we will see in more detail, the physical meaning of such ECs dynamics can be actually summarised by 't Hooft's words: ``\emph{there is a close relationship between a particle moving [very fast \textit{N.A.}] on a circle with period $T$ and the Quantum Harmonic Oscillator (QHO) with the same period}" \cite{'tHooft:2001ct,'tHooft:2001ar}. This correspondence is a cornerstone of the ECs theory: QHOs are the building blocks of second quantised fields, and thus of the whole Quantum Field Theory (QFT).  We will use it as starting point to build our ECs theory. \emph{From the covariant generalisation of this EC we will in fact derive all the axioms of QM and the Feynman path integral}, as well as all the other fundamental aspects of QM. Indeed  ECs constitute the elementary oscillators from which ordinary QFT can be derived. 

  It must be noticed however that  $\mathcal H$  has negative eigenvalues (corresponding to $n = -1, -2, -3, \dots$) for ECs --- similarly to CA. This apparent problem has a simple solution in ECs theory.  We will prove that, in general, these negative modes exactly describe anti-particles in ECs theory (in agreement with  't Hooft's conjecture for fermionic CA), so that this is absolutely not an issue for relativistic (bosonic and fermionic) ECs. Furthermore $\mathcal H$ is always positively defined in the non-relativistic limit of ECs theory. %For these reasons ECs can be regarded as elementary quantum oscillators whose dynamics reproduce, as we will see, relativistic particles quantum behaviours. 

ECs evolution  can be expressed as a sum of Dirac deltas, \ie as a sum of classical paths, by means of the Poisson summation $\sum_{n \in \mathbb Z} f(n) e^{-i n \alpha} = 2 \pi \sum_{n \in \mathbb Z} F(\alpha + 2 \pi n')$, where $F$ is the Fourier transform of $f$ --- similarly to 't Hooft CA evolution.  Without loss of generality we assume for a moment that we assume for a moment a unitary EC, represented by the hat symbol $\hat \Phi$. By unitary EC we mean that  all the Fourier coefficients are unitary: $\alpha_n \equiv 1,  \forall n$. Let us consider the evolution $\hat \Phi(\Delta t)$ of an EC from an initial time $t_i$ to a final time $t_f$, where $\Delta t = t_f - t_i$.   Hence,    
\bea\label{sum:class:path:free}
\hat \Phi(\Delta t) &=& \sum_{n \in \mathbb Z} e^{-i n \omega \Delta t} = (2 \pi) \sum_{n' \in \mathbb Z} \delta\left( \omega\Delta t + 2  \pi n'\right) \nn  \\ &=& T \sum_{n' \in \mathbb Z} \delta(\Delta t +  n' T)\,.
\eea

This describes all the possible classical paths, labelled by the winding number $n'$, linking the initial and final times ``on a circle'' of period $T$. That is, the EC evolution is given by the interference of all the possible degenerate classical solutions allowed  by the PBCs of the ECs physical state $\Phi(t) = \Phi(t + T)$, or equivalently by the evolution law  $ t_i  \rightarrow  t_f + \text{mod}~T $. In agreement with the notation introduced above and considering that the cyclic behaviour concerns the EC under consideration, it is more appropriate to express the ECs evolution law in the Hilbert space notation, e.g. $| t_i \rangle  \rightarrow  | t_f + \text{mod}~T \rangle$ (see also 't Hooft notation).

Remarkably, the relativistic generalisation of eq.(\ref{sum:class:path:free}) will naturally lead to the equivalence between the ECs classical evolution and the Feynman path integral of ordinary QM.

It is also important to bear in mind  that an EC, essentially, can also be regarded as describing a classical (closed) string defined on the (``ontic'') time $t$ and vibrating with period $T$, \ie with fundamental angular frequency $\omega = 2\pi / T$.  As we will argue in sec.(\ref{God:dice}), if the time period is very small \emph{w.r.t.} the observer timekeeper, a ``particle on a circle'' can be actually formalised at an effective level as a particle  in a ``time box'' of ordinary QM. Similarly to a particle in a box or a vibrating string the ``quantisation'' is given by the BCs (the rigorous formulation of the theory is actually based on a formalism analogous to that of theories of with compact extra-dimensions and string theory, see for instance \cite{Dolce:2009ce}). 

We will see that the PBCs are the sole quantisation condition in ECs theory (no other quantisation conditions are necessary).  In the analogy with a vibrating string, the EC eigenstates $\phi_n(t)$ denote the harmonics, the EC energy spectrum $\omega_n$ denotes the positive and negative vibrational eigenfrequencies, and the Schr\"odinger equation is related to the square root of the string evolution law in time. It is not a chance that the Hilbert space was historically conceived in classical physics to describe harmonic systems. That is, contrarily to the common opinion, the Hilbert space notation can be used to describe (statistically) basic classical physical systems such vibrating strings. The effective description of an EC as classical strings vibrating in time with period $T$ (compact time coordinate with PBCs) is essentially the original approach to ECs theory used in foundational papers \cite{Dolce:2010ij,Dolce:2010zz,Dolce:2009ce}. 

Another analogy that will be extensively used in this paper, especially for the generalisation of the results obtained from free ECs to interacting ECs (\ie interacting particles), is that every free EC can be regarded as a relativistic reference clock: ``a relativistic clock is a phenomenon passing periodically through identical phases'' according to A. Einstein \cite{Einstein:1923,Pais:einstein}. The EC covariant formulation  and the geometrodynamical description of particle interactions, including gauge invariance, will be derived  in terms of relativistic clock modulations, in close analogy with Einstein's original derivation of  General Relativity (GR).  Remarkably, with this formalism we will be able to derive gauge interactions and QED directly from the geometrodynamics of the ECs coordinates, without postulating gauge invariance, thanks to the intrinsically compact nature of the relativistic spacetime coordinates of ECs theory.  

\section{Relativistic Elementary Cycles}\label{Relativistic:EC:model}

In this section we give the rigorous definition of ECs. The model is obtained by  generalising to a covariant form the time dynamics described in the introduction above. We will obtain in a very natural way the undulatory mechanics at the base of modern relativistic QM. Essentially we will investigate relativistic modulations of elementary cyclic dynamics, in analogy with relativistic clocks and the relativistic Doppler effect. 

\subsection{Rest Elementary Cycles  and the definition of mass}\label{EC:MASS}

We have anticipated that an EC is an elementary cyclic system. Let us imagine to observe such an EC in a generic inertial reference frame $ \mathtt{S}$, also denoted by the vector $\vec k$ --- it will be identified with the fundamental spatial momentum of the elementary particle described by the EC. According to relativity, the  EC time period $T(\vec k)$ and fundamental energy $\omega(\vec k)$ must be frame dependent. Furthermore in every reference frame they must satifies the phase harmony relation for the  time coordinate in that reference frame $\omega(\vec k) T(\vec k) = 2 \pi $ (they will be the zero component of contravariant and covariant four-vectors, respectively). 

On one hand QM tells us through the Planck constant $h$ that the energy is determined by the periodicity of a ``periodic phenomenon'' \cite{Broglie:1924} (\eg of a wave, of a phasor or, more in general of an EC) according to the phase harmony relation $\omega(\vec k) T(\vec k) =   2 \pi  $. On the other hand relativity tells us through the speed of light $c$ that the mass is fixed by the rest energy $\omega(0) = m$. Hence, by considering both relativistic and quantum physics, we have that in general the mass of a particle must be identified  to a rest periodicity $T(0)$, \ie to the so-called Compton periodicity $T_C ~\dot = ~ T(0)$, according to the Compton relation $T_C = 2 \pi/ m$. It is understood that in this paper the Compton periodicity (as well as the Compton wave-length) is intended in a general way, not necessarily limited to electrons. Every particle has its characteristic Compton periodicity $T_C = 2 \pi / m$ depending on its mass $m$.  

The mass $m$ of an EC is thus determined by the EC time period $T(0)$ in its rest frame according to $m = \frac{2 \pi}{T(0)}$, in agreement with relativistic and undulatory mechanics. This is the EC analogous of the Compton time $T(0) = T_C$ of an elementary particle of mass $m$. Essentially an EC of rest period $T_C$ encodes the so-called Compton clock or  de Broglie internal clock \cite{Broglie:1924,1996FoPhL,Lan01022013,Penrose:cycles} of an elementary particle of mass $m = \frac{2 \pi}{T_C}$.  We will show that the classical-relativistic periodic dynamics of an EC of rest period $T_C$ corresponds to the quantum dynamics of an elementary bosonic particle of mass $m$ (rest energy). Such a definition of mass allows ECs theory to encode undulatory mechanics directly into the spacetime geometrodynamics.

It is interesting to notice that such a description of rest mass offers a fascinating  way out to GR paradoxes, see for instance recent 't Hooft paper ``light is heavy''  \cite{thooft:heavy}. A massless particle, \emph{e.g.} a photon, moving at the speed of light on a circle of Compton length can be equivalently described at an effective level as an elementary system of rest mass $m = \frac{2 \pi}{T_C}$. This effective description of rest mass can be actually tested experimentally. It is the mechanism of generation of the effective mass of the elementary charge carriers (\ie the electrons) in carbon nanotubes, as described in detail in \cite{Dolce:SuperC,Dolce:EPJP}. This represents one of the possible applications of ECs theory.   

In this paper we will mainly concern about bosonic particles. Nevertheless in sec.(\ref{fermions}) we will show how a particular case of the ECs internal cyclic dynamics (that we will address as \emph{twisted}) can be imposed to reproduce the essential properties of  fermionic particles according to the Dirac equation. As aspected the resulting description will have much in common  with the \emph{zitterbewegung} in which the intrinsic periodicity associated to the Dirac dynamics is actually  determined by the mass in analogy with the Compton relation.  
  
  \subsection{Free Elementary Cycles spacetime dynamics}\label{COV:ECS}

In close analogy with de Broglie derivation of undulatory relativistic mechanics \cite{Broglie:1924}, if an EC of rest periodicity $T_C$ is observed in a generic inertial reference frame $\vec k$, the Lorentz transformation of the EC Compton time $T_C$  implies a spatial periodicity $\lambda(\vec k)$ in addition to the EC time periodicity $T(\vec k)$ in that reference frame.  

In particular the Lorentz transformation of the EC Compton time is $T_C = \gamma T(\vec k) -  \gamma \vec \beta \cdot \vec \lambda(\vec k)$, where $\gamma = 1 / \sqrt{1-\vec \beta^2}$ is the Lorentz factor. Indeed a consistent covariant description of cyclic dynamics implies that a spatial periodicity $\vec \lambda (\vec k)$ must be associated to the EC, in addition to the time periodicity $T(\vec k)$. This wave-length defines the \emph{(``ontic'') 3D space} $\vec x$ of the EC. It is therefore possible to introduce the EC spacetime period $\lambda_\mu = \{T, - \vec \lambda\}$ (we omit the $\vec k$ dependency) in the (``ontic'') spacetime $x_\mu = \{t, - \vec x\}$ defined by the EC.

 The Lorentz transformation given above  leads to the covariant phase harmony condition in the EC ``ontic'' spacetime. That is, we have the relativistic invariant $m T_C  = \gamma m T - \gamma m \vec \beta \cdot \vec \lambda  = \omega T - \vec k \cdot \vec \lambda = \omega_\mu \lambda^\mu = 2 \pi$, where actually $\omega_\mu = \{\gamma m, -\gamma \vec \beta m \} = \{\omega, -\vec k\}$ is the fundamental EC four-momentum according to the relativistic laws.  That is the four-momentum associated to the EC fundamental mode ($n=1$) is $\omega_\mu$. 

Similar to the definition of the EC fundamental energy $\omega (\vec k)$, we can now identify $\vec k$ as the fundamental momentum of the EC in that particular reference frame: it is in fact related to the spatial periodicity (wave-length) by the de Broglie relation $k_i = 2 \pi / \lambda^i$ with $i = 1, 2, 3$.

 By denoting the Lorentz transformation with $\Lambda^\mu_\nu$, in the new reference frame $x'^\mu = \Lambda^\mu_\nu x^\nu$ the resulting EC spacetime period and fundamental four-momentum are $\lambda'^\mu = \Lambda^\mu_\nu \lambda^\nu$ and $\omega'_\mu = \Lambda_\mu^\nu \omega_\nu$, respectively. Thus the EC satisfies in every inertial reference frame the invariance of the phase harmony condition $ \omega_\mu \lambda^\mu = \omega'_\mu \lambda'^\mu=  2 \pi$. 
 That is, $ \omega_\mu$ and $\lambda^\mu$ are dual quantities. They are two faces  of the same coin. They are in fact covariant and contravariant four-vectors, respectively.  
 
 As discovered by de Broglie  \cite{Broglie:1924} the relativistic undulatory mechanics resulting from a periodic phenomenon of rest periodicity $T_C$ can be used to represent a relativistic  particle of mass $m$ and momentum $\vec k$. Indeed we have found a correspondence with the \emph{undulatory mechanics} at the base of modern QM. For instance, in the reference frame $\vec k$, the fundamental harmonic $n = 1$ of the EC, denoted by the bar symbol, has therefore the familiar form of a relativistic wave $$\bar \phi_{\vec k}(x)= e^{-i (\omega(\vec k) t - \vec k \cdot \vec x )} = e^{-i \omega_\mu x^\mu}\,.$$
 Every EC can be regarded as a ``de Broglie periodic phenomenon'' \cite{Broglie:1924,1996FoPhL}.  In 't Hooft's terminology, these transformations of  reference frames define a class of Lorentz ``changeable'' of the EC of rest period $T_C$ that we will represent in a Fock space, sec.(\ref{FOCK:HILBERT}).  
%\emph{Notice the correspondence to undulatory mechanics} as originally formulated by de Broglie \cite{Broglie:1924}.
 %CA can be therefore regarded as ``de Broglie periodic phenomena'', \cite{Broglie:1924,1996FoPhL}. 
 
Now it is easy to see that  $T(\vec k)$ transforms from inertial reference frame to inertial reference frame according to the relativistic constraint $\frac{1}{T_C^2} = \frac{1}{\lambda^\mu} \frac{1}{\lambda_\mu} = \frac{1}{T^2} - \sum_{i=1}^3\frac{1}{\lambda^i}\frac{1}{\lambda_i}$. By means of the phase harmony relation (\ie through the Planck constant), it in fact corresponds to the relativistic relation $m^2 = k_\mu k^\mu = k_0^2 - |\vec k|^2$. 

For reasons that will be clarified later, here we have introduced the notation $k^\mu = \{k_0, -\vec k\}$ where $k_0 = \pm \omega(\vec k)$ and $\pm$ denotes the positive and negative frequencies, respectively. This notation is that typically used in the formalism of ordinary QFT.  Hence, we find  that the EC fundamental energy satisfies the relativistic dispersion relation of a relativistic particle of mass $m$: $\omega(\vec k) =  2 \pi /   T(\vec k)=  \sqrt{m^2 + \vec k^2}$.    
  
  Our covariant EC model shows that the relativistic generalisation of the evolution law of a free EC in the ``ontic'' spacetime can be written as  $| x^\mu_i \rangle \rightarrow | x^\mu_f + \text{mod}~\lambda^\mu \rangle$. That is a covariant generalisation of 't Hooft's CA evolution law in the ``ontic'' time. In a generic inertial reference frame the spacetime period $\lambda^\mu$ of the EC evolution is determined by the EC fundamental four-momentum $k_\mu$ according to undulatory mechanics: $k_\mu \lambda^\mu = 2 \pi$.  It actually represents the four-momentum of the  elementary particle described by the EC. 
  
  We will denote the physical state of an EC of fundamental momentum $\vec k$ as $\Phi_{\vec k}(x)$.  Its explicit form will be given in the next subsection. The EC evolution law is characterised by the relativistic contravariant PBCs for the physical state $\Phi_{\vec k}(x) = \Phi_{\vec k}(x + \lambda)$ in the reference frame $ \mathtt{S}$,	  where  we have suppressed the Lorentz index in the argument. 

\subsection{Elementary Cycle energy spectrum and free physical state} \label{EC:LorentzInV}

In the previous subsection we have derived the dispersion relation of the EC fundamental energy $\omega(\vec k)$ and the corresponding relativistic modulation of EC period $T(\vec k)$. However, due to the intrinsic periodicity (PBCs), an EC --- similarly to a CA --- is characterised by a whole energy spectrum which, for a free EC of fundamental momentum $\vec k$, is the harmonic energy spectrum $\omega_n(\vec k) = n \omega(\vec k)$. 

 The combination of the EC harmonic energy spectrum above and the relativistic transformation of the EC time periodicity $T(\vec k) = 2 \pi / \omega(\vec k) = 2 \pi / \sqrt{m^2 + \vec k^2}$, see above, leads to the EC energy spectrum dispersion relation $\omega_n(\vec k) = n \omega(\vec k) =  {2 \pi n} /{T(\vec k)} = n \sqrt{m^2 + \vec k^2}$ (with $n \in \mathbb Z$). 
  Notice that it coincides with the energy spectrum of an ordinary normally ordered, second quantised scalar field describing a particle of mass $m$ --- in QFT every scalar mode of energy $\omega(\vec k) = \sqrt{m^2 + \vec k^2}$ has normally ordered spectrum  $\omega_n(\vec k) :=: n \omega(\vec k)$, similarly to a QHO.  
  In analogy with QFT notations,  we can now write $n \in \mathbb N$ and associate the negative frequencies to the harmonics of the negative solution of $k_0 = \pm \omega(\vec k)$. 
  
Similarly to the energy spectrum, in the free case the harmonic momentum spectrum resulting from the EC spatial periodicity  $\vec \lambda$ is ${k_n}_i = n  k_i = 2 \pi n / \lambda^i$, with $i = 1, 2, 3$. As a consequence of the PBCs $\Phi_{\vec k}(x) = \Phi_{\vec k}(x + \lambda)$ the four-momentum spectrum of a free EC is thus given by the quantisation condition ${k_{\mu}}_n \lambda^\mu = {k}_n \cdot \lambda =  2 \pi n$, similarly to the Bohr-Sommerfeld quantisation. This is in perfect correspondence with ordinary QFT. For instance photons have massless dispersion relation $\omega = |\vec k|$ so that $\omega_n(\vec k) = n \omega_n(\vec k) $ implies $\vec k_n = n \vec k$.

It is important to bear in mind that the EC (``ontic'') spacetime period is the contravariant Lorentz projection of the EC Compton period: $\lambda^\mu = \Lambda^\mu_0 T_C$. Notice that, despite the fact we have periodicities in time and space,  the fundamental topology of the relativistic EC is still that of the circle, $\mathbb S^1$. An EC can be actually represented by a one-dimensional classical closed-string vibrating in a four-dimensional spacetime. This implies that both the EC energy and momentum spectra are denoted by the same quantum number $n$. In other words, as for ordinary QFT, the two spectra are not independent: the momentum spectrum is the  Lorentz projection of the energy spectrum. By means of the relativistic equations of motion, once that the mass is fixed, it is sufficient to know the periodicity (\ie the energy) in a given reference frame to derive the corresponding wave-length $|\vec \lambda|$ (\ie the spatial momentum) in the ``ontic'' space  \cite{Dolce:2009ce}.

  In this way we have proven that the free (scalar) EC of fundamental momentum $\vec k$ is effectively described by the wave packet of all the harmonics allowed by its spacetime periodicity $\Phi_{\vec k}(x) = \Phi_{\vec k}(x + \lambda)$: 
  \bea\label{field:mode}
\Phi_{\vec k}(x) &=& \sum_{n \in \mathbb Z} \alpha_n(\vec k) e^{-i {\omega}_{n} \cdot x}   = \sum_{n \in \mathbb Z} \frac{a_n(\vec k) {e^{-i {\omega}_{n} \cdot x}}}{\sqrt{(2\pi)^3}\sqrt{2 \omega_n({\vec k})}}   \\ \nn &=& \sum_{n \in \mathbb N} \left[  \frac{a_n(\vec k) e^{-i {k}_{n} \cdot x}}{\sqrt{(2\pi)^3}\sqrt{2 \omega_n({\vec k})}}  + \frac{a_{-n}(-\vec k) e^{i {k}_n \cdot x}}{\sqrt{(2\pi)^3}\sqrt{2 \omega_{-n}({- \vec k})}} \right] \,,
\eea
%\beq\label{field:mode}
%\Phi_{\vec k}(x) &=& \sum_{n \in \mathbb Z} \alpha_n(\vec k) e^{-i {\omega_{\mu}}_{n} x^\mu}   = \sum_{n \in \mathbb Z} \frac{a_n(\vec k) {e^{-i {\omega_{\mu}}_{n} x^\mu}}}{\sqrt{(2\pi)^3}\sqrt{2 \omega_n({\vec k})}}   \\ \nn &=& \sum_{n \in \mathbb N} \left[  \frac{a_n(\vec k) e^{-i {k_{\mu}}_{n} x^\mu}}{\sqrt{(2\pi)^3}\sqrt{2 \omega_n({\vec k})}}  + \frac{a_{-n}(-\vec k) e^{i {k_{\mu}} _n x^\mu}}{\sqrt{(2\pi)^3}\sqrt{2 \omega_{-n}({- \vec k})}} \right] \,,
%\eeq
where we have suppressed Lorentz indexes and we have  normalised over an infinite number of EC spatial periods, \ie infinite volume, analogously to QFT.  

We will identify the classical-relativistic EC physical state eq.(\ref{field:mode}) with the normally ordered second quantised mode of momentum $\vec k$ associated to an ordinary scalar field of mass $m = \omega(0)$.  In oder words the physical state $\Phi_{\vec k}(x)$ will turn out to describe a scalar quantum particle of mass $m = \omega(0)$ in the inertial reference frame $\vec k$. 

\subsection{The free Elementary Cycle scalar action} \label{EC:ACTION:SCAL}
 Here we prove that the free EC physical state $\Phi_{\vec k}(x)$, reported in eq.(\ref{field:mode}), and describing a quantum bosonic free particle of mass $m$ and momentum $\vec k$, is the classical solution of the action of a one-dimensional string vibrating in spacetime with periodicity  $\lambda^\mu$ and rest period $T_C = 2 \pi / m$. Indeed we have already pointed out the correspondence --- which will be formalised in the next section ---  between the EC free physical state $\Phi_{\vec k}(x)$ and the mode of momentum $\vec k$ of a quantised free Klein-Gordon field of mass $m$.

We must bear in mind that a free EC is characterised by a global spacetime period $\lambda^\mu$. Indeed --- see discussion in \cite{Dolce:licata} --- we can paraphrase Newton's first law: a free EC, having by definition constant energy-momentum $k_\mu$, is characterised by persistent (\ie constant and global) spacetime periodicity $\lambda^\mu$, according to the phase harmony condition.
 In the free case the spacetime period must be necessarily global (persistent periodicity), in the sense that it does not depend on the spacetime point at which the EC evolution is evaluated. It only depends on the reference frame according to the relativistic laws. 
 
 \emph{Vice versa}, as we shall see in sec.(\ref{INT}), the case of interacting ECs is characterised by \emph{local} ``ontic'' spacetime periods, \ie the EC period must depend on the spacetime point on which the interacting EC is located in order to encode  the local variation of four-momentum associated to the interaction. 

As pointed out at the end of sec.(\ref{COV:ECS}), the evolution law of a free EC,  $| x^\mu_i \rangle \rightarrow | x^\mu_f + \text{mod}~\lambda^\mu \rangle$, such that $k_\mu \lambda^\mu = 2 \pi$, is equivalently represented by  the contravariant PBCs $\Phi_{\vec k}(x) = \Phi_{\vec k}(x + \lambda)$ for the  free EC physical state $\Phi_{\vec k}(x)$ (we have suppressed the Lorentz index).  
In other words, the physical state of a free EC $\Phi_{\vec k}(x)$ can be equivalently defined as the solution of a scalar action defined in the flat, cyclic spacetime of period $\lambda^\mu$ in order to encode cyclic dynamics of rest periodicity $T_C$. The EC (``ontic'') spacetime coordinates are therefore represented as compact with compactification length $\lambda^\mu$ and contravariant PBCs at the boundaries in a four-dimensional Minkowskian (flat) metric. The ECs formalism has indeed many similarities with that of extra-dimensional theories and String Theory. 

The action defining a free scalar EC is therefore 
\beq\label{CA:action}
\mathcal S_{EC} = \int_{\lambda^\mu} d^4 x \mathcal L(\partial_\mu \Phi_{\vec k}, \Phi_{\vec k})\,,
\eeq   
where global contravariant PBCs are assumed at the boundaries: $\Phi_{\vec k}(x) = \Phi_{\vec k}(x+\lambda)$. For the free case, these PBCs are global, in the sense that they do not depends on the spacetime point on which the EC is located. In every point $x$ of the EC evolution the EC has (instantaneous) spacetime periodicity $\lambda^\mu$. 

The  PBCs are contravariant in the sense that they vary with the EC reference frame  according to the relativistic phase harmony. 
EC formulation eq.(\ref{CA:action}) is manifestly covariant. Under the Lorentz transformation of coordinates $x'^\mu = \Lambda^\mu_\nu x^\nu$, the action eq.(\ref{CA:action}) turns out to be $ \mathcal S_{EC} = \int_{\lambda'^\mu} d^4 x' \mathcal L(\partial_\mu \Phi_{\vec k'}, \Phi_{\vec k'}) $. The resulting boundary transforms in fact in a contravariant way, according to the global phase harmony condition $k'_\mu \lambda'^\mu = 2 \pi$. In the new inertial reference frame, the four-momentum is actually transformed to $k'_\mu = \Lambda_\mu^\nu k_\nu$ whereas the spacetime period is transformed to $\lambda'^\mu = \Lambda^\mu_\nu \lambda^\nu$. We can say that the Lorentz transformation corresponds to a local rotation of the boundary of the EC. 

 Due to the transformed PBCs, the quantisation condition transforms as  ${k'_{\mu}}_n \lambda'^\mu = 2 \pi n$, so that the (global) energy spectrum dispersion relation can be actually rewritten as ${k'_{0}}_n = {k_{0}}_{n}(\vec k') = n \sqrt{m^2 + \vec k'^2} $ ($n \in \mathbb N$). 
The PBCs of the action eq.(\ref{CA:action}), $\Phi_{\vec k}(x) = \Phi_{\vec k}(x+\lambda)$, yield  the quantised EC spectra described in the previous subsection. 

It is well known from string and extra-dimensional theories that PBCs (as well as Neumann and Dirichlet BCs) are admitted by relativistic bosonic actions, in the sense that they fulfil the variational principle at the boundary \cite{Dolce:2009ce}. This compatibility has a fundamental relevance for the consistency of the ECs theory. % \cite{Dolce:cycles,Dolce:cycle,Dolce:tune,Dolce:ADSCFT,Dolce:FQXi,Dolce:2010ij,Dolce:2010zz,Dolce:2009ce}.  

 The free bosonic EC is effectively an homogeneous one dimensional vibrating string (\ie vibrating in the Compton world-line with periodicity $T_C$) vibrating in spacetime with four-period $\lambda^\mu$.  The general EC solution is of the type $\bar \phi_{\vec k}(x) \propto e^{-i k_{ \mu} x^\mu }$ according to eq.(\ref{field:mode}), such that $k_\mu \lambda^\mu = 2 \pi$. By scalar action we mean that the corresponding Euler-Lagrange equation is the ordinary Klein-Gordon equation $(\partial_\mu \partial^\mu - m^2) \bar \phi_{\vec k}(x) = 0$ --- its explicit form  is given in  \cite{Dolce:tune,Dolce:ADSCFT,Dolce:2009ce}. Notice that these equations of motion actually encode a rest periodicity $T_C = 2 \pi / m$. Thus, due to the PBCs, the solution of  eq.(\ref{CA:action}) is the scalar EC eq.(\ref{field:mode}), \emph{q.e.d.} 

As we shall see, this functional formalism will be particularly convenient, for instance, when we will describe interactions. Notice also that in ECs  the problematics ``edge states'' of 't Hooft's CA models \cite{Hooft:2014kka} are manifestly vanishes as can be easily seen thanks to the explicit assumption of contravariant PBCs. Such a problematic aspect of CA is therefore solved by ECs physics. 

\section{Equivalence to canonical Quantum Mechanics: the free case}\label{QM:free}

 We are now able to prove the exact correspondence between the classical dynamics of a free EC  and the canonical QM of an elementary free bosonic particle. This means that we will exactly derive, for the free case (and then for the interacting case), all fundamental axioms of QM, the commutations rations, the Heisenberg uncertainty relation, the Feynman path integral, the Dirac quantisation rule, the Bohr-Sommerfeld quantisation, the WKB approximation, and so on. 
 
 Our strategy is the one learnt from Newton: we first consider the ieal case of free elementary system and then we generalise to interactions. In sec.(\ref{INT}) the exact correspondence will be then exactly extended  to the case of interacting systems, composite systems and thermal systems, by developing the formalism of ECs spacetime geometrodynamics, Euclidean time periodicity and tensor products of Hilbert spaces, respectively.  
 
 Notice that the \emph{quantum} behaviours will be directly derived from the EC relativistic \emph{classical} dynamics, without any further quantisation condition except  intrinsic periodicity (PBCs). 

\subsection{Axioms of Quantum Mechanics from Elementary Cycles}\label{AXIOMS}

We are now able to derive the axioms of QM from ECs classical dynamics in the free case. Additional details  are given in \cite{Dolce:licata}. 

\textbf{\emph{i) { Axiom of the states}}}

In the covariant formulation of ECs theory introduced above, see eq.(\ref{field:mode}), the EC momentum eigenstates ${\phi_{  \vec k}}_n (\vec x) \, \dot = \, e^{ i \vec k_n \cdot \vec x}$ constituting the EC physical state $\Phi_{\vec k}$, as a direct consequence of the ECs spacetime evolution law, \ie of the PBCs, form a complete, orthogonal set. A free EC indeed naturally defines a corresponding Hilbert space of basis $| n_{\vec k} \rangle$, such that $\langle \vec x | n_{\vec k} \rangle \, \dot = \, {\phi_{\vec k}}_n (\vec x)$, with induced inner product defined as $\langle n_{\vec k} | n'_{\vec k} \rangle\, \dot = \, \int d^3 x e^{- i \vec k_n \cdot \vec x} e^{ i \vec k_{n'} \cdot \vec x } / (2 \pi)^{3} = \delta_{n_{\vec k}, n'_{\vec k}}$. The completeness relation is $\sum_n | n \rangle \langle n | = 1$ and the overlap between the position operator and the eigenstate of the momentum operator is $\langle \vec x | \vec k \rangle = e^{i \vec k \cdot \vec x}$. 

Hence the free EC is described by a point in the corresponding Hilbert space 
\bea |\Phi_{\vec k}\rangle &=& \sum_{n_{\vec k} \in \mathbb Z} \alpha_n(\vec k) |n_{\vec k} \rangle \nn \\ &=& \sum_{n \in \mathbb N} \left [\alpha_n(\vec k) |n_{\vec k} \rangle + \alpha_{-n}(-\vec k) |-n_{-\vec k} \rangle \right]\eea such that $\langle x |\Phi_{\vec k}\rangle = \Phi_{\vec k}(x) $; \emph{q.e.d.}

\textbf{\emph{ii) Axiom of the observables}} 

In this Hilbert space notation, we find that the EC dynamics define, through its quantised energy and momentum spectra, corresponding  Hamiltonian and momentum operators $\mathcal H$ and $\vec{\mathcal P}$ such that $\mathcal H(\vec k) |n_{\vec k} \rangle\,\dot =  \, \omega_n (\vec k) | n_{\vec k} \rangle$ and \\ $\vec {\mathcal P}  |n_{\vec k} \rangle \, \dot = \, \vec k_n (\vec k) | n_{\vec k} \rangle$, respectively. These operators, implicit in the EC formulation  eq.(\ref{field:mode}), are manifestly Hermitian due to the PBCs. 

The ``observables'' of an EC can be therefore always written as functions of Hermitian operators. The eigenvalues of these operators in turn describe the only possible values of the related physical quantities admitted for the EC eigenvalues (e.g. a free relativistic EC has quantised energy $\omega_n({\vec k})$, eigenvalues of the Hamiltonian operator $\mathcal H$, in analogy with the discrete frequencies of a classical vibrating string);   \emph{q.e.d.} 

\textbf{\emph{iii) Axiom of the motion}} 

From the  free EC solution eq.(\ref{field:mode}) it easily follows  that every EC eigenmode (harmonic) satisfies the equation $i \partial_\mu {\phi_{ \vec k  }}_n (x) = {k_{\mu}}_n   {\phi_{ \vec k  }}_n(x)$. In the related Hilbert space formalism  this means that  the time evolution of the EC eq.(\ref{field:mode}) turns out to be described by the  ordinary Schr\"odinger equation $$i \partial_t |\Phi_{\vec k}(x)\rangle = \mathcal H(\vec k) |\Phi_{\vec k}(x)\rangle$$ (the Hamiltonian in this case is not time dependent because, actually, we are in the free case). Similarly, the spatial evolution is given by $i \vec \partial_{\vec x} |\Phi_{\vec k}(x)\rangle = - \vec{\mathcal P}(\vec k) |\Phi_{\vec k}(x)\rangle$. In particular the unitary time evolution operator is $\mathcal U(d t) =e^{- i \mathcal H dt}$ and the spatial evolution operator is $\mathcal U(d \vec x) =e^{ i \vec{\mathcal P} \cdot d \vec x}$;   \emph{q.e.d.} 

\textbf{\emph{iv) Axiom of the measurement}} 

 Besides the mathematical relevance of the Born rule which we are going to derived here, in QM the axiom of the measurement has a particularly important interpretational and conceptual meaning which we will interpret in terms of ECs physics here and in sec.(\ref{God:dice}).  

To derive the Born rule  we must again consider that --- similarly to a covariant generalisation of 't Hooft's CA  --- an EC describes statistically and in a covariant way the dynamics of a single ``particle moving [very fast \emph{A/N}] on a circle''.   If the EC time period $T$ is very small \emph{w.r.t.} the time resolution of the observer's timekeeper, the only possible description of the ECs dynamics is statistical. 

Actually, the time scale of the ECs periodicities are determined by the mass of the corresponding particles. Since $\omega(\vec k) \geq m$, the upper bound of the EC time period $T(\vec k)$ is given by the Compton period $T_C$. This means that, even considering light particles\footnote{In the case of neutrinos the internal periodicity is very slow and, actually, the neutrinos periodicities (oscillations) are experimental manifestations of neutrinos masses.} such as electrons (or even electrodynamic phenomena),  the cyclic EC dynamics are always faster than any modern timekeeper resolution. The electron Compton time is about $10^{-21}$ s whereas the modern timekeepers resolution is ``only'' $10^{-17}$ s (it is however increasing very fast towards the electron Compton time, see \cite{Dolce:EPJP} for possibles indirect observations of the Compton clock). 

Let us consider the example of a rolling die, further discussed in sec.(\ref{God:dice}). A \emph{rolling die} can be actually regarded as an  EC whose time dynamics are on a periodic temporal lattice of 6 sites with very small time interval with respect the observer temporal resolution denoted by $\delta t$ (it can also be regarded as a CA with 6 states on a circle and permutations among neighbour). Let us suppose that the period $T$ is unknown. If the die rolls slowly or it is observed with sufficient resolution in time, the observer can resolve the die motion. However, if the die rolls very fast \emph{w.r.t.} the observer resolution in time (or if it is observed under a stroboscopic light, see Elze's stroboscopic quantisation \cite{Elze:2005gv}), only a statistical prediction of the outcomes is possible. 

For instance, let us suppose that when the timekeeper time is $ t_1$ the die shows a given face, and at time $t_1 + T $ it shows again the same face. The observer can only say that the frequency is $\omega_ 1=  \omega$, or $ \omega_ 2=  2 \omega$, or $\omega_ n = n  \omega$ with $n \in \mathbb Z$. Thus its evolution is that of a ``periodic phenomenon'' described by the superposition of the eigenstates $e^{- i \omega_n t}$, forming a complete, orthogonal set with eigenvalues $\omega_n = n \omega$. Furthermore the fast rolling die can be represented, as long as we do not observe it, as the superposition of the states of its six faces (similarly a very fast flipping coin is the superposition of the state $|\text{head}\rangle$ and the state $|\text{tail}\rangle$). 
Due to the finite resolution of the timekeeper the die time period $T = 2 \pi / \omega$ can only be determined with an experimental uncertainty $\delta t$. The observer uncertainty in time implies an simultaneous uncertainty in the angular frequency $\delta \omega$ (\ie in the ``energy''). This simultaneous uncertainty is actually described by the Heisenberg uncertainty relation, as we will show in sec.(3.1). 

In addition to the analysis given for the previous axioms and in sec.(\ref{Relativistic:EC:model}), this example wants to illustrate in a naive way that, in analogy with 't Hooft's analysis of CA, the evolution of a die rolling very fast can be described in a corresponding Hilbert space with associated Hamiltonian operator and Schr\"odinger equation, in close analogy with QM. According to ECs theory, this correspondence can be interpreted in the following way. Similarly to the example of the die, if the system is characterised by very fast cyclic dynamics, as for ECs, \emph{w.r.t.} the observer resolution in time, only a statistical description of the outcomes can be given.  This statistical description yields a Hilbert space notation and the other correspondences to QM (frequency eigenstates, Schr\"odinger equation, unitary evolution, etc). %As well show below, the experimental indetermination in time $\Delta t$ implies an indetermination on the angular frequency $\Delta \omega$ , these two indeterminations are turns out to be related  by the Heisenberg uncertainty relation  $\Delta \omega \Delta t \geq \pi / 2$. 

Finally, in order to derive the axiom of the measurement and the Born rule it is convenient to consider the example of an electric current. As well-known the motion of electrons in an electric circuit is typically described statistically due to the large number of electrons constituting the electric current even though they move very slowly. Such a statistical description is given by means of a ``wave function'' (\ie by a phasor), which is the analogous of the EC physical state, describing the density of electrons $\rho(x)$, \ie the density of charges, and satisfying the continuity equation. 

Obviously, such a statistical description can be extended to a current of neutral particles. In this case $\rho(x)$ simply describes the density of particles. Furthermore, such a  statistical description can be generalised to ECs --- or CA ---, \ie to a  single ``particle moving [very fast] on a circle''. In the case of an EC, even though we have a single ``particle on a circle'', \ie in our circuit, the statistical description is necessary due to the fact that the particle moves very fast. If the period is very small \emph{w.r.t.} the experimental resolution in time, it will be not possible to determine the exact position of the particle at a given time, so we can only describe the particle motion statistically. That is described by a wave-function defining  the probability and current densities, and satisfying the continuity equation.

The EC physical state  $\Phi_{\vec k}(x)$ is actually a wave-packet of solutions of the Klein-Gordon equation. It therefore satisfies the continuity equation  $\partial_t \rho(x) = - \vec \partial_{\vec x} \cdot \vec j(x)$ where $\rho(x) = |\Phi_{\vec k}(x)|^2$ describes the probability density to find in $x$ the single particle ``on a circle'' associated to the EC, in analogy with the phasor of an electric current. %Typically,  in the case of charged particles in a circuit, $\rho(x)$ describes the density of electric charges, but this can be generalised to neutral particles or, in the case of ultra-fast dynamics, to a single particle in the circuit (circle).   
Similarly, $\vec j(x)$ describes the current of probability associated to the EC (for the sake of simplicity, since the particle must be necessarily stopped in the detector when observed we can assume here a non-relativistic continuity equation \cite{Nikolic:2008sn}). 

We conclude that for an EC the current of probability is constituted by a ``single particle on a circle'', so that the integral of the probability density $\rho(x)$ over the whole space (\ie over an infinite number of periods along the ``circle'') is unitary: $$ \int |\Phi_{\vec k}(x)|^2 d^3 x \equiv 1\,.$$ We conclude that the statistical description of an EC, \ie ``a particle moving [very fast] on a circle'' leads exactly to the ordinary Born rule of QM; \emph{q.e.d.}

\subsubsection{Comments about axiomatic Quantum Mechanics} 

\emph{We have demonstrated mathematically the complete, exact equivalence between the free ECs classical dynamics and the axiomatic formulation of  QM for the free (bosonic) case}.  This result will be generalised to interactions in sec.(\ref{INT:QM}).  Obviously the exact equivalence with the axioms of QM is very important. The axioms of QM constitute the base from which all the known results of quantum physics can be derived. Here we have derived them from the classical ECs dynamics. Hence we are allowed to state  that --- at least --- EC is consistent with all known results of QM.

Besides the generalisation to the interacting case, we will also derive from ECs classical dynamics all the other (secondary) aspects such as the Feynman path integral (obtained independently from the axioms above), the commutation relations (relevant for the Dirac quantisation, the Fock space and the Heisenberg relation), the product of Hilbert spaces (necessary to describe composite systems and Bell's experiment),  the spin-statistics, the quantisation of statistical systems, etc.   
 
Since  in an EC the periodic motion along the ``circle'' (\ie the circuit) is constrained to satisfy the PBCs $\Phi_{\vec k}(t) = \Phi_{\vec k}(t + T)$ (and thus discrete eigenmodes), we have found an additional justification of the fact that an EC, \ie  a ``particle moving [very fast] on a circle'', can be effectively represented as a classical one dimensional (scalar) string vibrating with period $T$, and related harmonics.

\subsection{Commutation relations  from Elementary Cycles}\label{COMM:REL} 

In order to derive the commutation relations of QM directly from ECs cyclic dynamics let us evaluate the expectation value of the partial derivative of an Hermitian observable $\partial_{\vec x} \mathcal F(\vec x)$ among arbitrary initial and final ECs physical states --- notice that the expectation value describes the ``average'' value associated to an EC physical observable, \eg the energy, in perfect correspondence with ordinary QM.   

Through integration by parts and keeping in mind our definition of the momentum operator $\vec{\mathcal P}$ as well as of the inner product associated to the EC, it is easy to see that \cite{Dolce:2009ce,Dolce:licata} 
\bea \langle \Phi_{fin}| \vec \partial_{\vec x} \mathcal F(\vec x) | \Phi_{in}\rangle &=& i \langle \Phi_{fin} | \vec{\mathcal P} \mathcal F(\vec x) - \mathcal F(\vec x)  \vec{ \mathcal P} | \Phi_{in}\rangle \nn \\ &-& [\Phi_{fin}(\vec x) \mathcal F(\vec x) \Phi_{in}(\vec x) ]_0^{\vec \lambda}\,.
\eea  
It is important to notice the fundamental role of the PBCs in this demonstration: the boundary term of this expectation value vanishes as  direct consequence of the periodicity $\vec \lambda$ for the EC ``ontic'' spatial coordinate $\vec x$. Hence we exactly \emph{obtain}, for arbitrary initial and final  EC physical states $\Phi_{in}$  and  $\Phi_{out}$ (in this form the demonstration can be easily generalised to the interacting case), the same commutation relations of canonical QM: 
\beq\label{comm:rela:gen}
[ \mathcal F(\vec x),  \vec{\mathcal P} ] = i \vec \partial_{\vec x} \mathcal F(\vec x)\,. 
\eeq 
Indeed, by assuming $\mathcal F(\vec x) = x_i$  we find 
\beq\label{comm:rela:space}
[ x_i,  \mathcal P_j ] = i \delta_{i,j} \,, 
\eeq
 where $i,j = 1, 2, 3$. In a similar way it is possible to derive the commutation relation between time and Hamiltonian operator: $[ t,  \mathcal H ] = i $. 
 
 From these fundamental commutation relations of the canonical quantities, the commutation relation for all the derived observables can be inferred.   For more details see for instance \cite{Dolce:2009ce}. We conclude that \emph{the commutation relations of QM are implicit in ECs periodic dynamics}. In turn the so-called Dirac quantisation rule or the second quantisation are implicit as well, see below. 
 
 Paraphrasing Dirac quantisation rule we have proven that that, if the commutation relation of the physical observables $A$ and $B$ of a (free) classical particle is described by the Poisson bracket $\{A, B\}_P$, the related EC is described by the commutation relation $$[\mathcal A, \mathcal B] = i \{A, B\}_P$$ where $\mathcal A$ and $\mathcal B$ are the Hilbert operators associated to the EC observables, in perfect agreement with ordinary QM.  

 Notice that this demonstration is the straightforward generalisation to ECs physics of Feynman's derivation of the commutation relations from the path integral \cite{Feynman:1942us,Feynman:1948art} proving the equivalence between the path integral formulation of QM and the axiomatic one. Thus, if --- as it is --- mathematics is not an opinion (if, according to Feynman ``the same equations have the same solutions''), our demonstration proves that ECs theory is equivalent to both axiomatic and Feynman formulations of QM. Indeed we will also be able to derive, in an independent way \textit{w.r.t} demostration given above, that the Feynman path integral directly from ECs classical dynamics, in either the free and interacting cases, see par.(\ref{FPI:EC}).  

\subsection{Heisenberg uncertainty relation  from Elementary Cycles}

The Heisenberg uncertainty relation is, in general, a direct consequence of the commutation relations eqs.(\ref{comm:rela:gen}-\ref{comm:rela:space}). The latter has been derived from ECs dynamics, hence we can safely say that the Heisenberg uncertainty relation is implicit in ECs theory, in perfect analogy with ordinary QM. 

Alternatively --- as further crosscheck --- the Heisenberg uncertainty relation can be intuitively inferred directly in terms of ECs periodic dynamics. We consider, for instance, that only the modulo square $|\Phi_{\vec k}(x_i)|^2$ has a physical meaning (Born rule). Due to the periodic dynamics, the phase of an EC physical state is defined modulo phase factors. This is a direct consequence of the PBCs or equivalently of the evolution law (thus it can be generalised to 't hooft CA).  Without loss of generality here we only consider phase factors $n \pi$ and $n=1$ because different phase factor give weaker uncertainty relations. 

Without loss of generality let us  consider  the $i$-th ``ontic'' spatial coordinate of period $\lambda_i$. As already said, due to the fast periodic dynamics, an observer can only describe the EC statistically. In particular the observer can only interpret the invariance of the EC phase by factors $\pi$'s as a simultaneous indetermination in the momentum and position: $|e^{i k_i x^i }| =|e^{i (k_i x^i + \pi)}| =  |e^{i [(k_i + \delta k_i)  x^i]}| = |e^{i [k_i  (x^i + \delta x^i)] }| $ where $\delta k_i  x^i= \pi $ and $k_i  \delta  x^i = \pi $ ($k_\mu$, or equivalently $\lambda^\mu$, and $x^\mu$ are unknown to the observer). 

Now, by considering that the ``ontic'' spatial coordinate is periodic, \ie $0 < x^i + \text{mod}(n' \lambda^i) \leq  \lambda^i$, with $n' \in \mathbb N$ (for the sake of simplicity we can assume $x^i \in (0, \lambda^i]$ similarly to angular variables), and that the phase harmony condition is $ k_i  \lambda^i = 2 \pi$, we find that the simultaneous indeterminacy between momentum and space is described by the ordinary Heisenberg relation: $$  \delta k_i \delta x_i = \frac{\pi^2}{ k_i  x_i} \geq \frac{\pi^2}{k_i \lambda_i} = \frac{\pi}{2}\,.$$ Generalising this demonstrations we have the other Heisenberg uncertainty relations $ \delta k_j \delta x_i \geq \delta_{i,j} \pi / 2$ and, for the temporal component, $\delta \omega \delta t \geq \pi / 2$ \cite{Dolce:cycles,Dolce:2009ce,Dolce:2009cev4,Nielsen:2006vc}, \emph{q.e.d}. %An alternative interpretation can be given by mean of the Feynman Path Integral description of EC evolution, see below.

\subsection{Second quantisation from Elementary Cycles}\label{sec:quant}

It is well known that the commutation relations of the QHO ladder operators are a direct consequence of the commutation relations of QM eqs.(\ref{comm:rela:gen}-\ref{comm:rela:space}). Since the latter are implicit in ECs physics, the commutation relation of the ladders operators are  implicit in ECs cyclic dynamics as well. The full derivation of the QHO in ECs theory is given in \cite{Dolce:Dice2012,Dolce:cycle,Dolce:Dice,Dolce:2009ce,Dolce:2009cev4} and summarised in sec.(\ref{nonrel:prob}). 

In terms of the correspondence between EC and QHO this means that, by means of the definition of the position and momentum operators in ECs theory, to the Fourier coefficients $a_n(\vec k)$ and  $a_{-n}(-\vec k)$ of eq.(\ref{field:mode}) is possible to associate the corresponding ladder operators defined as,respectively,  \bea \hat a(\vec k) &=& \sqrt{\frac{\omega(\vec k)}{2}} x +  \frac{i}{\sqrt{2 \omega(\vec k)}} | \vec {\mathcal P} | \nn \\  \hat a^\dagger(\vec k) &=&   \sqrt{\frac{\omega(\vec k)}{2}} x -  \frac{i}{\sqrt{2 \omega(\vec k)}} | \vec {\mathcal P} |\,Ê.\eea
%$\hat a(\vec k) | n \rangle = a_n(\vec k) | n \rangle$ and, according to the normalisation chosen in eq.(\ref{field:mode}),   $\hat a^\dagger(\vec k) | n \rangle = a_{-n}(-\vec k) | n \rangle$, and $\hat a^\dagger(\vec k) \hat a(\vec k) | n \rangle = n | n \rangle$.  
Due to the commutation relations of QM derived above directly from the effective description of the ECs ultra-fast cyclic dynamics in the Hilbert notation, these ladder operators play the role of  the creation and annihilation operators of ordinary QM. That is, they satisfy the commutation relation $[ a(\vec k),  a^\dagger(\vec k') ] = \delta(\vec k - \vec k')$. Indeed they describe the creation and annihilation of the EC $n$-th harmonic mode. Then, it is also possible to define the ECs ``vacuum'' state $|0\rangle$ from which all the possible ECs harmonics are created, as $| n_{\vec k} \rangle = \frac{{a^\dagger}^n (\vec k)}{\sqrt{n !}} | 0 \rangle$. Similarly the normally ordered Hamiltonian of the EC can be rewritten as $\mathcal H (\vec k) = \omega (\vec k)  a^\dagger(\vec k) a(\vec k) $.

 The free solution of an EC of momentum ${\vec k}$, denoted by the physical state $\Phi_{\vec k}(x)$, is therefore equivalent to the (normally ordered) \emph{second quantised} mode of momentum $\vec k$ of an ordinary  Klein-Gordon field $\Phi_{KG}(x)$ of mass $m$. A second \emph{quantised} scalar field $\Phi_{KG}(x)$ of mass $m = 2 \pi /T_C$ is  in fact the  EC physical state $\Phi_{\vec k}(x)$  integral over all the possible fundamental momenta $\vec k$:   $\Phi_{KG}(x) = \int d^3 k \Phi_{\vec k}(x)$. That is, in analogy with QFT we have that the physical state operator associated to eq.(\ref{field:mode}) is 
 \beq 
 \Phi_{\vec k}(x) = \frac{1}{\sqrt{(2\pi)^3 2 \omega({\vec k})}} \left [ a(\vec k) e^{-i k \cdot x} +  {a^\dagger(\vec k)} e^{+i k \cdot x} Ê\right ] \, . 
 \eeq 
 As we will see, the mathematical tool which actually describes such an integration of physical states of an EC is the Fock space. 
 
 By using 't Hooft terminology we say that a second quantised field is the ensemble of all the Lorentz ``changeables'' obtained by transforming the EC to all the possible inertial reference frames. This shows the equivalence between the scalar EC dynamics and second quantised free Klein-Gordon fields. 
 
 In the case of neutral scalars we see that the positive EC modes are indistinguishable from the negative ones. Actually, neutral bosonic particles and antiparticles are identical in ordinary QM. Therefore,  the EC Hamiltonian can be regarded as positively defined. The demonstration goes like that showing that in the Kaluza-Klein theory there are not tychionic modes despite the fact that Kaluza-Klein modes can have positive and negative frequencies \cite{Dolce:tune,Dolce:2009ce,Dolce:ADSCFT}. This shows that --- at least in the neutral case --- the non-positively defined Hamiltonian operator in ECs theory is not an issue, solving such problematic aspects of CA models.  We will see that in general, as conjectured by 't Hooft, the negative solutions correspond to antiparticles.    

\subsection{Equivalence between Elementary Cycles classical evolution and Feynman path integral}\label{FPI:EC} 

In addition to the exact equivalence with the axioms of QM, we now rigorously prove that \emph{the classical evolution of an EC is equivalent to the quantum evolution prescribed by the ordinary Feynman Path Integral} for the corresponding elemenetary particle \cite{Dolce:cycles,Dolce:tune,Dolce:ADSCFT,Dolce:2009ce,Dolce:licata,Dolce:SuperC,Dolce:EPJP,Dolce:Dice2014,Dolce:TM2012,Dolce:Dice2012,Dolce:cycle,Dolce:ICHEP2012,Dolce:FQXi,Dolce:Dice,Dolce:2010ij,Dolce:2010zz,Dolce:2009cev4}. Again, the covariant PBCs of the ECs theory play a central role in the demonstration. The demonstration will be given for the free case and then  generalised to interactions in sec.(\ref{INT:QM}). 

As we have already seen in sec.(\ref{AXIOMS}), in the Hilbert space formalism the evolution of a free EC in its (``ontic'') time is given by the operator $\mathcal U(dt) = e^{- i \mathcal H dt} $ which is an unitary, hermitian operator. So the evolution from an initial time $t_i$ to a final time $t_f$  can be written as the product of the elementary time evolutions of infinitesimal duration $\epsilon$:
$$\mathcal U(t_f; t_i) =  \prod_{j=0}^{N-1} \mathcal U(t_f + t_{j+1}; t_i - t_j - \epsilon)$$ 
with $N\epsilon = t_f - t_i$, $N \rightarrow \infty$ and $\epsilon \rightarrow 0$. 

By plugging the orthogonality relation associated to the EC inner product in between these elementary time evolutions we obtain that the EC evolution between two generic spacetime points $\mathcal Z = \mathcal U(\vec x_f, t_f; \vec x_i, t_i)$ is the product of elementary spacetime evolutions: 
\begin{equation}	\label{FPI:elem.paths}
\mathcal Z \! = \!\! \int \!\! \left  (\prod_{m=0}^{N-1}  \!\! d^3 x_m \!\! \right ) \!\! \mathcal  U(\vec x_f, t_f; \vec x_{N-1}, t_{N-1}) %U(\vec x_{N-1}, t_{N-1}; \vec x_{N-2}, t_{N-2}) 
\dots \mathcal  U(\vec x_1, t_1; \vec x_i, t_i)\,,
\end{equation} 
where the EC elementary spacetime evolutions are: 
\beq\label{single:elem:path}
\mathcal  U(\vec x_{m+1}, t_{m+1}; \vec x_m, t_m) = \langle \hat\Phi_{\vec k}| e^{-i [\mathcal H(\vec k) \Delta t_m - \vec {\mathcal P} \cdot  \Delta \vec x_m]} | \hat\Phi_{\vec k} \rangle\,,
\eeq
and $ \Delta t_m = t_{m+1} - t_m$, $\Delta \vec x_m =  \vec x_{m+1} - \vec x_{m}$, and $| \hat\Phi_{\vec k} \rangle = \sum_{n \in \mathbb Z} | n_{\vec k} \rangle$ (unitary EC physical state, $\alpha_n \equiv 1$, $\forall n$).

By construction, the phase of these elementary evolutions (\ie the phase of the EC physical state)  defines an action which turns out to be formally the classical action $ \mathcal S^{Classic}$ of the free classical-relativistic particle of mass $m$ and momentum $\vec k$ associated to our EC:
 $$ \mathcal S^{Classic}[t_f; t_i] = \int_{t_i}^{t_f} dt (\vec {\mathcal P} \cdot \dot{\vec x}  - \mathcal H )\,. $$
 In fact, from the phase of the EC physical state, \ie from the phase of the elementary evolutions eq.(\ref{single:elem:path}), we obtain $  \vec{\mathcal P} \cdot \vec \Delta \vec x_m -\mathcal H \Delta t_m  = (\vec{\mathcal P} \cdot  \vec{ \dot x}_m - \mathcal H) \Delta t_m = \mathrm L^{Classic} \Delta t_m = \Delta \mathcal S^{Classic}_m,$ where the classical Lagrangian is $\mathrm L^{Classic} = \vec{\mathcal P} \cdot  \vec{ \dot x}_m - \mathcal H$ with $ \vec{ \dot x}_m = \frac{\Delta \vec{ x}}{\Delta  t_m}$. 
 Notice that, contrarily to the EC action eq.(\ref{CA:action}), this new classical action associated to the EC is defined on a non-compact spacetime (it is not subject to PBCs), it is defined on the ordinary relativistic spacetime. The PBCs are encoded on the quantised spectra of the Hilbert operators.

 Finally, by putting all these elements together we obtain the remarkable result that \emph{the EC  classical evolution is exactly described by the ordinary Feynman Path Integral of a relativistic scalar particle of mass $m$}:
\beq\label{FPI:free}
\mathcal Z = \int \mathcal D^3 x e^{i \mathcal S^{Classic} [t_f; t_i]}\,.
\eeq
Thus we have proven that, in the free case, the classical evolution of an EC is equivalent to the quantum evolution prescribed by the Feynman path integral.  The interpretation of this exact correspondence will be discussed in sec.(\ref{inter:FPI}). 

By following the same steps of ordinary QFT, that is, by redefining the Hamiltonian operators and coordinates in terms of fields, it is now possible to generalise the Feynman path integral functional to QFT (e.g. to scalar fields).  This substitution has a particular physical interpretation in ECs theory: an EC can be regarded as a string vibrating in spacetime with corresponding fundamental periodicities, or equivalently as vibrations of spacetime itself so that the spacetime coordinates can be substituted with fields (physical states) encoding the harmonics modes of the spacetime vibrations: $x \rightarrow \Phi_{\vec k}(x)$.

 \subsubsection{Further proofs of the  exact correspondence between Feynman path integral and Elementary Cycles evolution}\label{further:proof:FPI}

We have just proven that the classical evolution of an EC is exactly described by the Feynman path integral. Notice that we have already provided, with an independent demonstration, the equivalence with axiomatic QM. 

As a further confirmation, among the others, we now independently prove that, \emph{vice versa}, the ordinary Feynman path integral of a relativistic free particle yields the characteristic cyclic evolution in spacetime of an EC. We will derive the covariant evolution law of a free relativistic EC, characterised by cyclic dynamics, directly from the ordinary Feynman path integral of a free relativistic particle. That is, we prove that ECs physics is already contained implicitly in Feynmann formulation of QM. In addition to the demonstrations reported below we mention that this correspondence  can be also derived graphically with interesting analogies to Feynman chessboard as reported in, e.g., \cite{Dolce:cycle,Dolce:2009ce}, as well as to the Feynman educational description of QED, \cite{feynman1990q}.  

The ordinary Feynman path integral \emph{can be always expressed as an integral (sum) of Dirac delta functions}. These turn out to describe all the degenerate classical paths with different winding numbers associated to the ECs cyclic dynamics.  The proof is very simple and it is based on the Poisson summation already introduced in this paper. It was first reported on sec.(4.1) of the first versions (arXiv versions $1$ to $4$) of the foundational paper {\cite{Dolce:2009ce}. The form reported here, though it is given for the free case, can be generalised to interacting particles, as we will see. 

In ordinary QM the Feynman path integral describing the ordinary quantum evolution of a free relativistic particle from an initial spacetime point $x_i$ to a final spacetime point $x_f$ is formally given by eq.(\ref{FPI:free}). It is  well known that the ordinary Feynman path integral can also be written as  eq.(\ref{FPI:elem.paths}), 
%\begin{equation}
%\mathcal Z =  \int \left ( \prod_{j=0}^{N-1}  d^3 x_j \right )  U(\vec x_f, t_f; \vec x_{N-1}, t_{N-1}) %U(\vec x_{N-1}, t_{N-1}; \vec x_{N-2}, t_{N-2}) 
%\dots U(\vec x_1, t_1; \vec x_i, t_i)
%\end{equation}
where the elementary Feynman spacetime evolutions are formally given by eq.(\ref{single:elem:path}), with $ \Delta t_m = t_{m+1} - t_m$, $\Delta \vec x_m =  \vec x_{m+1} - \vec x_{m}$, and $| \hat\Phi_{\vec k} \rangle = \sum_{n \in \mathbb Z} | n_{\vec k} \rangle$, in perfect analogy with the EC evolution described above --- the symbols in this context are referred  to ordinary QM and $n \in \mathbb Z$ obviously means that we are considering particles and antiparticles.  

In ordinary QM as well as in ECs theory, a free relativistic bosonic particle has normally ordered harmonic energy spectrum
$
\mathcal H (\vec k) |n_{\vec k}\rangle =  \omega_n (\vec k) |n_{\vec k}\rangle = n \omega(\vec k) |n_{\vec k}\rangle$ with $n \in \mathbb Z$ (i.e. we consider both particle and anti-particles). The general dispersion relation is $ \omega^2(\vec k) = \vec k^2 + m^2$. This implies a corresponding harmonic momentum spectrum $\vec {\mathcal P} |n_{\vec k}\rangle = \vec k_n |n_{\vec k}\rangle = n \vec k |n_{\vec k}\rangle
$ as well known for instance for photons whose massless dispersion relation is $\omega(\vec k) = |\vec k|$, so that the normally ordered energy spectrum $\omega_n = n \omega$ actually implies the harmonic momentum spectrum $\vec k_n = n \vec k$. 

By applying the Poisson summation $\sum_{n \in \mathbb Z} e^{-i n y} = 2 \pi \sum_{n'  \in \mathbb Z}\delta(y + 2 \pi n')$  we now find that the ordinary QM associates a sum of Dirac deltas  to the elementary spacetime quantum evolutions of a relativistic particle 
\begin{eqnarray}\label{ele:path:cycls}
\mathcal  U(\vec{x}_{m+1}, t_{m+1}; \vec x_m, t_m) = \sum_{n_m \in \mathbb Z} e^{-i n_m [\omega(\vec k) \Delta t_m - \vec k \cdot  \Delta \vec{x}_m]} \nonumber \\
 = 2 \pi  \!\! \sum_{n'_m \in \mathbb Z} \!\! \delta \left(\omega(\vec k) \Delta t_m - \vec k \cdot \Delta \vec x_m + 2 \pi {n'_m}\right ) \,. ~~~~~~~
\end{eqnarray}
These Dirac deltas actually describe classical cyclic paths characterising an EC, as we shall discuss below in more detail. 

By plugging eq.(\ref{ele:path:cycls}) in the Feynman path integral written as in eq.(\ref{FPI:elem.paths}) and by using the Dirac delta property $\int d^3 x_m \delta (\vec x_{m+1} - \vec x_m) \delta (\vec x_m - \vec x_{m-1}) = \delta (\vec x_{m+1} - \vec x_{m-1}) $ we finally demonstrate that \emph{the ordinary Feynman path integral of a free relativistic particle is expressed by the sum (integral) of Dirac deltas associated to classical cyclic spacetime paths}
\begin{widetext}
\begin{eqnarray}\label{FPI:inv:dem}
\mathcal Z &=&   \int \left ( \prod_{j=0}^{N-1}  d^3 x_j \right ) (2 \pi)^N   \sum_{n'_0, n'_1, \dot, n'_{N-1} \in \mathbb Z} \!\!\!\!\! \delta \left(\omega(\vec k) \Delta t_{N-1} - \vec k \cdot \Delta \vec x_{N-1} + 2 \pi {n'_{N-1}}\right ) \delta \left(\omega(\vec k) \Delta t_{N-2} - \vec k \cdot \Delta \vec x_{N-2} + 2 \pi {n'_{N-2}}\right )  \times \nonumber \\ & \times & \delta \left(\omega(\vec k) \Delta t_{N-3} - \vec k \cdot \Delta \vec x_{N-3} + 2 \pi {n'_{N-3}}\right ) \dots  \delta \left(\omega(\vec k) \Delta t_{0} - \vec k \cdot \Delta \vec x_{0} + 2 \pi {n'_{0}} \right ) \nonumber \\ 
%&=&(2 \pi)^{N}\!\!\!\!\!\!\!\!\!\!\! \sum_{n'_0, n'_1, \dot, n'_{N-1} \in \mathbb Z} \!\!\!\!\!\!\!\!  \delta \left(\omega(\vec k)  (t_f - t_i)   - \vec k \cdot  (\vec x_f - \vec x_i) + 2 \pi {n'_0} + 2 \pi {n'_1} + \dots + 2 \pi {n'_{N-1}} \right ) \nonumber \\
&=&(2 \pi)^{N} \sum_{n'_0, n'_1, \dot, n'_{N-1} \in \mathbb Z}  \delta \left(\omega(\vec k)  (t_f - t_i)   - \vec k \cdot  (\vec x_f - \vec x_i) + 2 \pi ( {n'_0} +  {n'_1} + \dots +  {n'_{N-1}}) \right ) \nonumber \\
 &=& (2 \pi)  \sum_{n' \in \mathbb Z} \delta \left(\omega(\vec k) (t_f - t_i) - \vec k \cdot  (\vec x_f - \vec x_i) + 2 \pi {n'} \right )\,.
\end{eqnarray}
\end{widetext}

The Feynman path integral tells us that  the quantum evolution of a free relativistic particle of four-momentum $k_\mu$ is given by the sum (integral) of all the possible cyclic classical paths of spacetime period $\lambda^\mu$ between the initial and final spacetime points. 

These elementary spacetime loops are exactly those prescribed by ECs physics, as can be explicitly read out from both EC evolution law and PBCs --- see next subsection for more details. 
The EC describing a particle of four-momentum $k_\mu$ has global  spacetime periodicity $\lambda^\mu = \{T, \vec \lambda\}$, such that $T(\vec k) = 2 \pi / \omega({\vec k}) $ and $\lambda^i = 2 \pi / k_i$. That is, the ordinary Feynman path integral eq.(\ref{FPI:inv:dem})  describes all the possible spacetime loops of period  $\lambda^\mu$ between the initial and final points, according to the covariant EC evolution law
\begin{equation}\label{CA:Ev:law}
| x^{\mu}_i \rangle \rightarrow |  x^{\mu}_f + \text{mod} ~~ \lambda^\mu \rangle\,.
\end{equation} 
In fact, in the free case, by using the Poisson summation, the EC physical state can be explicitly written as a sum over Dirac delta functions  
 \begin{eqnarray}
  \Phi_{\vec k}(x_f, x_i) = \sum_{n \in \mathbb Z} a_n e^{i n \omega(\vec k)(t_f - t_i) - i n \vec k \cdot (\vec x_f - \vec x_i) } \nn \\ =   2 \pi \!\!  \sum_{n' \in \mathbb Z} \!\! {a'}_{n'} \delta[\vec k \cdot (\vec x_f - \vec x_i) - \omega(\vec k)(t_f - t_i) + 2 \pi n']\,, ~~~~
 \end{eqnarray}
  where ${a'}_{n'}$ are the transformed Fourier coefficients according to the Poisson summation, \cite{Dolce:2009ce,Dolce:Dice}. We conclude that a unitary EC physical state $\hat \Phi_{\vec k}$ (such that $\alpha_n \equiv 1 \Rightarrow {\alpha'}_{n'} \equiv 1$, $\forall n, n'$) reproduces the same result obtained independently from the Feynman path integral eq.(\ref{FPI:sum:deltas}).  

The demonstration of this correspondence, in the form given above,  is general. It can be generalised to the interacting case. However, in the free case, it is particularly easy to cross-check, once more, the consistency of this result. As the Hamiltonian and momentum operators are global in the free case,  the total evolution of a free relativistic bosonic particle, in ordinary QM \cite{peskin1995introduction} as well as in EC theory, is  
\begin{eqnarray}\label{FPI:sum:deltas}
 \mathcal Z &=& \mathcal U(\vec x_f, t_f; \vec x_{N-1}, t_{N-1})  \nonumber \\ &=&    \langle \hat\Phi_{\vec k}| e^{-i [\mathcal H(\vec k) ( t_f - t_i) - \vec {\mathcal P} \cdot  ( \vec x_f - \vec x_i)]} | \hat\Phi_{\vec k} \rangle \nonumber \\
 &=& \sum_{n \in \mathbb Z} e^{-i [\omega_n(\vec k) ( t_f - t_i) - \vec k_n \cdot  ( \vec x_f - \vec x_i)]} \nonumber \\ &=& 2 \pi   \sum_{n'} \delta \left(\omega(\vec k) (t_f - t_i) - \vec k \cdot (\vec x_f - \vec x_i) + 2 \pi {n'} \right )\,. ~~~~~
 \end{eqnarray}
  This proves that the ordinary Feynman path integral describes the classical evolution of a corresponding EC, and \emph{vice versa}. We will generalise this result to interactions. Now we shall interpret its physical meaning. 
  
  Thanks to this fascinating exact correspondence it is in fact possible to give an elegant interpretation, classical in the essence, of the ordinary Feynman path integral in terms  of  the degenerate classical solutions associated to the contravariant PBCs of the free relativistic EC. %All these results can be easily derived graphically \cite{Dolce:cycle,Dolce:2009ce}, and have interesting correspondences to Feynman chessboard  as well as to the Feynman educational description of QED, \cite{feynman1990q}.
 
 \subsubsection{Interpretation of the equivalence with the Feynman formulation}\label{inter:FPI}

As anticipated above, the exact equivalence between EC classical evolution end the Feynman path integral has a very intuitive explanation in terms of classical EC dynamics.

Due to the PBCs, the classical least action principle (variational principle) applied to the EC action eq.(\ref{CA:action}) yields an infinite number of degenerate solutions corresponding to the periodic paths described by the Dirac deltas in eq.(\ref{FPI:sum:deltas}), \ie the spacetime loops of period $\lambda^\mu$. These are implicit in the covariant EC evolution law eq.(\ref{CA:Ev:law}). 

 The evolution of an EC in its ``ontic'' spacetime can be in fact regarded as the evolution on a cyclic geometry. There are an infinite set of paths linking two arbitrary points on a cylindric geometry. Notice that the initial and final points of the EC evolution are not necessarily separated by integer numbers of spacetime periods; they can assume any possible value.  This infinite set of classical cyclic paths are labeled by the winding number $n' \in \mathbb Z$. That is, the evolution from $\Phi_{\vec k}(x_i)$ to $\Phi_{\vec k}(x_f +  \lambda)$ is degenerate \emph{w.r.t.} the evolution from $\Phi_{\vec k}(x_i)$ to $\Phi_{\vec k}(x_f + n' \lambda)$ (where we have suppressed the Lorentz index).

 %Above we have  proven that the EC evolution (\ref{FPI:free}) is given by the interference of all the degenerate classical solutions associated to the EC action (\ref{CA:action}). 
  
Indeed, we have seen that the sum of all the degenerate classical solutions (expressed as Dirac delta functions) associated to the EC action eq.(\ref{CA:action}) is equivalent to the Feynman path integral and \emph{vice versa}. That is the meaning of eq.(\ref{FPI:elem.paths}) and eq.(\ref{FPI:inv:dem}). Hence, the interference of the classical EC degenerate solutions reproduces the Feynman variations around the path of the corresponding classical particle, \ie of the classical path associated to the action $S^{Classic} [t_f; t_i] = \int_{t_i}^{t_f} dt (\vec {\mathcal P} \cdot \dot{\vec x}  - \mathcal H )$ appearing in the phase of the path integral. This classical action is determined by the phase of the covariant EC physical state.  

It is easy to infer in particular from the free case  that if the final point, \emph{w.r.t.} the initial one, is on the path of the corresponding classical particle, the interference among the EC degenerate cyclic paths is  constructive, whereas the interference becomes less and less constructive as the final point moves away from the classical particle path \cite{Dolce:2009ce,Dolce:2009cev4,Dolce:licata,Dolce:cycles}. It reveals that the classical particle path corresponds to the maximal probability (constructive interference) associated to the EC evolution. Finally, notice that this degeneracy of classical paths implicitly contains the Heisenberg uncertainty principle.

Our description of the Feynman path integral as interference of periodic classical paths is an example --- and there are many --- of the interesting physical aspects implicit in ordinary QM that become manifest in ECs (or CA)  formulation. 

It is worth noting a novel aspect \emph{w.r.t.} the ordinary interpretation of the Feynman path integral. According to the ordinary Feynman interpretation of QM \cite{Feynman:1942us},  the classical variational principle must be relaxed in ordinary QM and the evolution of a quantum particle is give by the sum over all the (classical and non-classical) paths linking the initial and final spacetime points $x_i$ and $x_f$. On the contrary, the demonstration of the Feynman path integral given above, sec.(\ref{further:proof:FPI}), shows that only the classical periodic paths are relevant (in both cases the time interval of the evolution is divided in slices of infinitesimal duration, then integrated over the spatial coordinate $x_m$ of each slice). 
The demonstration given in sec.(\ref{further:proof:FPI}) is absolutely general, and shows that the paths relevant to the Feynman path integral are exactly the classical periodic paths prescribed by the ECs theory, \ie the spacetime loops of period $\lambda^\mu$. Due to the PBCs, these periodic paths are the classical degenerated paths resulting from the classical variational principle applied to the EC action eq.(\ref{CA:action}), so that one of the great advantages of ECs formulation of QM is that \emph{it preserves the full validity of the classical variational principle in QM}. It proves an exact correspondence between cyclic evolution, classical in the essence, and quantum evolution. 

Finally, we conjecture that such a over-counting of paths in the ordinary interpretation of the Feynman path integral \emph{w.r.t.} the interpretation emerging from ECs theory could be responsible for the infinite terms (divergences) that must be renormalised in the ordinary approach. In other words, we conjecture that, by evaluating the Feynman path integral according to the prescriptions of the EC theory, one could obtain the finite expression for measurable quantum quantities avoiding the renormalisation process.           
 
\section{Interacting relativistic Elementary Cycles}\label{INT}

So far we have exclusively considered free ECs, characterised by global spacetime periodicities, which turn out to describe the quantum behaviour of corresponding  free particles, \ie particles with constant four-momenta. In this section we must bear in mind that interactions, \ie local variations of four-momenta, imply local modulations of the ECs spacetime periodicities. 

To achieve the correct description of interactions we must also remember that (similarly to the Compton clock or de Broglie internal clock) an EC can be regarded as a moving relativistic reference clock (time period) and ruler (wave-length). The duration of the clock period and the length of the ruler is determined locally by the amount of energy and momentum associated to the interacting EC in that point. During interactions these conjugated quantities vary from point to point, \ie locally, depending on the interaction scheme considered, in such a way that the local phase harmony is satisfied locally. This provides  a fundamental link to the geometrodynamical description of interactions typical of gravitational interaction in GR, resembling original Einstein's derivation (roughly speaking ``relativity is about clocks and rulers''). In analogy with GR, the local modulations of periodicities associated to interactions will be encoded in local geometrodynamics of the ``ontic'' spacetime coordinates.   
 
  In this section we will exclusively describe interactions at a classical-relativistic level. This means that we will neglect quantum corrections by consider only the fundamental eigenmode ($n=1$) of the ECs  physical states. We have already seen in the free case, for instance see eq. (\ref{field:mode}), that the fundamental mode corresponds to the non-quantised scalar field mode of momentum $\vec k$, \ie to a classical particle. Here we denote by a prime the quantities associated to interactions: for instance the physical state of the interacting EC is $\Phi'_{\vec k'}$.  
  
  In sec.(\ref{INT:QM}) we will derive the quantum behaviour of  interacting ECs by simply considering all the ECs harmonics allowed by the PBCs. The quantum corrections are encoded in the higher modes of the ECs vibrations. The resulting ECs dynamics will be equivalent to the quantum dynamics of the corresponding interaction scheme. In this way, in sec.(\ref{INT:QM}) the equivalence with QM will be fully generalised to the interacting case.  

%Interactions, \ie the local and retarded variations of four-momentum characterising a given interaction scheme, are encoded by corresponding local and retarded modulations of the EC periods, due to the phase harmony. 

Let us consider a free EC of constant momentum $ k_\mu$, \ie global spacetime periodicity $\lambda^\mu$ such that $ k_\mu \lambda^\mu = 2 \pi$, and let us compare it with the case in which  a generic interaction is switched on. Also, let us denote the resulting, locally varying four-momentum in the spacetime point  $x'$ of the interacting EC evolution as $k'_\mu (x')$. The local variation of four-momentum \emph{w.r.t.} the free case can be described by a tetrad ${e^\mu}_{\mup}(x')$, which therefore uniquely encodes the interaction scheme under consideration. That is, if we switch on a generic interaction we must replace the local four-momentum $k_\mu$ of the free EC with a local one according to $k'_\mup(x') = {e^\mu}_{\mup}(x') k_\mu$.  This is the generic local momentum of the interacting elementary classical-relativistic particle associated to the EC. 

We show that this generic interaction scheme is described by replacing locally the Minkowskian ``ontic'' spacetime $d s^2 = d x_\mu d x_\nu \eta^{\mu \nu}$ of the the free EC with the local metric  $d s^2 = d x'_{\mu'} d x'_{\nu'} g^{{\mu'} {\nu'}}$ encoding the interaction itself. In other words, we prove that interactions (including gauge interactions!) are encoded in the corresponding local metric tensor $g_{\mup \nup} = {e^\mu}_{\mup} {e_{\nup}}^{\nu} \eta_{\mu \nu}$. In analogy with GR the effect of interaction is to locally transform the flat spacetime $\mathtt{S}$ of the free EC to the new manifold $\mathtt{S}'$ characterising the interacting EC. The physical quantities labelled by the prime symbol denotes the manifold $\mathtt{S}'$ associated to interacting EC. The locally transformed metric can be either flat (local rotation) or curved, these will correspond  to describe gauge interactions and gravitational interaction, respectively. 

To retrieve such a geometrodynamical description of interactions let us consider the local transformation (deformation) of the free EC ``ontic'' spacetime coordinates $x^\mu \rightarrow x'^{\mu'}(X) = x^\mu {\Gamma_\mu}^{\mup}(x)_{x=X}$, such that the tetrad introduced above is ${e_{\mu}}^{\mup} = \frac{\partial {x'}^{\mu'}}{\partial x^\mu}$ (for the sake of simplicity we neglect Christoffel symbols,  relevant for gravitational self-interaction  and non-abelian gauge theories \cite{Dolce:tune,Birrell:1982ix}). Under this local redefinition of coordintates the free EC action eq.(\ref{CA:action}) is transformed to 
\beq\label{CA:action:int}
\mathcal S_{EC} = \int_{\lambda'^\mup(X)} d^4 x' \sqrt{-g} \mathcal L({e_\mu}{^\mup}\partial_\mup \Phi'_{\vec k'}, \Phi'_{\vec k'})\,.
\eeq
This action describes the interacting EC in the point $x'=X$. It will describe the corresponding quantum elementary particle subject to a generic interaction scheme.

Notice the local contravariant transformation of the boundary in the EC action eq.(\ref{CA:action:int}). As usual, PBCs are assumed at the local boundary $\lambda'^\mup(X) = \lambda^\mu {\Gamma_\mu}^{\mup}(x)|_{x=X}$. This means that the EC located in $X$ has locally modulated spacetime period $\Phi'_{\vec k'}(X) = \Phi'_{\vec k'}(X+\lambda'(X))$. We also notice that the metric tensor $g_{\mup \nup}$ of the new manyfold $\mathtt{S}'$ is the one introduced above in terms of the  tetrad   ${e^\mu}_{\mup}$.

When interaction is switched on, the phase harmony relation for the free EC $k_\mu \lambda^\mu = 2 \pi $ turns out to be replaced by the local phase harmony relation $k'_\mup \tau'^\mup = 2 \pi $ where $\tau^\mup(x')$ is the \emph{instantaneous spacetime period} of the EC, such that $\tau'^\mup(x') = \lambda^\mu {e_\mu}^\mup(x')$. It transforms as $dx'^\mup$ whereas $\lambda'^\mup (x)$ transforms as $x'^\mup$. Indeed  $\tau'^\mup$ is a space-like tangent four-vector which in general does not coincide with $\lambda'^\mu$ (\eg in the free case the period coincides with the instantaneous periodicity: $T^\mu = \tau^\mu$).

To show that this actually describes interaction we notice that the EC solution of eq.(\ref{CA:action:int}) turns out to have  in general locally modulated spacetime period. The fundamental solution $n=1$ (as well as the generic solution) of the Euler-Lagrange equation of  eq.(\ref{CA:action:int}) has in fact the form of a locally modulated wave $\bar \Phi'_{\vec k'}(x') \propto e^{-i \int^{{x'}^{\mu'}} d y^\mup k'_\mup(y)} $. Indeed it satisfies the equation  $i \partial'_{\mu} \bar \Phi'_{\vec k'} = k'_\mup(x') \bar \Phi'_{\vec k'}$ of instantaneous periodicity $\tau^\mup(x')$, \ie of local periodicity $\lambda'^\mup(x')$. It is easy to foresee  that it will yield the Schr\"odinger equation for interacting ECs. 

Hence we have shown that the local transformation of coordinates from $\mathtt{S}$ to $\mathtt{S}'$ actually yields the local variation of EC four-momentum $k'_\mup(x') = {e^\mu}_{\mup}(x') k_\mu$ originally assumed for our interacting EC. It transforms as $\partial_\mup$ and coincides with the local four-momentum of the corresponding classical-relativistic bosonic particle of mass $m$ interacting under our interaction scheme; \emph{q.e.d.}

In ECs theory interactions are equivalently encoded by both the local deformations of the EC ``ontic'' spacetime $g_{\mup \nup} $ and the local deformations of  boundary $\lambda'^\mup(x')$ of the EC action eq.(\ref{CA:action:int}). Thus in ECs theory the local boundary provides an holographic description of the particle dynamics, in analogy with the \emph{holographic principle} \cite{Stephens:1993an} and the holographic description of extra-dimensional theories \cite{ArkaniHamed:2000ds,Casalbuoni:2007xn}.  In 't Hooft terminology we can say that the interacting EC is the geometrodynamical ``changeable'' obtained through local deformations of the ``ontic'' flat spacetime of a free EC. 

\subsection{Gravitational interaction and Elementary Cycles geometrodynamics} 

To illustrate the meaning of the ECs geometrodynamical description of interactions we first consider an EC in a weak gravitational (Newtonian) potential $V(\vec x) = - G M_\odot / |\vec x|$. In this particular case the locally transformed metric of the interacting EC is curved, but as we will see this is not the only possible way to transform the ECs spacetime. 
Assuming a Newtonian potential, the fundamental energy of the EC located at distance $|\vec x|$ from the gravitational centre of mass $M_\odot $ varies, \emph{w.r.t.} the free case, as $\omega \rightarrow \omega' = (1 + G M_\odot / |\vec x|) \omega$. By considering the EC local phase harmony relation in the ``ontic'' time,  $\omega'  = 2\pi/ T'$ (in this weak case we can approximate $\tau'^\mup \simeq \lambda'^\mup$), the EC time period varies as $T \rightarrow T'  = (1 - G M_\odot / |\vec x|) T$. Therefore an EC, similarly to a clock, runs slower inside a gravitational well. This correctly describes two fundamental aspects of GR: time dilatation and gravitational red-shift. 

Furthermore, by considering the transformation of EC momentum due to gravitational interaction $|\vec k| \rightarrow |\vec k'| = (1 - G M_\odot / |\vec x|) |\vec k|$, the corresponding modulation of EC spatial period is $|\vec \lambda| \rightarrow |\vec \lambda'| = (1 - G M_\odot / |\vec x|)^{-1}|\vec \lambda| $. We find that the deformation of EC ``ontic'' metric encoding the weak Newtonian interaction is actually  the Schwarzschild metric $d s^2 = (1 - G M_\odot / |\vec x|) d t^2 - (1 - G M_\odot / |\vec x|)^{-1} d |\vec x|^2 - |\vec x|^2 d \Omega^2 $.  

We have thus obtained linearised gravity by simply combining the Newtonian gravitational potential and undulatory mechanics \cite{Ohanian:1995uu}. It is know in literature that, by considering self-interactions, this approach actually leads  to a consistent derivation of the whole GR. Actually original Einstein's approach to GR was based on relativistic clocks local modulations and the corresponding geometrodynamical description. 

By considering self-interaction (\ie we have to consider the Christoffel symbols in the expansion of $e^\mu_a(x)$ and $\tau'^\mup \neq \lambda'^\mup$) it is possible to obtain the Einstein equation $\mathcal R^{\mup \nup} = - 8 \pi G \mathcal \mathcal T^{\mup \nup}$ where $\mathcal R^{\mup \nup}$ is the Ricci tensor and $\mathcal T^{\mup \nup}$ is the ordinary stress-energy-tensor. By giving dynamics to the metric tensor we can thus write the Hilbert-Einstein action (modulo boundary terms) for EC
\beq\label{CA:action:Gra}
\mathcal S_{EC} %\!\!
= %\!\! 
\int_{\lambda'^\mu(X)}  \!\!\!\!\!\!\!\!\!\!\!\! 
d^4 x' \!\! \sqrt{-g}\!\left[ \frac{ - g^{\mup\!\nup\!} \mathcal R_{\mup \nup}}{16 \pi G} + \mathcal L({e_\mu}^{\mup} \partial_\mup \Phi'_{\vec k'}, \Phi'_{\vec k'})\right]\,.
\eeq
In other words the Ricci tensor describes how the EC ``ontic'' spacetime instantaneous period $\tau'^\mup$ is curved locally due to gravitational interaction. Actually, the curvature of the EC ``ontic'' spacetime depends on the amount of EC energy-momentum, \ie on the EC spacetime instantaneous period, in the point $X$ in analogy with the ordinary interpretation of GR. 

Such geometrodynamical description of gravitational interaction has been directly derived from the EC phase harmony relation, as for undulatory mechanics or wave-particle duality, in which the local energy-momentum is fixed by the local spacetime instantaneous period of the ``periodic phenomenon''.  

Furthermore our analysis clearly proves  that the ECs ``ontic'' spacetime coordinates, intrinsically periodic, represent perfectly consistent sets of relativistic spacetime coordinates. Indeed ECs mimic the behaviours of relativistic clocks and rulers. This is due to the fact that relativity only concerns with the differential structure of spacetime, \ie the metric, without giving any particular prescriptions about the BCs of spacetime. On the other hand BCs has characterised QM since its earliest days, see discussion in sec.(\ref{ONTICSPACETIME}) and \cite{Dolce:Dice,Dolce:2010ij,Dolce:2010zz,Dolce:2009ce}. 

\subsection{Derivation of gauge interaction from Elementary Cycles geometrodynamics}\label{GAUGE:CLASS}  

In ordinary QFT gauge interactions (electroweak and strong interactions) are postulated, despite early attempts to derive them directly from relativistic spacetime geometrodynamics (e.g. Weyl, Nordstr\"om, Kaluza, Einstein, etc). A surprising, unprecedented property of ECs physics is that gauge interactions are fully derived directly from the spacetime geometrodynamics of the theory. That is, gauge interactions are inferred in perfect correspondence with gravitational interaction in GR. This property of ECs theory reveals a deep relationship between gauge interactions and GR. The geometrodynamical description of gauge interaction that we are going to present has been originally proven in full mathematical details in \cite{Dolce:tune}. 

We now choose a particular class of local transformations of the EC ``ontic'' spacetime coordinates, such that the metric remains flat, contrarily to the gravitational case described above, whereas the boundary $\lambda'^\mup$ is locally rotated (transformed). For the sake of simplicity we only consider the case of electromagnetism, which corresponds to unitary local rotations   $U(1)$ of the boundary of the free EC action. Notice that the Lorentz transformation described in sec.(\ref{EC:LorentzInV}) is a global rotation of the ECs boundary: this consideration is very important to interpret the physical meaning of gauge interactions and their relationship with Lorentz group \cite{Dolce:tune}. As a consequence, in spite of the fact that the metric remains flat, the local rotation of the EC boundary implies, through the PBCs, local instantaneous modulations of EC spacetime instantaneous periodicity $\tau'^\mup(X)$ which in turn corresponds to local variations of four-momentum, \ie to an interaction which is equivalent to electromagnetic interaction as we are going to see. 

The local rotation of the EC boundary reproducing electromagnetism is induced by the local transformations of spacetime coordinates $x^\mu \rightarrow x'^{\mu'}(X) = x^\mu {\Lambda_\mu}^{\mup}(x')_{x'=X}$ where ${\Lambda^\mu}_\mup(x')\, \dot = \, \delta^\mu_\mup - e {\Xi^\mu}_\mup (x')$. The parameter $e$ will be identified with the electric charge. The tetrad of the transformation is such that ${e_\mu}^\mup(x') = \frac{\partial x'^a}{\partial x^\mu} = \delta^\mu_\mup -  e {\xi^\mu}_\mup(x') $ where ${\xi^\mu}_\mup(x')  \in U(1)$ is a peculiar unitary subclass of Killing vectors on the spacetime defined by the interacting EC. For reasons that will be clarified below, they will addressed as ``polarised'' rotations. 

Under this transformation the EC ``ontic'' spacetime in fact transforms locally from a flat metric to another flat metric. That is,  the assumption of the Killing vectors guarantee that $\sqrt{-g'} = 1$. Notice that this transformation of coordinates  has no effect in ordinary QFT where, actually, the BCs play no roles in the derivation of the field solutions and propagators: in QFT the field solution (\eg the Klein Gordon field) is the most general solution of the equations of motion. This is why in ordinary QFT gauge interaction must be postulated  and it cannot be derived from spacetime geometrodynamics.  

Let us prove that the interaction corresponding to these ECs  geometrodynamics is equivalent to the ordinary electromagnetic interaction \cite{Dolce:tune}. 
The effect of this transformation on an EC located in $x'=X$ is a local modulation of the  spacetime periodicity.  Indeed, though the spacetime remains flat, the local EC action turns out to have a local rotation of the boundary  ${\lambda'}^\mup(X) = \delta^\mup_\mu \lambda^\mu -  e \lambda^\mu {\Xi_\mu}^\mup (x')|_{x'=X}$.

In fact, as a consequence of this transformation of coordinates, the action eq.(\ref{CA:action:int}) describing our interaction scheme in this case is on a flat metric with locally rotated boundary
\beq\label{CA:action:QED}
\mathcal S_{EC} = \int_{\lambda'^\mup(X)} d^4 x' \mathcal L({e_\mu}^\mup \partial_\mup \Phi'_{\vec k'}, \Phi'_{\vec k'})\,.
\eeq

The resulting local modulation of EC instantaneous period is thus $\tau'^\mup(x) = \delta^\mup_\mu \lambda^\mu -  e \lambda^\mu {\xi_\mu}^\mup (x)$.  
It is now convenient to introduce the vectorial field defined as $A_\mup(x)\, \dot = \, {\xi^\mu}_\mup(x)k_\mu$. As it can be  checked from the local phase harmony relation $ k_\mup(x) \tau^\mup(x) = 2 \pi $, in this case  the resulting local variation of EC four-momentum takes the familiar form $k'_\mup(x) = k_\mup - e A_\mup(x) $ typically associated to gauge interaction. Our interaction scheme is therefore formally described by the  ordinary \emph{minimal substitution} of electromagnetism. 

The fundamental EC solution ($n=1$ denoted by the bar symbol or the generic EC solution) resulting from the local PBCs of eq.(\ref{CA:action:QED})  has local instantaneous periodicity $\tau'^\mup(x)$, that is, as we have seen, it has  the form of a locally modulated wave $\bar \Phi'_{\vec k'}(x')\propto e^{ie  \int^{x'^\mup} A_\mup(y)  d y^\mup} e ^{-i k_\mup x'^\mup}$ which can be written as  $\bar \Phi'_{\vec k'}(x') = U(x')\bar  \Phi_{\vec k}(x')$ where $U(x') = e^{i e \int^{x'^\mup} A_\mup(y)  d y^\mup}$. 

Notice that the local modulation of EC period  $U(x')$ \emph{w.r.t.} the free case describes the gauge connection, \ie the Wilson line, of ordinary electromagnetism. Thus the gauge connection $U(x')$, which must postulated in ordinary QFT in order to derive gauge invariance, in the EC theory has been directly derived from the relativistic geometrodynamics of the ECs spacetime dimensions as modulating term of the EC spacetime period. Hence, the so-called \emph{internal transformation} of gauge interaction is directly obtained as transformation of the EC solution associated to  local ``ontic'' spacetime geometrodynamics: $\delta \Phi_{\vec k} =  \Phi'_{\vec k'} - \Phi_{\vec k} = i e A_\nup \Phi_{\vec k} \delta x'^\nup$.  \emph{This proves that actually electromagnetism (or in general gauge interactions) can be derived from spacetime geometrodynamics in perfect correspondence with gravitational interaction. }

It is possible to see even more explicitly that these geometrodynamics excactly describes ordinary gauge interactions. It is in fact quite inconvenient to work with an action whose boundary $\lambda'^\mup(x')$ varies from point to point, as in eq.(\ref{CA:action:QED}). It is convenient to rewrite the EC action eq.(\ref{CA:action:QED}) in an equivalent form, in such a way that it has as solution the same interacting EC physical state $\Phi'_{\vec k'}$ of local instantaneous periodicity $\tau'^\mup(x')$ described above, but whose boundary is kept global (constant), for instance, to  $\lambda^\mu$ as for the free action eq.(\ref{CA:action}). 

Such an equivalent action  with global boundary and locally modulated solution can be easily written by considering that, as well-known, ``covariant derivatives allow for a background independent description of physics'' and thus for a description independent from the BCs --- the background in our case is locally rotated causing a local rotation of the boundary whereas the metric stays flat. 

Actually, as shown in \cite{Dolce:tune}, covariant derivatives can be used to ``tune'' the local periodicity of a a locally modulated field solution to the global periodicity imposed by the fixed boundary of the new action. 
It is straightforward (for a rapid check substitute  $\Phi_{\vec k}  = U^{-1} \Phi'_{\vec k'} $ in the free EC action)  to prove that the action eq.(\ref{CA:action:QED}) can be equivalently written as 
\beq\label{CA:action:YM}
\mathcal S_{CA} = \int_{\lambda^\mup} d^4 x' \left[-\frac{1}{4} F^{\mup\nup} F_{\mup \nup} + \mathcal L(D_\mup \Phi'_{\vec k'}, \Phi'_{\vec k'})\right]\,,
\eeq
where the $D_\mup = \partial_\mup - i e A_\mup(x)$ is the covariant derivative of the gauge interaction. Please refer to \cite{Dolce:tune} for a detailed description. 

In the Lagrangian density $\mathcal L(D_\mup \Phi'_{\vec k'}, \Phi'_{\vec k'}) = \mathcal L(U^{-1} D_\mup \Phi'_{\vec k'}, U^{-1} \Phi'_{\vec k'})$ both terms have fixed periodicity $\lambda^\mu$, as $U^{-1} D_\mup \Phi'_{\vec k'} = \partial_\mu \Phi_{\vec k} $ and $U^{-1} \Phi'_{\vec k'} = \Phi_{\vec k} $.  The global PBCs of (\ref{CA:action:YM}) are therefore satisfied despite the fact that the solution $\Phi'_{\vec k'}$ has locally modulated periodicity. 

Notice that in general the only terms determining the periodicity of the EC physical state $\Phi'_{\vec k'}$ are the derivative terms of the Lagrangian density. Only the derivative terms are relevant for the BCs  (these are the only terms generating, through integration by parts, boundary terms when we vary the action). In the non derivative terms the modulation of local periodicity $U^{-1}$ is not relevant. Obviously this is clear manifestation of gauge invariance. 

The introduction of the covariant derivative in (\ref{CA:action:YM}) is directly related to the necessity to satisfy the PBCs  of the theory and thus the variational principle at the boundary points. Hence the two actions eq.(\ref{CA:action:QED}) and eq.(\ref{CA:action:YM}) have the same solution $\Phi'_{\vec k'}$ so that they describe the same physics. In particular they actually describe the same local modulation of periodicity $\lambda'^\mup(x')$ that, as we have seen, describes electromagnetic interaction. 

In eq.(\ref{CA:action:YM}) we have also included the field strength $F_{\mup \nup} = \partial_\mup A_\nup - \partial_\nup A_\mup$ in order to give dynamics to the vectorial field $A_\mup$, in analogy with the dynamical term of the metric $g^{\mup\nup}$ in Hilbert-Einstein action eq.(\ref{CA:action:Gra}). The form of the kinetic term $F^{\mup \nup} F_{\mup \nup}$ is obliged by the fact that only in this form the local period of $A_\nup$, in general different from $\lambda^\mu(x)$, can be always ``tuned'' to the global PBCs imposed by the action eq.(\ref{CA:action:YM}). That is indeed a gauge invariant term, see more details in \cite{Dolce:tune}. 

Furthermore it implies that $A_\mup$ satisfies the Maxwell equations. Since the equations of motion of $A_\mup$ (the Maxwell equations) restrict the general form of ${\xi^\mu}_\mup$   we can actually say that the unitary Killing vector  describing the geometrodynamcis of electromagnetism is ``polarised'' as anticipated above. 

 There are many others fascinating details worth to mention fully confirming the equivalence between flat ECs geometrodynamics and gauge theories  \cite{Dolce:tune}. Among them we could mention for example that, in our description, gauge invariance is manifestly a consequence of the holonomy of the ECs theory, \ie of the general fact that the ``boundary of the boundary is zero''. 
 
 By  adding a boundary term to the local boundary $\lambda^\mup(x')$ of the EC action we must obtain an invariance of the theory. 
 Such   ``boundary of the boundary'' can be obtained by adding a total derivative term to the local boundary; that is, by adding  a local term proportional to $\partial_\mup \theta(x')$ to  
  %${\xi^\mu}_\mup(x') \rightarrow {{\xi''}^\mu}_\mup(x') + \lambda^\mu \partial_\mup \theta(x')/2\pi$ to 
  ${\xi^\mu}_\mup(x')$. As a consequence of the local phase harmony and the definition of $A_\mu$, the resulting local transformation of $k'_\mup$ implies the (gauge) transformation $A'_\mup(x') = A_\mup(x') - e \partial_\mup \theta(x') $.  Due to the definition of $A_\mu$, it actually appears in an integral on the phase of the locally modulating term $U(x')$. The resulting ``boundary of the boundary'' term corresponds to the local phase invariance $\Phi''_{\vec k}(x') = e^{-i\theta(x')}\Phi'_{\vec k}(x')$.  
  
  Obviously such holonomy of the EC geometrodynamics actually describes gauge invariance. The ECs theory is therefore defined modulo gauge orbits.  Thus we have explicitly proven that such  interacting EC described by action eq.(\ref{CA:action:YM}), or equivalently by action eq.(\ref{CA:action:QED}), has gauge invariance $U(1)$. 
 
 Remarkably,  a similar geometrodynamical analysis of the correspondence between gauge interactions and  gravitational interactions has been successfully applied to mathematically prove --- for the first time in literature, as far as known by the author --- the central correspondence of Maldacena conjecture, also known as AdS/CFT  correspondence or gauge/gravity duality \cite{Dolce:ADSCFT,Dolce:ICHEP2012}.
 
\subsection{Toward a formulation of fermionic Elementary Cycles}\label{fermions}

Here we discuss an attempt of generalisation of the ECs free bosonic dynamics to free fermionic ones. Obviously, considering that the periodic dynamics characterising ECs are determined by the Compton periodicity, we expect to find deep correspondences to the \emph{zitterbewegung} description of Dirac solution. As discovered by Schr\"odinger, the Dirac equation implies that fermions are characterised by intrinsic cyclic dynamics, related to the Compton period, which in turn lead to intuitive semiclassical interpretations of peculiar fermionic behaviour aspects such as spin and intrinsic magnetic momentum.  

Here we adopt an approach  suggested by Hestenes' formalism of spacetime algebra for the \emph{zitterbewegung}  \cite{Hestenes:zbw:1990} (combined with some ideas of Penrose's twistor theory).  Let us considerer the following global redefinition of the free EC ``ontic'' spacetime coordinates: $x^{\mu} \rightarrow x^{\alpha \dot \beta } = x^{\mu} \gamma^{\alpha \dot \beta}_{\mu} \equiv x \!\!\! \slash$ where $\alpha$, $\beta$ are spinorial indexes and $\gamma^\mu$ is the Dirac matrix. The aim is to  ``twist'' the cyclic spacetime geometry of a scalar EC in order to encode the Dirac equation into the resulting geometrodynamics. Again, we have assumed a global transformation because we want first to describe free fermions, \ie global ``twisted'' periodicity, and then generalise to interactions. 

%The Minkowski metric of the free scalar EC ontic spacetime $d s^{2} = d x_{\mu} d x_{\nu} {\eta^{\mu\nu}}$ is replaced by $ d s^{2} \mathbf 1_{4\times 4} =  dx\!\!\! \slash  dx \!\!\! \slash \equiv  d x_\mu d x_\nu {\eta^{\mu\nu}} \mathbf 1_{4\times 4}  $  where we have used the Clifford algebra property $\lbrace \gamma^\mu\gamma^\nu \rbrace   = 2 \eta^{\mu \nu } \mathbf 1_{4\times 4}$. The EC  four-momentum $K_\mu$ and the operator $\partial_\mu$ are transformed to  $k_{\mu} \rightarrow k_{\dot  \beta \alpha} = \gamma_{\dot \beta \alpha}^{\mu} k_{\mu} \equiv k \!\!\! \slash\,$ and $\partial_{\mu} \rightarrow \partial_{\dot  \beta \alpha} = \gamma_{\dot \beta \alpha}^{\mu} \partial_{\mu} \equiv \partial \!\!\! \slash\,$, respectively. 

Under this local ``twist'' of cyclic spacetime coordinates,  the Minkowski metric of the free scalar EC ``ontic'' spacetime $d s^{2} = d x_{\mu} d x_{\nu} {\eta^{\mu\nu}}$ is therefore replaced by  $ d s^{2}  = {\text{Tr}}{} (dx\!\!\! \slash  dx \!\!\! \slash) / 4 \equiv {\text{Tr}}{}(\gamma^\mu\gamma^\nu) d x_\mu d x_\nu  / 4  $  where we have used the Clifford algebra property ${\text{Tr}}(\gamma^\mu\gamma^\nu) = 4 \eta^{\mu \nu}$. The EC  four-momentum $k_\mu$ and the operator $\partial_\mu$ are transformed to  $k_{\mu} \rightarrow k_{\dot  \beta \alpha} = \gamma_{\dot \beta \alpha}^{\mu} k_{\mu} \equiv k \!\!\! \slash\,$ and $\partial_{\mu} \rightarrow \partial_{\dot  \beta \alpha} = \gamma_{\dot \beta \alpha}^{\mu} \partial_{\mu} \equiv \partial \!\!\! \slash\,$, respectively. 

According to this ``twist'' of the EC coordinates, the fundamental mode (or a generic mode) of a bosonic EC physical state  $\bar \Phi_{\vec k}(x) \propto e^{-i k_\mu x^\mu}$, satisfying $i\partial_\mu \bar \Phi_{\vec k}(x) = k_\mu \bar \Phi_{\vec k}(x) $ and $(\partial^2 - m^2) \bar \Phi_{\vec k}(x)$, must be replaced by the generic mode of the EC fermionic physical state which (we have suppressed prime indexes and symmetrised the phase) has the form $\bar \Psi_{k \!\!\! \slash}(x) \propto \chi e^{-  \frac{i}{4} \text{Tr} ( k\!\!\! \slash \cdot x\!\!\! \slash ) }$ where $\chi$ is a spinorial basis. 

The resulting fermionic EC physical state turns out to satisfy   $i \gamma^{\mu} \partial_{\mu} \bar \Psi_{k \!\!\! \slash}(x) = k\!\!\! \slash ~ \bar \Psi_{k \!\!\! \slash}(x)$ and thus the Dirac equation $(i\gamma^\mu \partial_\mu - m)\bar \Psi_{k \!\!\! \slash}(x) = 0$. Actually, the Dirac equation is the ``square root'' of the Klein-Gordon equation. In analogy with  the discussion in the derivation of the  ``axiom of the motion'': $(\partial^2 + m^2)\bar  \Psi_{k \!\!\! \slash}(x) = (-i \partial \!\!\! \slash - m)(i \partial \!\!\! \slash - m) \bar \Psi_{k \!\!\! \slash}(x) = 0 $.    The EC spacetime periodicity (which in the global case is equivalent to the instantaneous periodicity) is ``twisted'' according to
$\lambda^{\mu} \rightarrow \lambda^{\alpha \dot \beta} = \lambda^{\mu} \gamma^{\alpha \dot \beta}_{\mu} \equiv \lambda \!\!\! \slash~$.   % The EC phase harmony $ k_\mu \lambda^\mu = 2 \pi$ becomes $ \text{Tr}(k\!\!\! \slash \cdot \lambda \!\!\! \slash ~) = 8 \pi$. 

According to this analysis the free Dirac dynamics are related to a ``twist'' of the EC cyclic dynamics  in order to pass from bosonic to fermionic dynamics. In other words the topology of the EC spacetime determines the type of particle associated to it. Such  ``twisted'' geometrodynamics can be associated to the \emph{zitterbewegung}, according to Hestenes's spacetime algebra approach (with some analogies to twistor theory).  

The  \emph{zitterbewegung} is directly inferred from the complex phase factor of the fermionic wave-function \cite{Hestenes:zbw:1990}. It characteristic period is half the Compton period of  the bosonic EC. %   $T_Z = T_C / 2$. % ( after a  \emph{zitterbewegung} period it acquires a minus sign). 
Such geometrodynamics associated to the fermionic EC physical state can also be regarded as induced by  anti-periodicity% in the orbifold $\mathbb S^1 / \mathcal Z_2$
 \footnote{The BCs allowed by the variational principle to a fermionic action are non trivial. For the scope of this paper they can be regard as anti-PBCs}, which in turn  leads to the Pauli exclusion principle, \ie to the spin-statistics, similarly to field theory at finite temperature where the fermions are characterised by anti-PBCs in the Euclidean time, as we  will discuss in sec.(\ref{therm:QM}.). 
 
Similarly to  the bosonic case, one can introduce creation and annihilation operators. As also pointed out by 't Hooft \cite{Hooft:2014kka} (which actually proposes an alternative way to derive fermionic dynamics in CA models, which can be generalizated to ECs theory and explicitly tested in graphene physics \cite{Dolce:Dice2014}) the negative modes associated to fermionic ECs are consistently interpretable as holes in the Dirac sea. This solves the problem of the negativity of the Hamiltonian operator in the relativistic fermionic case as well. In general,  the negative modes in the EC theory describe antimatter whereas, as we will see, ECs Hamiltonian operators are always positively defined in the non-relativistic limit. 

The ECs description of antimatter is confirmed experimentally by % , as well as the fact that the effective electron mass is determined by intrinsic cyclic dynamics,  
 carbon nanotubes whose cylindric geometry implies that the elementary charge carriers (\ie the electrons) actually behave as ECs on a lattice: they acquire an effective mass fixed by their rest periodicity (the carbon nanotube diameter), the negatives modes associated to these cyclic dynamics correspond to holes in the Dirac sea (the Fermi sea of graphene physics), and the Dirac dynamics (pseudo-spin) emerges from the graphene sublattice in agreement with 't Hooft derivation of fermionic dynamics in CA models \cite{Dolce:Dice2014,Dolce:cycles,Dolce:SuperC,Dolce:EPJP}.  

The generalisation of the free ECs fermionic dynamics to interactions can be achieved by following the same steps of the bosonic case.  Interacting fermionic ECs are described by local modulations of the \emph{zitterbewegung}. In particular we can describe classical electrodynamics by assuming the following local modulation of the free \emph{zitterbewegung}, \ie of the ``twisted'' periodicity described above:  $\gamma_\mu \rightarrow \gamma'_{\mup}(x') =  \gamma_{\mu} {e^\mu}_\mup(x') = \gamma_{\mu} (\delta^\mu_\mup -  e {\xi^\mu}_\mup(x')) $ where the polarised Killing vector ${\xi^\mu}_\mup(x') \in U(1)$ defines the electromagnetic field $A_\mup (x') =  {\xi^\mu}_\mup(x') k_\mu$ similarly to the bosonic case. In this way the equations of motion of the fermionic EC interacting electromagnetically turns out to be  $(i\partial \!\!\! \slash - e A \!\!\! \slash - m)\Psi'_{k' \!\!\!\!\!\! \slash}\,\,(x') = 0$ where $k'_\mup (x') = k_\mup - e A_\mup(x')$, in agreement with ordinary QED.

\section{Interaction: generalised equivalence to QM}\label{INT:QM}

So far we have described interactions at a classical level, neglecting quantum corrections. We have successfully described interactions as local modulations of spacetime periodicity, but  we have only considered the fundamental vibrational mode, labeled by $n=1$, for the interacting case. In order to describe the quantum behaviours of interacting particles we must bear in mind that the constraint of intrinsic periodicity is the quantisation condition of the ECs theory. 

As we have seen for the free case,  all the possible vibrational eigenmodes allowed by the EC periodicity (PBCs) must be consider in order to describe the quantum dynamics of a particle. %By considering the covariance of the local ontic spacetime period $\lambda'^\mup(X)$, in a generic interaction spacetime point $x=X$ 
Indeed an interacting EC can be regarded as particle in a locally deformed periodic ``spacetime box'' of locally modulated period $\lambda'^\mup(x)$.  Thus, along the evolution of an interacting EC, say in $x'=X$, the quantisation is locally given by the local PBCs $\Phi'_{\vec k}(X) =  \Phi'_{\vec k}(X+ \lambda'(X))$.  Contrarily to the free case, in the interacting case the resulting spectra  are in general non-harmonic due to the deformed ``ontic'' spacetime $g^{\mup\nup}$ (similarly to an non-homogeneous classical vibrating string).  

The equivalence between interacting ECs physics and ordinary quantum description of interacting elementary particles (axioms of QM, commutation relations, Feynman path integral, etc) is obtained by generalising the demonstrations given for the free case from global periodicity to local periodicity.

As already said, in the free case, the generic EC physical state, solution of the free action eq.(\ref{CA:action}) has the form eq.(\ref{field:mode}). From the free EC physical state $\Phi_{\vec k}$, the global PBCs at $\lambda^\mu$, \ie $\Phi_{\vec k}(x) = \Phi_{\vec k}(x + \lambda)$, implies the harmonic quantisation condition ${k_ \mu}_n \lambda^\mu = 2 \pi n$ in every point $x$ of its free evolution, \ie globally. It represents the  generalisation to relativistic free particles of the Bohr-Sommerfeld quantisation condition. From this we have obtained the energy spectrum dispersion relation  of the normally ordered second quantised free field: $\omega_n(\vec k) = n \sqrt{m^2 + \vec k^2}$. 
 
 In the interacting case the eigenmodes of the EC physical state, solution of the equations of motion of the action eq.(\ref{CA:action:int}),  have the generic form of a modulated wave $\phi'_{\vec k'}(x') \propto e^{- i \int^{x'} d y^\mup  k'_\mup(y)}$. As proven in sec.(\ref{INT}), $k'_\mup$ is the four-momentum of the interacting elementary particle described by our locally modulated EC.  The local period $\lambda^\mu(x)$ imposed as constraint to the EC physical state by means of the local PBCs of (\ref{CA:action:int}), implies the local quantisation condition $\oint_{x'} d y^\mup {k'_{\mup}}_n (y) = 2 \pi n $ in the point $x'$ of its evolution characterised by interaction. 
 
 In simple words, the only possible vibrational modes of an interacting EC are those with integer numbers of cycles along a local period, \ie closed orbits, similarly to a non-homegenous vibrating string. This determines the EC local quantised spectrum and the local  momentum eigenstates ${\phi'_{\vec k'}}_n(x') = e^{i \int^{x'} d \vec y \cdot \vec k'_{n}(y)}$. We will see that this description correctly reproduces the quantum interactions from QED to non-relativistic Schr\"odinger problems. 
 
 As for the free case, despite the local character of the periodicity, it is easy to see that in the interactiong case the local momentum eigenstates form locally a complete, orthogonal set, so that they define a \emph{local} Hilbert space of basis $| n \rangle$, such that locally $\langle \vec x | n \rangle\, \dot =\, {\phi'_{\vec k'}}_n (\vec x)$. The related local inner-product is $\langle n' | n \rangle = \delta_{n,n'} $. Thus the evolution of an interacting EC is represented by a point in the corresponding local Hilbert space $|\Phi'_{\vec k'} \rangle = \sum_{n \in \mathbb Z} \alpha_n | n \rangle$. With this we have generalised the axiom of the states. 
 
We are finally able to show that the classical-relativistic cyclic dynamics of interacting ECs are equivalent to the ordinary QM of the corresponding interacting particles. The equations of motion of the interacting ECs are equal to those prescribed by QM for interacting particles. 

It is convenient to introduce a four-momentum Hilbert operator. For the free case, we define the four-momentum operator as $\mathcal P_\mu = \{\mathcal H, - \vec{\mathcal P}\}$. 
According to our geometrodynamical description of interaction, see sec.(\ref{INT}), the local four-momentum operator of the interacting EC is given by the transformation $\mathcal P'_\mup(x) = {e^\mu}_{\mup}(x) \mathcal P_\mu $, where $\mathcal P'_\mup = \{ \mathcal H', - \vec {\mathcal P}' \}$ defines the local Hamiltonian and momentum operators for the interacting case. 

They are Hermitian operators due to the PBCs of the theory, and describe the spectra of the interacting EC in its local Hilbert space. Indeed the EC four-momentum local spectrum is $\mathcal{P'}_\mup(x) | n \rangle = {k'_{\mup}}_n (x)| n \rangle $. We have generalised the axiom of the observables. 

The EC local periodicity implies that the EC evolution is given locally by $i \partial_\mup |\Phi'_{\vec k'} (x)\rangle = \mathcal P'_\mup |\Phi'_{\vec k'} (x)\rangle$. Its time component describes the ordinary time dependent Schr\"odinger equation of  interacting quantum systems  $i \partial_t |\Phi'_{\vec k'} \rangle = \mathcal H'(t)|\Phi'_{\vec k'} \rangle$. By construction it turns out to be written in terms of the same time dependent Hamiltonian operator prescribed by ordinary QM for the corresponding interaction scheme as we are going to show below. We have generalised the axiom of the motion.

With these results at hands,  by following the same steps described in sec.(\ref{AXIOMS}) and by paying particular attention to the role of the local PBCs, it is now straightforward to check  the correspondence of the interacting ECs dynamics to the axiom of the measurement and, in particular to the Bohr rule. 

Similarly it is straightforward to generalise the demonstration given in sec.(\ref{COMM:REL}) and check the validity of the  commutation relations for interacting ECs $[ x_i,  \mathcal P'_j ] = i \delta_{i,j}$.  We conclude that \emph{there is an exact equivalence between ECs dynamics, classical in the essence, and the axiomatic QM and Dirac quantisation prescription in both the free and interacting cases.}  %Similarly to  sec.(\ref{QM:free}) we also obtain the local commutation relations for interacting EC $[ x_i,  \mathcal P'_j ] = i \delta_{i,j}$. 

The generalisation to interactions of the exact equivalence to the Feynman path integral  follows easily by considering that, similarly to the free case, see (\ref{single:elem:path}), the local infinitesimal spacetime evolution operator of our modulated EC is $\mathcal U(d x_m) =  e^{-i (\mathcal H' d t_m - \vec {\mathcal P'} \cdot  d \vec x_m)} $. 

According to our description of interactions, its phase locally defines  the action $\mathcal {S'}^{Class}$, which actually corresponds to the action  of the classical-relativistic particle associated to the EC and interacting under the same generic interaction scheme: $ -\mathcal H' d t_m + \vec{\mathcal P}' \cdot  d \vec x_m = (\vec{\mathcal P}' \cdot   \dot {\vec x}_m - \mathcal H') d t_m = \mathrm L' d t_m = d \mathcal {S'}^{Class}_m$, where ${\mathrm L'}^{Class}  = \vec{\mathcal P}' \cdot   \dot {\vec x}_m - \mathcal H'$ (written in terms of Hilbert operators), so that $\mathcal {S'}^{Class}[t_f, t_i] = \int_{t_i}^{t_f}  {\mathrm L'}^{Class} d t$ . As anticipated, $\mathcal H'$ is therefore the ordinary time dependent Hamiltonian associated to the interaction scheme under consideration. 

The elementary spacetime evolutions of interacting ECs, being written in terms of infinitesimal spacetime intervals, have the same form as for the free case, see eq.(\ref{single:elem:path}), but the constant  Hamiltonian and momentum operators of the free case must be now vreplaced by the local ones of the interacting case, \ie $\mathcal H'$ and $\vec{\mathcal P}'$. 

The product of integrals $\int \mathcal D x$ resulting from eq.(\ref{FPI:elem.paths}) is not trivial in the interacting case (as in the ordinary Feynman formulation) due to the fact that the Hilbert space defined by the interacting EC is local. It takes into account  that in every point of the interacting EC evolution (as for an ordinary interacting quantum particle)  a different, local, complete and orthogonal set of eigenfunctions is defined. 

\emph{In this way we have generalised the equivalence to the Feynman path integral eq.(\ref{FPI:free}) to the interacting case}. In short we have the remarkable result that \emph{the classical evolution of an interacting EC is equivalent to the ordinary Feynman Path Integral for that interacting scheme } 
\beq
\mathcal Z = \int \mathcal D^3 x e^{i \mathcal {S'}^{Class}[t_f, t_i]}\,,
\eeq
where ${\mathcal S'}^{Class}$ is, by construction, the action of the corresponding interacting classical-relativistic particle of mass $m$.

%All the known aspects of QM follows from the axioms of QM and the Feynman path integral, and  the EC evolution is exactly described by these two formulations of QM, we can safely say that EC theory is consistent with all the known and can be used to formaliaspects of QM. 

\subsection{Equivalence to Quantum ElectroDynamics} \label{Equi:QED}
As we have proven in sec.(\ref{GAUGE:CLASS}), classical electromagnetism is directly inferred from the geometrodynamics of the EC ``ontic'' spacetime, without postulating gauge invariance. These geometrodynamics are unitary (polarised) rotations of the EC ``ontic'' spacetime boundary with flat metric. We recall that the local modulation of the EC instantaneous periodicity associated to electromagnetism is $\tau^\mu(x) = \delta^\mup_\mu \lambda^\mu -  e {\xi_\mu}^\mup (x)\lambda^\mu$. Through the local phase harmony this corresponds to the minimal substitution of classical electromagnetism $k'_\mu(x) = k_\mu - e A_\mu(x) $, as soon as we consider our definition of vectorial field $A_\mup \, \dot = \, {\xi^\mu}_\mup k_\mu$.  

In particular we have seen that the resulting dynamics of the EC fundamental state (\ie the EC mode $n=1$) are formally described by the ordinary classical (non-quantised) Yang-Mills action of classical electromagnetism eq.(\ref{CA:action:YM}).   
To derive QED, \ie to extend our description of electromagnetism to QM, we must as usual consider all the possible harmonics allowed to the EC by its local periodicity imposed as constraint by means of the local PBCs: the quantisation condition in ECs theory is the constraint of local intrinsic periodicity.  

In ECs theory, the quantised dynamics of electromagnetism (QED) can be easily inferred by using the formalism of the Hilbert space. According to the equivalence of interacting EC dynamics to QM,  the local four-momentum operator resulting form the  geometrodynamics that, as we have proven, are associated to electromagnetism is $ \mathcal P'_\mup(x) = e_\mup^\mu (x) \mathcal P_\mu = \mathcal P_\mup - e A_\mup(x) $, where $A_\mup$  must now be interpreted as a Hilbert operator. 

From the phase of the interacting EC it follows that the resulting Lagrangian describing this particular case of EC interaction takes the familiar form $ \mathrm L^{EM} = \mathrm L^{free} + e A_\mu J^\mu$ where $J^\mu = d x^\mu / d t$. In other words it turns out to be formally the Lagrangian of an ordinary relativistic particle interacting electromagnetically. Such bosonic description can be extended to ordinary QED by using the formalism of fermionic EC described in sec.(\ref{fermions}).  

By substituting $ \mathrm L^{EM}$ in the action eq.(\ref{FPI:inter}) we finally have that \emph{the full classical-relativistic evolution of such an  EC with $U(1)$ local rotation of the  boundary is exactly described by the ordinary (bosonic)  QED} \cite{Dolce:tune}:
\beq\label{FPI:inter}
\mathcal Z = \int \mathcal D^3 x e^{i \int_{t_i}^{t_f} d t (\mathrm L^{free} + e A_\mu J^\mu) }\,.
\eeq
Notice that in EC theory QED has been directly derived from constrained classical-relativistic geometrodynamics, without postulating gauge invariance, and without relaxing the classical variational principle, and without imposing any quantisation condition except intrinsic periodicity.

ECs description of QED  is particularly convenient as it allows us to easily stress out fundamental quantum aspects of electromagnetism.  For instance, in the Hilbert space formalism, the local modulations associated to these geometrodynamics \emph{w.r.t.} the free case are given by the ordinary scattering matrix of QED $\mathrm S(x) = e^{i e \int^{x^\mu} d x^\mu A_\mu}  $, which is actually the operator associated to the gauge connection $U(x')$ describing the local modulations \emph{w.r.t.} the free case. It is in fact easy to see that the electromagnetic interacting EC in the Hilbert space formalism is given by $|\Phi'_{\vec k'} (x)\rangle = \mathrm S(x) |\Phi_{\vec k}(x)\rangle$.  

From the gauge connection $U(x')$, the PBCs of the interacting EC solution $\Phi'_{\vec k'} (x)$, together with those of the free EC $\Phi_{\vec k}(x)$, directly implies the Dirac quantisation condition for magnetic monopoles: $e \oint_{x'} d y^\mu A_\mu(y) = e g_n = 2\pi n$. 
It has been proven in a number of papers \cite{Dolce:SuperC,Dolce:EPJP,Dolce:Dice2014} that with similar arguments it is also possible to derive, for example, all the fundamental aspects of superconductivity directly from first principles of QM (\ie from intrinsic periodicity) rather than from microscopic, empirical considerations about the materials typical of the ordinary BCS (Bardeen-Cooper-Schrieffer) description, as describe in \cite{Dolce:SuperC,Dolce:EPJP,Dolce:Dice2014,Weinberg:1996kr}. We will give some more detail in the next subsection. 

Similarly ECs allow for a straightforward derivation of the peculiar quantum behaviour of electrons in graphene systems \cite{Dolce:EPJP}.  In particular,  electrons in a carbon nanotubes with $N$ carbon atom along the diameter behaves as  a ECs whose ``ontic'' compact word-line (of Compton length) is on a lattice of $N$ sites (\ie relativistic CA of $N$ sites). In this way it is possible to test explicitly that the negative EC modes correspond to holes in the Dirac sea \cite{Dolce:Dice2014}, in full confirmation of 't Hooft's conjecture. 

\subsection{Elementary Cycles at finite temperature}\label{therm:QM}

To fully appreciate the exact correspondence of ECs dynamics and ordinary QM, and the power of ECs theory in describing non-trivial quantum phenomena, it is not possible not to mention the straightforward description of statistical quantum systems --- field theory at finite temperature --- allowed by the theory. For the sake of simplicity we will only consider time component of Minkowskian and Euclidean  persistent periodicities  characterising respectively  pure quantum systems and to thermal systems at the equilibrium, \ie  isolated ECs and ECs at finite temperature, respectively. 

The  Minkowskian and Euclidean time periodicities  can also be addressed as  purely quantum and thermal periodicities. Summarising we will see that they have opposite physical meaning: Minkowskian periodicity describes the purely recursive phenomena characterising  QM (perfect coherence), the Euclidean periodicity describes the purely dissipative phenomena characterising systems at finite temperature (thermal dissipation). Every quantum system at finite temperature is described by the competition of these two aspects. 

It is well known that the quantisation of classical thermal systems (statistical systems) at temperature $\mathcal T$ is achieved by imposing the constraint of Euclidean time periodicity of duration $\beta = 1 / k_B \mathcal T$, where $k_B$ is the Boltzmann constant and $k_B \mathcal T$ is the thermal energy. 
 This quantisation prescription --- Matsubara theory --- as well as Wick's rotations and analytical continuation --- are often believed to be a mere ``mathematical tricks'', without physical motivations. 
 
 On the contrary, such a mathematical trick has a deep physical motivation which becomes manifest in ECs theory. Actually,  in the ECs theory  the quantisation of classical elementary system of energy $\omega$ is obtained by imposing the intrinsic Minkowskian time periodicity $T = 2 \pi / \omega$ as constraint. Furthermore ECs physics provides a simple explanation, in a unified view, of the correspondence between the partition function of statistical mechanics and the path integral (already derived from ECs dynamics) of QM. The partition function of a quantum statistical system is a direct consequence of cyclic dynamics in the Euclidean time as the Feynman path integral is the direct consequence of cyclic dynamics in Minkowskian time. Indeed, the Matsubara theory (field theory at finite temperature) represents an (further) indirect confirmation of the equivalence between ordinary QM (in allt its fundamental aspects) and cyclic dynamics.  
 
We have seen that an EC is characterised by a persistent periodicity in time $T = 2 \pi / \omega$ such that the physical state is constrained to satisfy PBCs $\Phi_\beta (t) = \Phi_\beta (t + T)$, where we have omitted the label $\vec k$ denoting the dependency on the spatial momentum and we have  introduced a label $\beta$ indicating that the ECs is at finite temperature $\mathcal T$.  By using a terminology close to condensed matter textbooks we can address the Minkowskian periodicity of a free ECs described so far as the condition of ``perfect quantum coherence'' of pure quantum systems (at zero temperature).  Clearly the free isolated ECs described so far refer to quantum particles at zero temperature: they form  perfectly coherent states. 

In simple words, the perfect quantum periodicity (Minkowskian periodicity) characterising ECs is referred to the ideal case of zero temperature (isolated systems: pure quantum systems characterised by perfect coherence), exactly as the uniform rectilinear motion is an ideal case of isolated systems in classical mechanics (isolated systems,  no interactions and in particular no friction according to Newton's first principle). 

We typically do not see the perfect coherence of QM due to the effect of the  thermal noise in ordinary systems, as much as we typically do not typically see objects in pure uniform rectilinear motion due to the effect of friction. Nevertheless QM, and in particular quantum phenomena in condensed mater, see next subsection, can be inferred from the ideal case of perfect intrinsic periodicity as much as classical physics can be inferred from the ideal case of perfect isolated systems.   

Let us now prove that, for an EC is at finite temperature, the Boltzmann probability implies the Euclidean periodicity of finite temperature field theory. 
We recall that, due to the Minkowskian time periodicity (discrete Fourier transformation), the time component of the ECs physical state at temperature $\mathcal T$ is $\Phi_\beta (t) = \sum_{n \in \mathbb Z} {a_\beta}_n \phi_n(t) / \sqrt{(2 \pi)^3 2 \omega_n}$. The energy spectrum is $\omega_n = n \omega$ and the eigenmodes are $\phi_n(t)= e^{-i \omega_n t}/\sqrt{2 \pi}$, see sec.(\ref{CA:EC}). In the thermal case the  Fourier coefficients ${a_\beta}_n$ are determined by the Boltzmann probability and by the Born rule inferred from ECs dynamics in sec.(\ref{AXIOMS}). That is, in the thermal case,  the probability to populate a level (EC harmonic) of energy $\omega_n$ is  proportional to ${a_\beta}_n \propto e^{- \omega_n / k_B \mathcal T} = e^{-  n \omega \beta} = e^{- 2 \pi n \beta  / T}$. 

Notice that, in analogy with our geometrodynamics description of interactions such physical state at finite temperature $\Phi_\beta$  can be directly obtained by assuming the following substitution of the time variable $t \rightarrow t' = (t - i \tau)_{\tau = \beta} $ in the unitary physical state  $\hat \Phi(t - i \beta) =  \Phi_\beta (t)$ --- neglecting normalisation factors.    

From the definition of Hamiltonian operator for a free EC we can associate to ${a_\beta}_n$ the operator $e^{- \beta \mathcal H}$ such that $e^{- \beta \mathcal H}| n \rangle = {a_\beta}_n | n \rangle$. From the trace of this operator we can now define the partition function $Z$ of a free EC at temperature $\mathcal T$: $Z = \text{Tr}(e^{-\beta \mathcal H}) = \sum_n \langle n | e^{-\beta \mathcal H} | n\rangle = 1 / (1 - e^{ \beta \omega}) = 1 / (1 - e^{2 \pi \beta / T}) $. For instance, the mean energy of an EC at temperature $\mathcal T$ is $E = \omega N $ where $N = 1 / (e^{ 2 \pi \beta / T} - 1)$.  

 In all these relations we notice that the Euclidean and the Minkowskian periods $\beta$ and $T$ appear in competitions each other, \ie in the ratio $\beta / T$. On one hand the phasor of a free EC has the generic form $e^{- i n \omega t}$. It describes an perfect periodic phenomenon of persistent periodicity, which is the peculiar character of pure isolated quantum systems (perfect coherence) and, in particular, of quantum systems at zero temperature. On the other hand  the thermal coefficient has the generic form  $e^{- 2 \pi n \beta  / T}$. It describes the dumping, \ie a dissipation, of the Minkowskian periodicity of pure quantum phenomena resulting from the thermal diffusion (gaussian law). The Minkowskian periodicity $T$ encodes a perfect quantum coherence whereas  $\beta$ encodes the thermal \emph{dissipation} associated to  the chaotic collisions responsible for the thermal noise.   
 
This competition between pure time periodicity (Minkowskian periodicity) and thermal noise (Euclidean periodicity) can be easily understood if we consider the nature of temperature in statistical systems. Temperature is the manifestation of the so called thermal noise, \ie of the chaotic collisions among particles, \ie among ECs. Clearly, as the ECs time periods are determined by their energies, the continuous collisions (sudden variations of energies) resulting from the thermal noise leads to a dumping (dissipation) of the ECs Minkowskian periodicities which is continuously broken by the thermal noise. Such thermal dumping is therefore represented by an exponential decay of the ECs periodicities. It can be actually obtained by replacing the (Minkowskian) time in the phasor $e^{- i n \omega t} = e^{- i 2 \pi n  t / T}$ with an imaginary (Euclidean) time in order to get the Boltzmann factors  $e^{- 2 \pi n \beta  / \mathcal T}$.

  These considerations are interesting to clarify the physical meaning  of the mathematical trick of the Wick's rotation and analytical continuation.  Indeed, by applying a Wick rotation to a field theory, or other theories based on undulatory mechanics such as ECs theory, we pass from pure coherent phenomena (no dissipation of periodicity) characterising QM and encodes in the imaginary exponential (phasor) $e^{- i n \omega t} = e^{- i 2 \pi n  t / T}$ (imaginary exponential) to the dissipative phenomena characterising systems at finite temperature (thermal noise) and encoded by the dumping factor  $e^{- 2 \pi n \beta  / \mathcal T}$ (real exponential). Summarising we have the following correspondence: QM describes pure recursive phenomena, upon Wick's rotation, describes dissipation, \ie thermal phenomena \cite{deCordoba:2013cda,licata:2015}. 

 In order to encode the effect of the dumping of the periodic behaviour associated to the thermal noise, the temperature $\mathcal T$ can be therefore parametrized in ECs physics as an Euclidean time of value $ \beta$. Obviously such Euclidean time coordinate $\tau$ doesn't flow contrarily to the Minkowskian time coordinate such that $\tau = \beta$. 
Furthermore, the partition function introduced above tells us that such Euclidean time has an intrinsic periodicity $\beta$ \cite{Zinn-Justin:2002ru,Kapusta:1989tk}.
 
  \emph{Vice versa},  it is now straightforward to generalise the demonstration of the Feynman path integral from the cyclic dynamics of duration $t_i - t_f$ associated to an intrinsic Minkowskian periodicity $\Phi(\vec x, t) = \Phi(\vec x,  t+ T)$, see sec.(\ref{FPI:EC}), to find that  cyclic Euclidean dynamics of duration $\beta$ and periodicity  $\Phi(\vec x, 0) = \Phi(\vec x, i \beta)$  are described by the ordinary partition function of ordinary quantum statistical mechanics, instead of the Feynman path integral.
  
  Formally, by following the same demonstration in sec.(\ref{FPI:EC}) with Euclidean time, it is easy to prove that the cyclic classical evolution of an Euclidean EC is described by the   partition function of ordinary quantum statistical mechanics
\beq\label{PART:FUNC}
Z = \int \mathcal D \vec x e^{-\mathcal S^{Class}[\beta,0]}
\eeq
where, in perfect correspondence to  our derivation of the Feynman path integral in sec.(\ref{FPI:EC}), $\vec x$ has spatial periodicity $\vec \lambda_\beta$ resulting from the Euclidean periodicity $\beta $ exactly as the same term in the ordinary path integral eq.(\ref{FPI:free}) has spatial periodicity $\vec \lambda$ resulting from the Minkowskian periodicity $T$; $\mathcal S^{Class}[\beta]$ is the classical action corresponding to a time interval of duration $\beta$:  $\mathcal S^{Class}[\beta] = \int_0^\beta L^{Class} d t$. 

ECs theory reveals a perfect correspondence between Euclidean and Minkowskian periodicity. We have proven that the Feynman path integral eq.(\ref{FPI:free}) describes the evolution of duration $t_f$ and $t_i$ characterised by Minkowskian cyclic dynamics of period $T= 2 \pi / \omega$. Similarly the partition function eq.(\ref{PART:FUNC}) describes the evolution of duration $\beta$ characterised by Euclidean cyclic dynamics of period $\beta$.  

  Once again we have found an explicit confirmation of the fact that quantum dynamics, in this case of statistical systems, are obtained by constraining the classical dynamics, in this case in the Euclidean time.  In short \emph{these arguments reveal the physical origin of the correspondence between Feynman path integral and partition function}. More in general they reveals the deep relationship, upon Wick's rotation, of QM and thermodynamics: both are statistical theories emerging from Minkowskian and Euclidean cyclic dynamics, respectively. 
  
  Since thermal dynamics are essentially consequence of the statistical behaviour of composite systems of elementary particles, and the latter are Minkowskian periodic phenomenon of persistence periodicity if observed in an isolated state (or in their reference frames), we can conjecture that elementary particles are systems at zero entropy whereas composite system have an high content of entropy for statistical reasons (see chaotic evolution of composite systems of ECs). This conciliates the  irreversibility of the arrow of time, which has a statistical motivation as we have just seen, with the relativistic description of time which is reversible and intrinsically mixed with space through  Lorentz transformations.

  The analogous of the energy eigenvalues $\omega_n = n \omega =  2 \pi n / T$ of an isolated quantum system directly derived from  the Minkowskian periodicity, are the Matsubara frequencies $ 2 n \pi / \beta$ directly derivable from the condition of Euclidean periodicity. Thanks to ECs physics, we have actually obtained in a natural way the ordinary description on quantum statistical systems, \ie of  quantum field theory at finite temperature, see \cite{Kapusta:1989tk,Zinn-Justin:2000dr}. Notice that, in the case of fermions the PBCs of the bosonic particles must be replaced by anti-PBCs as already discussed, in order to encode the Dirac-Fermi statistics.

Thus ECs theory yields a unified description between the partition function of the Boltzmann formulation of statistical mechanics and the path integral of Feynman formulation of QM. The quantisation of statistical systems, which is obtained by imposing intrinsic periodicity in the Euclidean time, is nothing but a direct consequence of the fact ordinary QM is the manifestation of a intrinsic periodicity in the Minkowskian time, as prescribed by ECs theory. 
 
 \subsubsection{Quantum coherence, collective phenomena and superconductivity}
 
Summarising the description above on one hand we have that the Minkow\-skian time periodicity (imaginary exponential)  describes the perfect coherence of pure quantum systems.  On the other hand, the Euclidean time periodicity (real exponential) describes the dissipative phenomena associated to the thermal noise.    The classical Minkowskian cyclic dynamics yields the Feynman path integral of QM, sec.(\ref{FPI:EC}) as the classical Euclidean cyclic dynamics yields the partition function describing the quantum behaviour of statistical systems.  From these considerations  it is now easy to infer, in a very straightforward and novel way,  fundamental aspects of condensed matter such as superconductivity directly from first principles of QM (\ie intrinsic periodicity) rather than from phenomenological aspects like in BCS theory, which is actually an empirical theory. All these aspects are investigated in \cite{Dolce:Dice2014} and similar papers \cite{Dolce:SuperC,Dolce:EPJP}. 

In classical mechanics the natural state of elementary isolated objects is persistent linear motion, \ie constant energy (and momentum). According to de Broglie constant energy means persistent periodicity. Hence the natural state of elementary isolated systems is perfect Minkowskian intrinsic recurrence. We have proven that, by promoting this to a postulate, Minkowskian period dynamics are equivalent to QM in all its aspects. This also means that, according to ECs physics, the natural state of elementary free systems is pure QM. The thermal (non-quantum) limit mechanics emerges from this ideal (quantum) case as a consequence of the thermal noise (Euclidean periodicity) as much as in classical mechanics the natural state of bodies is in uniform rectilinear motion and their tendency to stay at rest is due to the friction.

To elucidate some central concepts let us consider the Black-Body radiation. According to ECs theory, every component of the electromagnetic radiation with fundamental angular frequency $\omega(\vec k)$, \ie of periodicity $T(\vec k)= 2 \pi / \omega(\vec k)$, is an EC of periodicity $T(\vec k)$ and infinite Compton periodicity $T_C = T(0)$. Thus, through discrete Fourier transform, we have the Planck spectrum $\omega_n(\vec k) = 2 \pi n / T(\vec k)$ for each component. 

For those components of the electromagnetic radiation whose quantum (Minkowskian) periodicity $T(\vec k)$ is small \emph{w.r.t.} the thermal period $\beta$, \ie in the limit $T(\vec k) \ll \beta$, we have purely quantum behaviour  in the sense that the thermal noise is not sufficiently fast (\ie the thermal period $\beta$ is not sufficiently short) to break the ``perfect coherence'' of pure quantum systems, \ie the Minkowskian periodicity of ECs. The periodic behaviour characterising QM is preserved in this limit,in agreement with quantum phenomenology.  In this case only the lower harmonics of the EC are excited and if the periodicity is sufficiently low they can condensate on the fundamental mode ($n = 1$) generating collective phenomena such as the Bose-Einstein condensation.  The quantum character is preserved for the ultra-violet components of the electromagnetic radiation, as for a string vibrating with very short period the harmonics are well separated, see the analogy between the Black-Body radiation and the sound spectrum of  grand-piano given in \cite{Dolce:licata} for more details. 

In the infra-red region of the electromagnetic radiation the thermal (Euclidean) period $\beta$ is too short \emph{w.r.t.} the quantum (Minkowskian) periodicity characterising quantum system.s This means that the thermal noise is so intense that the pure periodic (quantum) behaviour is broken. In turn the phenomena associated with it (which are equivalent to the quantum phenomena according to ECs physics)  cannot take place. Due to the fast thermal scattering, the quantum periodicity  cannot autocorrelate to give rise to the perfect coherence and collective phenomena. Also, the vibrational modes, \ie the energy spectrum, can be approximated to a continuum, as for a string vibrating with very long or infinite periodicity which can vibrate with a continuum of harmonics since the BCs can be neglected in this case. Summarising we have obtained the ordinary Planck description of the Black-Body radiation, avoiding the ultra-violet catastrophe.   
 
From simple considerations about the periodic behaviour of ECs it is straightforward to derive, for instance, superconductivity in all its fundamental phenomenology such as the Meissner effect, the Josephson effect, the Little-Park effect, the gap opening and so on \cite{Dolce:SuperC,Dolce:EPJP,Dolce:Dice2014}.  If the temperature is sufficiently law the electrons in a conductor behave as (non-relativistic) ECs. They are constrained by a condition of (anti-)periodicity (see description of fermionic dynamics), \ie closed orbits along the circuit $\Sigma$, similarly to the electron in the atomic orbitals (Bohr atom) which actually can be regarded in a superconducting state (see next section). Furthermore the electrons interact electromagnetically so that under the condition of (anti-)periodicity, we have the Dirac quantisation condition for the electromagnetic field 
$e \oint_{\Sigma} d x^\mu A_\mu(x) = e g_n / 2 = n \pi $, where $\Sigma$ denotes the path along the closed circuit, \cite{Dolce:SuperC,Dolce:EPJP,Dolce:Dice2014}. See also equivalence to QED, sec.(\ref{Equi:QED}).
The factor $2$ in this equation is due to the fact that anti-periodicity can be described in an orbifold $\mathbb S^2 / \mathcal Z_2$ in analogy to extra-dimensional theories. 

In short, \cite{Dolce:cycles,Dolce:Dice2014,Dolce:SuperC,Dolce:EPJP}, the spatial components of the Dirac quantisation condition directly leads to the Meissner effect: $\int_{S_{\Sigma}} \vec B(t, \vec x) \cdot d S_{\Sigma} = \oint_{\Sigma} d \vec x \cdot  \vec A(t, \vec x) %= \oint  \nabla \theta (t, \vec x)
= n {\varphi_0}/{2}$ where $S_{\Sigma}$ is the area delimited by the circuit ${\Sigma}$ and $\varphi_0 = 2 \pi / e$ is the quantum unit of magnetic flux. Hence we have found that the magnetic flux through the circuit is quantised in units $\varphi_0$, so that the current cannot smoothly decay and there is not electric resistance, \ie we have no ordinary electric resistence. The Minkowskian periodicity of ECs theory also means that the local phase $\theta(x)$ of the electromagnetic gauge invariance in the conductor can only vary by finite steps $n \varphi_0 /2$. The local phase plays the role of the Goldstone which actually transform as a fermionic condensate of charge $-2e$ which corresponds to the Cooper pair. 

This also correctly describes   the effective breaking of the electromagnetic gauge invariance in superconductors \cite{Weinberg:1996kr,Dolce:SuperC}.   
Interesting enough, such a novel description of superconductivity based on purely geometrical arguments should imply a novel and more fundamental interpretation of the corresponding phenomenon in particle physics, \ie of the \emph{Higgs mechanism} in terms of spacetime geometrodynamics \cite{Dolce:SuperC,Dolce:EPJP,Dolce:Dice2014}.   This would eventually unify the meaning of mass in quantum mechanics (Compton periodicity), in GR (spacetime curvature) and in the Higgs mechanism (gauge symmetry breaking). 

 From the temporal component of the Dirac quantisation condition follows the Josephson effect. If we assume that the circuit contains a junction with a voltage difference $\Delta V$ at its ends, the BCs now must be applied to the ends of the junction so that we get $\int_{junct} A_0 d t = T_{Junct} \Delta V = \varphi_0 / 2$ corresponding to the fundamental Josephson frequency $f_{Junct} = 1 / T_{Junct}$. With similar simple arguments the whole phenomenology of superconductivity can be derived as proven in \cite{Dolce:cycles,Dolce:Dice2014,Dolce:SuperC,Dolce:EPJP}. % In this relation we have also considered that the local phase of the electrons $\theta  (x)$ can only vary of finite steps as can also be seen explicitly from the anti-PBCs of the electrons: $ e \theta  (x) = \theta  (x +  \lambda) + \text{mod} n \pi$.  

\subsection{Non-relativistic limit of the Elementary Cycles dynamics}

We have demonstrated the equivalence of ECs classical-relativistic dynamics and ordinary relativistic QM in both the free and interacting cases, in particular for QED, as well as for statistical systems. However there are two relevant aspects to clarify explicitly. The first one is to check that, actually, the non-relativistic limit of our theory reproduces non-relativistic QM. This correspondence is essentially a double check: non-relativistic QM is already implicit in the results obtained above, as it is a limit of relativistic QM.  The second aspect is the necessity of a more rigorous formal description in ECs theory of composite systems. We will show that, by simple arguments inferred from ECs physics, this implies the introduction of the tensor product of Hilbert spaces, which plays a crucial role for instance in the derivation of Bell's theorem and of the Fock space, and has an intuitive physical meaning in ECs theory.

%To derive the remaining aspects of QM from CA classical dynamics we now consider the non-relativistic limit. 
\subsubsection{Non-relativistic free particle and the classical particle limit}

The non-relativistic limit of a free particle is characterised by small spatial momentum \emph{w.r.t.} the rest energy (the mass $m$), $|\vec k| \ll m $ (here we only consider massive particles as the massless particles are always relativistic, see description of the Black Body radiation). The mass forms an infinite energy gap. In the non-relativistic limit it can therefore omitted yielding the non-relativistic energy dispersion relation: $\omega(\vec k) = \sqrt{\vec k^2 + m^2} = m + \frac{\vec k^2}{2 m} + \mathcal O(\vec k^2)$, so that  $\omega^{class}(\vec k) = \frac{\vec k^2}{2 m}$ is the classical dispersion relation for a non-relativistic EC.  

This limit can be obtained, on one hand, by assuming that the EC rest period (Compton period) is so small that it can be approximated to zero $T_C  \rightarrow 0$ (\ie $m \rightarrow \infty$). On the other hand, it can be obtained by assuming that the momentum is so small that  the EC spatial period tends to infinity: $|\vec k| \rightarrow 0 \Rightarrow |\vec \lambda| \rightarrow \infty$. 

Notice that the rest (Compton) periodicity represents the upper limit of the EC time period $T(\vec k) \leq T_C$. The internal temporal cyclic dynamics associated to the Compton clock can be therefore neglected for a free non-relativistic EC. In this limit, the EC can be effectively described by a non-compact 3D ``ontic'' space $\mathbb R^3$ as the compactification lengths (\ie the period) of the ``ontic'' time and space tend to zero and infinity, respectively. That is, we obtain the ordinary non-compact space $\mathbb R^3$ and universal (external) time coordinate $t$ of ordinary non-relativistic classical mechanics (Galileian relativism). 

The vanishing EC time period implies that the gap between the EC energy levels tends to infinity, $\omega(\vec k) = 2 \pi / T(\vec k) \rightarrow \infty$. Only the fundamental EC harmonic ($n = 1$) can be populated in this limit --- see for instance the  discussion about Boltzmann distribution in the previous section. The fundamental mode ($n=1$) as we have seen  describes in fact non-quantised physics. In other words the quantisation associated to the constraint of EC Compton periodicity is lost as there is not sufficient energy to populate the higher energy levels. On the fundamental mode, the energy $\omega^{class}(\vec k)$ varies in a continuous way with $\vec K$ according to the classical dispersion relation. That is, we have the ordinary classical variation of the energy without quantization. The EC physical state in this approximation becomes $\Phi_{\vec k} \propto e^{-i\omega_\mu x^\mu} \simeq e^{-i (m t + \vec k^2 t /2 m - \vec k \cdot \vec x) } \rightarrow \Phi^{Class}_{\vec k}(x) \propto e^{-i (\vec k^2 t /2 m  - \vec k \cdot \vec x) }$. Actually, we have obtained, starting from EC dynamics, the ordinary description of the non-relativistic free particles  of standard QM.

In general, by plotting the modulo square of the EC physical state $|\Phi_{\vec k}(x)|^2$ (it is convenient to subtract the Compton ultra-fast oscillating term $e^{-i m t}$ and to assume, for instance, the Fourier coefficients $a_n$ of a thermal state or a coherent state), it is possible to check graphically that the EC physical state is localised predominantly inside a region of Compton width ($\sim 2\pi / m$) along the path of the corresponding classical particle \cite{Dolce:2009ce,Dolce:2009cev4}. The Compton width tends to zero in the non-relativistic limit. In the massive case, the EC physical state, which relativistically behaves as a ``localised'' wave,  becomes a Dirac delta distribution centred on the classical path of the corresponding classical particle in the free non-relativistic limit. We have proven that in the non-relativistic limit a free EC behaves as a free classical particle.

From these considerations it is also possible to give, for instance, an interpretation of the double-slit experiment \cite{Dolce:2009ce}. A quantum particle gives self-interference only if the slits are closer than its Compton length. With such a spatial resolution better than the Compton length the EC can no longer be approximated to a Dirac delta, it reveals its wave (harmonic) nature. Similar arguments apply by increasing the resolution in time in order to pass from the non-relativistic to the relativistic description. The effect of the intrinsic periodicity cannot be neglected in this limit. By increasing the resolution in time or space it is possible to resolve more and more harmonics (we pass for a sinusoidal wave description to a description in which we can resolve the ``timbre'' of the wave, \ie its harmonic content) of the ECs physical state. This corresponds to particles creation and, more in general, to the fact that the number of particles is not an observable in relativistic QM according to the Heisenberg uncertainty principle. 

 In general it is easy to prove  that the single particle description emerges from EC dynamics as soon as the EC time period can be approximated to  zero. This also implies, for instance, that the electromagnetic waves with very high frequencies (UV region) has a massless corpuscular behaviour, \ie the electromagnetic radiation is composed by photons --- see Black Body radiation in previous section.   
 
\subsubsection{Non-relativistic Schr\"odinger problems and semi-classical quantum mechanics}\label{nonrel:prob}

Relativistic ECs dynamics are characterised by intrinsic periodic behaviours (of fundamental topology $\mathbb S^1$) which in turns exactly reproduces QM. In the non-relativistic limit, even though the Compton periodicity can be approximated to zero, the peculiar cyclic nature of an EC becomes again manifest as soon as the EC is bounded by a potential $V(\vec x)$, for instance in an infinite well or in a harmonic potential.  The intrinsic cyclic behaviour (PBCs) implies, as we have seen for relativistic ECs, that the EC physical state is a superposition of all the possible eigenstates (harmonics) corresponding to the possible closed orbits, \ie of all the orbits characterised by an integer number of recurrences, according to the local phase harmony (in analogy to a vibrating string in which the closed orbits are the possible harmonics).  %That is, as usual --- see the relativistic case --- the only possible ECs evolutions are  only those with closed spatial and temporal orbits.

An non-relativistic EC bounded in a potential is a generic superposition of eigenvectors $e^{-i [{\omega'}_n^{class}(\vec k') t - \vec k'_n \cdot \vec x] }$. The quantised spectra are determined by imposing the PBCs resulting from the bounding potential $V(\vec x)$. As already noticed for interacting ECs, sec.(\ref{INT}), in the relativistic case these PBCs leads to the relativistic quantisation condition ${\oint k'_{\mu}}_n d x^\mu = 2 \pi n$ which actually means closed spacetime orbits. From the spatial and temporal components of this condition we find that, in the non-relativistic limit, the EC intrinsic periodicity yields the Bohr-Sommerfeld quantisation condition $\oint \vec k'_n \cdot d \vec x = 2 \pi (n + \nu)$, and the energy quantisation condition $\oint \vec \omega^{Int}_{n, Class}(\vec k') d t = 2 \pi (n + \nu)$ where $\omega^{Class}_{n, Int}(\vec k') = \frac{\vec {k'}_n^2}{2 m} + V(\vec x)$.  

The EC   instantaneous time period $\tau^0 = T^{Class}(\vec k')= 2 \pi / \omega_{Int}^{Class}(\vec k')$ and the EC  instantaneous wave-length (spatial period) $\tau^i = \lambda^{i,Class}= 2 \pi / {k'}_{i,Int}^{Class}$ are therefore determined locally by the potential $V(\vec x)$, in analogy with the geometrodynamical description of EC relativistic interactions.  We have included a Morse factor $\nu$ (arbitrary global phase of an EC) which is determined by the BCs at the spatial infinite. It can also be regarded as a (``unphysical'' \cite{'tHooft:2001ct,'tHooft:2001fb,'tHooft:2001ar} ) twist  factor $2 \pi \nu$ in the PBCs along the orbits (which otherwise is arbitrary due to the phase invariance of the EC). 

It is interesting to mention here that, actually, the Casimir effect, which is commonly associated to the zero point energy, was originally calculated in terms of BCs in analogy to Van der Walls forces \cite{Jaffe:2005vp}. It is not a case that the modern method to calculate the Casimir forces in complicated geometries is based on the imposition of the corresponding BCs to the electromagnetic radiation, in full confirmation of EC approach to QM. 

In analogy with the WKB method of ordinary QM it is now possible to exactly solve, in complete agreement with ordinary QM, non-relativistic quantum problems \cite{Dolce:Dice2012}. According to our description of interactions the EC Hamiltonian operator is identical to that prescribed by ordinary QM for a potential $V(\vec x)$.  The non-relativistic EC classical evolution is therefore described by the ordinary Schr\"odinger equation for that interaction scheme $i \partial_t |\Phi'_{\vec k'} \rangle = [\frac{{\vec {\mathcal P'}}^2}{2 m} + V(\vec x)] |\Phi'_{\vec k'} \rangle$.  

The illustrative simplest cases are those which can be explicitly reduced to an harmonic behaviour in time or space. For the sake of simplicity we consider only one spatial dimension. In an infinite potential well of size $L$  the resulting  EC global spatial  periodicity (wavelength) $2L$ yields the harmonic quantisation of the momentum $|\vec k_n| = n / 2 L$. This harmonic spectrum leads, through the equations of motion, to the corresponding energy spectrum $\omega_n = n^2 / 8 m L^2$ (the EC ``ontic'' time is deformed). This quantised solution describes the classical EC closed orbits in time and space allowed by the infinite potential well.  In this simple case we have PBCs in flat ECs ``ontic'' spatial dimensions of length $2L$, just like a vibrating string. The only harmonics allowed for the EC on an infinite potential well of width $L$ are those with closed orbits of length $2L$.

 To solve QHO it is sufficient to consider the pendulum isochronism, that is the fact that all the orbits (all the energies) have global period $T$. This corresponds to PBCs in the flat EC  time of period $T=\tau^0$, which directly implies the harmonic energy spectrum $\omega_n = (n + \frac{1}{2}) 2 \pi /  T$. The zero point energy $\omega/2$ can be associated to a twist factor $\pi$ in the PBCs or, as already said about the Morse factor, to the BCs at the spatial infinity of the EC physical state. From the harmonic energy spectrum the momentum spectrum of the QHO follows as in ordinary QM. Notice that such a description of the QHO represents a further confirmation of the full correspondence between ECs physics and second quantisation.  Obviously the QHO can be equivalently described by means of ladder operators introduced in sec.(\ref{sec:quant}).  Also, this is a cross check of the equivalence between ECs and QFT, as the QHO is the building block of second quantised  fields.

 In a generic potential $V(\vec x)$ we must work in the analogy with non-homogeneous strings whose spectra are in general not harmonic. It is easy to check that possible to show that the EC quantisation prescription (the requirement of close orbits in space and time) leads, in a very straightforward way if compared with  the ordinary methods of QM, to \emph{the correct solution of all the possible Schr\"odinger problems}, also those characterised by non-trivial potentials such as the anharmonic quantum oscillator (we obtain $\omega_n = \frac{3}{4}\eps(2 n^2 + 2 n)$, where $\epsilon$ is the quartic correction $\epsilon x^4/l$ to the harmonic potential and $	\sqrt{2 \pi / m f}$), the linear  potential (we obtain $\omega_n =[3 \pi ( + 1/4)^{2/3})] m g^2 /2 $ where the linear potential is $ m g x$), and so on,  \cite{Dolce:Dice2012,refId0}. Ordinary problems such the Dirac delta potentials and  tunnel effect can be explicitly solved by means of BCs, as for ordinary QM, so that they fully confirm the full consistence of EC formulation of QM.  
 
  Below we will explicitly discuss the atomic orbitals. Notice that ECs theory brings new elements for an improved formulation of the Bohr-Sommerfeld quantisation and WKB method, which in this way can be applied to solve exactly (and not in an approximative way as believed) relativistic QM as well as quantum problems with more particles. In particular our analysis of the QHO shows that the whole QFT can have an equivalent, exact, completely semi-classical formulation. 

\subsubsection{Atomic orbitals, tensor product of Hilbert spaces and Fock space}\label{FOCK:HILBERT}
Similarly to Bohr's description of hydrogen atom, by means of the recipe given above to solve Schr\"odinger problems it is easy to prove that  the locally modulated closed spacetime orbits of an EC bounded in an atomic Coulomb potential imply the atomic energy spectrum $\omega_n = -{13.6 ~\text{eV}}/{n^2}$. Notice that, for what which concerns the quantisation of the atomic energy levels, contrarily to Bohr's derivation in ECs theory it is not necessary to restrict our choice to circular spatial orbits as the EC closed orbits are in spacetime of topology $\mathbb S^1$.  

 So far we have only considered the intrinsic periodicity characterising the ECs spacetime closed obits, but this is not the only possible periodicity for an EC. Recall that an EC is a one-dimensional periodic phenomenon (closed string) vibrating in the four-dimensional spacetime. Its fundamental topology is that of the circle $\mathbb S^1$. This periodicity implies a single quantum number $n$ labelling the possible spacetime closed orbits. As we have seen this is the ordinary principal quantum number of QM. It correctly describes the related quantisation of the energy and momentum in all the possible case, including the atomic orbitals as we have just seen. The energy and momentum spectra are both described by the same quantum number $n$. They correspond to the temporal and spatial periodicities projected by the intrinsic EC periodicity (e.g. by the Compton periodicity in the relativistic case). Indeed the energy and momentum spectra are related by the equations of motion.  
 
 Peculiar configurations of the ECs may have additional fundamental periodicities \emph{w.r.t.} the intrinsic ones of fundamental topology $\mathbb S^1$. In close analogy with the quantisation of the energy-momentum described so far, these possible additional fundamental periodicities imply other quantised quantities (\ie the conjugated quantities) labelled by additional quantum numbers, one for each fundamental periodic (angular) parameter.  
 
We may consider, for instance, isotropic potentials as the Coulomb potential, \ie potentials characterised by a spherical symmetry $\mathbb S^2$. In addition to the quantisation of the energy-momentum spectrum described by the principal quantum number $n$ and directly associated to the EC intrinsic periodicity $\mathbb S^1$, this additional spherical symmetry $\mathbb S^2$ implies the further periodic conditions of the two spherical angles for the EC ``ontic'' physical state, $\varphi \in ( 0 , 2 \pi]$ and $\vartheta \in ( 0 , \pi]$. Obviously this spherical symmetry yields the ordinary quantisation of the angular momentum and the further decomposition of the EC physical state $\Phi_{\vec k}$ in spherical harmonics, in perfect agreement with ordinary QM. The two angular variables associated to the spheric geometry imply the two additional quantum numbers typically denoted by $m$ and $l$. 

It is interesting to notice that the energy-momentum quantisation in terms of spacetime cyclic dynamics  can be also regarded as the spacetime analogous of the quantisation of the angular momentum of ordinary QM, which is a perfectly valid and universally accepted quantisation method based on the constraint of intrinsic periodicity, even in QFT.  \emph{In ECs physics it is spacetime itself which plays the role of an angular variable, and the quantisation of the energy-momentum is the analogous of the quantisation of the angular momentum}. ECs predicts that spacetime coordinates are angular coordinates.

In general to every fundamental angular variable parametrizing a physical system is associated a quantum number and the quantisation of the corresponding conjugated quantity. For instance, the principal quantum number $n$ describing the quantisation of the energy-momentum is associated the to EC intrinsic periodicity;  $m,l$ are associated to a spherical periodicity describing the quantisation of the angular momentum and so on.

 Every fundamental periodicity $\theta_1, \theta_2, \theta_3, \dots$ implies an independent  set of orthogonal and complete eigenfunctions (harmonics)  constituting the bases $|n_1\rangle$, $|n_2\rangle$, $|n_3\rangle$, $\dots$ of independent Hilbert spaces $\mathcal H_1$, $\mathcal H_2$, $\mathcal H_3$, $\dots$ labeled by the indexes $n_1$, $n_2$, $n_3$, $\dots$, respectively.  In short, a system characterised by more fundamental periodicities is described by the tensor product of the Hilbert spaces defined by each fundamental periodicity $\mathcal H = \mathcal H_1 \otimes \mathcal H_2 \otimes \mathcal H_3 \dots$.  

Such composition of fundamental periodicities is clearly illustrated by the example of the atomic orbitals. In order to determine the atomic orbitals, beside the closed spacetime orbits yielding the atomic energy levels labelled by $n$,  we now have to consider that the Coulomb potential has a spherical symmetry. This means that the EC physical state is constrained to have spherical symmetry. Such an additional periodicity $\mathbb S^2$, parametrized by the two spheric angles, implies that the EC physical state is a point in the tensor product of the two fundamental Hilbert spaces $|n, m,l \rangle = |n\rangle \otimes |m,l \rangle$, where   $|m,l \rangle$  denotes the spherical harmonics in the Hilbert space formalism. These two Hilbert spaces are associated to the closed EC orbits in spacetime, of fundamental topology $\mathbb S^1$, and to the closed spherical orbits associate to the spherical topology  $\mathbb S^2$, respectively. The fundamental topology of the atomic orbitals is therefore $\mathbb S^1 \otimes \mathbb S^2$ (neglecting the spin). 
%Thus, in general, the composition of many independent periodicities of the ontic CA coordinates are represented by the tensor product of the corresponding fundamental Hilbert spaces (the harmonics associated to each fundamental periodicity form independent complete sets of eigenfunctions). 

\emph{We have thus obtained the ordinary quantum description of the atomics orbitals}. 
Contrarily to the common opinion, it has been shown that a similar semi-classical description, if correctly applied as also prescribed by the ECs theory, can consistently describe the Zeeman effect and the other fundamental phenomenology of atomic physics \cite{refId0,Dolce:cycles,Dolce:Dice}.

The product of Hilbert spaces is also relevant as it is at the base of the Fock space. Two distinct relativistic ECs (``beables'' in 't Hooft's terminology), having two independent spacetime intrinsic periodicities,  are represented by the tensor product of the two corresponding Hilbert spaces  $|n_1 \rangle \otimes |n_2 \rangle$, with quantum numbers $n_1$ and $n_2$.  The fundamental topology of this composite system is thus $\mathbb S^1 \otimes \mathbb S^1$ (now we do not consider the spherical periodicity). 

Similar arguments can be generalised  to the description of the same EC observed in two different kinematical states, \ie observed from two different inertial reference frames $\vec k_1$ and $\vec k_2$. These are described by two Hilbert spaces, $\mathcal H_{\vec k_1}$ and $\mathcal H_{\vec k_1}$ of bases $|n_{\vec k_1} \rangle$ and $|n_{\vec k_2}\rangle$, respectively. The same EC observed from two different inertial reference frames indeed is characterised by two different complete and orthogonal sets of eigenfunctions, labelled by $n_{\vec k_1} $ and $n_{\vec k_2} $, respectively. The resulting Hilbert space $\mathcal H_{\vec k_1} \otimes \mathcal H_{\vec k_2}$ has basis $|n_{\vec k_1}, n_{\vec k_2} \rangle = |n_{\vec k_1} \rangle  \otimes |n_{\vec k_2} \rangle$. 

Clearly, by iterating this composition of Hilbert spaces associated to the same EC observed from all the possible reference frames, one obtains the Fock space. The description of an arbitrary large number $N$ of the possible kinematical states of the same EC is in fact represented by a Hilbert space $\mathcal H^{\otimes N} = \mathcal H_{\vec k_1} \otimes \dots \otimes \mathcal H_{\vec k_N}$ of basis $|n_{\vec k_1}, \dots, n_{\vec k_N} \rangle$. We this we have proven that an ECs is described by the ordinary Fock space of QFT. 

From this also follows that a quantum field of ordinary QFT describes an EC in all its possible kinematical states. We have actually seen in sec.(\ref{sec:quant}) that the physical state $\Phi_{\vec k}(x)$ can be identified with the component of momentum $\vec k$ of an ordinary quantum field. We have in fact also seen that the Klein-Gordon field  $\Phi_{KG}(x)$ is the integral of  the physical state $\Phi_{\vec k}(x)$ over all its possible momenta: $\Phi_{KG}(x) = \int d \vec k \Phi_{\vec k}(x)$. Besides the definitions of Fock space, the product of Hilbert spaces plays a fundamental role in the demonstration of Bell's theorem that we will shortly discuss in the next section.    
 
 Notice that, in addition to the exact equivalence to axiomatic QM, to the Feynman path integral, to second quantisation and so on,  \emph{the tensor product of Hilbert spaces was the very last ingredient to complete the exact mathematical equivalence between ECs cyclic dynamics, classical in the essence, and ordinary QM}.
 
We conclude that our results falsify the common opinion that quantum phenomena cannot be fully described by means of classical arguments. In general, it is believed that the semi-classical formulation of QM  is a limit of ordinary relativistic QM (QFT), and it can only provide an approximative description of non-trivial quantum phenomena.  On the contrary, our results clearly show that ordinary QM can be fully derived from classical arguments (including  the classical variational principle),  \ie from  the approach known as ``old formulation'', originally proposed by the fathers of QM (de Broglie,  Einstein, Bohr, Sommerfeld, etc...), if correctly applied as prescribed by ECs theory. The semi-classical approach based on ECs --- a modern version of de Broglie's ``periodic phenomena'' --- is more fundamental --- or at least equivalent --- to ordinary relativistic QM (QFT).  In a few words ECs theory shows that the fathers of QM were on the right track more than typically believed by we modern physicists.

\section{``God doesn't play dice''} \label{God:dice}
The key to interpret the exact matching between EC classical-relativistic cyclic dynamics and ordinary QM in all its fundamental aspects and phenomenology  can be found in the statistical description of a ``particle moving [very fast] on a circle'' --- see also  't Hooft's CA.  It is crucial to consider that, even considering simple quantum systems, for instance, based on QED, the characteristic time scales involved in quantum dynamics are always faster than the Compton clock of the electron\footnote{Even though photons have infinite Compton time, they necessarily need to be emitted or absorbed by electrons in sources or detectors, respectively. So the time scale of quantum electrodynamics are faster or of the order of the electron Compton clock.}, \ie they are faster than  $10^{-21}$ s (zettasecond), sec.(\ref{AXIOMS}).  

These time scales are in fact far beyond any modern timekeeper resolution. The present experimental techniques are not able to directly observe such extremely small time scales. Thus, as for a die observed without a slow-motion camera, we can only describe statistically the outcomes of these ultra-fast cyclic dynamics --- see Axiom of the measurement in sec.(\ref{AXIOMS}). 

We can now interpret the results obtained so far by saying that the extremely fast dynamics characterising ECs, if observed with our current experimental  resolution in time (not to mention the resolution at the times of the ``old'' QM), are equivalent to ordinary QM. That is, they are described by the Hilbert space formalism and they fulfil the Born rule, the Heisenberg relation, all the  axioms of QM, the Feynman path integral, the Dirac quantisation prescription, the Bohr-Sommerfeld quantisation, the thermal QM and all the exact correspondences described so far. 

In general, the results reported in this paper show that, as long as we observe ECs with non sufficiently accurate timekeepers, the effective statistical description associated with their ultra-fast periodic dynamics is not distinguishable from ordinary QM. \emph{Vice versa}, as for a dice player with a slow motion-camera, an observer with infinite or extremely good resolution in time --- say, 	``God'' --- would be able to reveal the underlying, potentially deterministic, ECs dynamics and, in principle, to predict the quantum dice outcomes. A die rolling very fast can be regarded as the temporal dynamics of an EC on a lattice, or equivalently a dice with a continuum of faces can be regarded as an EC. 

So, we can actually say: having infinite time resolution, \emph{God has no fun playing dice}; in the sense that there is no fun playing dice if the player can predict the dice outcomes by means of a slow-motion camera, or of an ultra-fast timekeeper, or --- ideally --- of an infinite resolution in time \cite{'tHooft:2001fb,Dolce:Dice}. 
The resolution of modern timekeepers however is improving very fast towards the electron Compton time scale $\sim 10^{-21}$ s (zettasecond), \cite{Margolis:2014}. This means that in a next future timekeepers could be able to directly test, by means of experiments, the underling ECs dynamics, and thus the possible new physics beyond QM predicted by ECs theory. 

ECs theory --- similar to CA theory --- must not be confused with the de Broglie - Bohm or similar interpretations of QM whose philosophy is to try to approximate the quantum behaviours through fine-tuneable parameters. EC theory is a new formulation of QM formally equivalent to the axiomatic, the Feynman and the Dirac formulations of QM, as proved in many previous papers and reported here. In particular no fine-tunable parameters of any sort has been introduced in ECs.

The aim of this paper is to give a general overview of the exact matching between ECs physics and QM. At this stage we do not discuss about predictions of new physics --- though they can be more or less directly inferred from our description of ECs. According to our arguments we can conjecture, for instance, that QED outcomes observed at ultra-fast time resolution should manifest sub-Compton recurrence patterns, such that the average values over an EC spacetime period would coincide with  ordinary QED outcomes.

ECs theory strongly points out the viability of a deterministic interpretation of QM. It must be noticed that the ECs dynamics do not involve hidden variables of any sort (nor fine-tunable parameters): QM is exactly obtained by imposing, as quantisation condition, contravariant BCs to relativistic dynamics without introducing any additional parameter or variable\footnote{Contrarily to ECs dynamics based on continuous spacetime coordinates, CA dynamics as conceived by 't Hooft are determined by fast permutations which may however involve ``invisible hidden variables'', \cite{Hooft:2014kka}. These however are only relevant to time scales of the order of the Planck time, they are absolutely not relevant to QM but they could be relevant to quantum gravity. So our considerations about Bell's theorem can be extended to CA.}. On the contrary Bell's theorem is based on the primary hypothesis of the existence of local hidden variables. 

In Bell's demonstration the correlation $\mathcal C$ between observables $\mathcal A$ and $ \mathcal B$ is in fact given by an integration over a hypothetical hidden variable $\upsilon$: $\mathcal C = \int A(\upsilon) B(\upsilon) \rho(\upsilon)  d\upsilon$, where $\rho(\upsilon)$ is a probability density. Hence Bell's theorem, or similar no-go theorems based on the hypothesis of hidden variables, cannot be applied to ECs theory. The restriction to determinism given by these theorems does not apply to ECs physics.  

Even though a thorough description of the EPR paradox in terms of EC dynamics is beyond the scope of this paper, a simple fact can be safely stated. If, as Feynman used to say, ``the same equations have the same solutions'', the equivalence of ECs relativistic dynamics and QM (including the description of the tensor product of Hilbert space described above) automatically implies that ECs physics violates Bell's inequalities exactly as mush as ordinary QM. 

ECs theory has no hidden variables of any sort. However ECs physics theory has a non-local nature due to the cyclic character of the spacetime coordinates of the theory. Actually, in quantum optics textbooks the PBCs $\Phi_{\vec k}(x) = \Phi_{\vec k}(x + \lambda)$ are typically named  ``complete coherence'' or ``perfect entanglement''. By knowing the free EC physical state in a given spacetime point,  its values at spacetime points separated by integer numbers of periods is  automatically determined  (without  ``conspiracy'' arguments \cite{Hooft:2014kka}). Similarly to relativistic clocks, the whole information of an EC is contained in a single spacetime period. 

Also, ECs theory must not be confused with interpretations of QM stocastic quantisation of random walk. Indeed it provides a quite opposite interpretation \emph{w.r.t.} these interpretations which, all in all, are by far less effective that ECs theory if compared with the results reported so far. As already said, according to experimental observations we have that pure quantum system are ideal systems characterised by perfect Minkowskian periodicity (``perfect coherence''). The randomness of QM could not be fundamental, \ie indeterministic, as it emerges from the extremely fast cyclic dynamics involved in quantum processes, which are shorter than the Compton time of the electron $10^{-21} s$.  The assumption of chaotic dynamics or randomness can therefore have some justification only in the effective description of quantum dynamics, see discussion about ``a particle moving very fast on a circle''.  In any case ECs theory fully confirm that the ordinary theory of QM is a purely statistical theory. 

Despite these non-local aspects ECs theory is an absolutely local theory for what which concerns relativity: the EC spacetime period $\lambda^\mu$ transforms in a local relativistic geometrodynamical way, similar to relativistic clocks and rulers. According to our geometrodynamical description of interactions, the local modulations of the ECs periods in a given spacetime region, \eg obtained  by switching on a generic interaction, propagates to other spacetime regions according to relativistic causality. The EC physical state value after an integerÊnumber of periods is automatically determined but, due to interactions, the position of these spacetime points varies locally and in perfect agreement with relativistic causality, see sec.(\ref{INT}). 

The local modulations of spacetime period is in fact related to the local variations of four-momentum by the phase harmony relation, and the local variations of four-momentum propagates in spacetime according to relativistic causality. The propagation of the EC modulations of periodicity is therefore described by the retarded and advanced potentials of the ordinary relativistic wave theory as shown in \cite{Dolce:2009ce}.  

The full compatibility of ECs theory with relativistic causality and locality can also be seen by noticing that an EC is essentially a classical-relativistic string vibrating in spacetime, \ie it is constituted by harmonics which are in fact ordinary relativistic waves governed by advanced and retarded solutions. 

Actually, as pointed out in other papers \cite{Dolce:2009ce,Dolce:2009cev4,Dolce:cycles,Dolce:tune}, one of the beauties of ECs theory is that it turns out to be the full relativistic generalization of the theory of sound. It considers relativistic vibrations not only in space but also in time. An EC is the relativistic analogous of a ``sound'' source that can vibrate also in time and not only in space. We can actually say that ECs theory  provides a fascinating ``harmonic'' description of our quantum universe.

\section{Where is the boundary of spacetime?: From ``ontic'' spacetime to relativistic spacetime}\label{ONTICSPACETIME} 

In this paragraph we discuss the most original aspect of ECs theory, that is the absolutely novel description of the relativistic spacetime predicted by the theory. In other words, the most fascinating prediction of this study is the following: \emph{QM tells us above any reasonable doubt that spacetime has an intrinsically cyclic nature, contrarily to the emphatically non-compact nature typically considered in the ordinary interpretation of relativity}. \emph{A compact nature of spacetime is the price to pay for a unified description of relativistic and quantum mechanics, as well as of gravitational and gauge interactions.}  Yet, such an alternative formulation of spacetime is completely compatible with the whole relativistic physics and, at the same time, it allows us to derive QM  directly from relativistic geometrodynamics, in a unified view. 

An EC can be regarded essentially as classical-relativistic vibrations of relativistic spacetime.
We start by noticing that the ECs cyclic spacetime coordinates  enter into the equations exactly as the relativistic spacetime coordinates of ordinary QFT and relativity, as can be noticed for instance from eq.(\ref{field:mode}) or eq.(\ref{CA:action:Gra}). Hence the (``ontic'') spacetime coordinates of ECs theory and ordinary relativistic spacetime can be in principle identified. 

The ECs ``ontic'' coordinates are perfectly consistent sets of relativistic coordinates, even though they are angular (cyclic) coordinates (with cyclic or angular coordinates we mean coordinates defined on a compact flat manyfold and PBCs, in analogy to, \eg, the Kaluza-Klein cyclic extra dimension). The differential spacetime structure of an EC is the exactly same as for ordinary relativity. Special and General Relativity only concern about the differential structure of relativistic spacetime. It is a fact that relativity doesn't give any particular prescription about the boundary of spacetime. In other words, relativity is not able to answer to the question: ``where is the boundary of spacetime?''\footnote{The analysis of the BCs of relativistic theories is the foundational point of ECs theory as described in \cite{Dolce:2009ce}. The variational principle applied to relativistic actions prescribes particular restrictions to the BCs. For instance, as well-known from instance in string theory and extra dimensional theory, for a bosonic action ---without additional boundary terms --- the possible choices are Neumann BCs, Dirichlet BCs, PBCs and so on. In general the BCs prescribed by ECs theory, such as PBCs in the bosonic theory, are those compatible with the relativistic actions.} \cite{York1986}.  EC theory complete spacial and general relativity with BCs (see Cauchy problem in mathematics) allowing the unified description of relativistic and quantum dynamics. 

ECs can be regarded as relativistic clocks (and rulers). Local contravariant PBCs are assumed in every relativistic spacetime point, depending on the content of energy-momentum in that point. Notice that the requirement of covariance for the BCs is essential to obtain a consistent relativistic theory. Without this requirement one obtains problematic relativistic theories like Closed Timelike Curves (CTC) theory \cite{Cooperstock2005}.  In other words spacetime is compactified in a cyclic geometry, similarly to extra-dimensional theory or string theory. The fundamental difference is that in those theories the inevitable extra-dimensions must compactified because they are problematics. On the contrary, in ECs theory there are not extra-dimensions, it is the ordinary four-dimensional spacetime which is compactified in a contravariant local way in order to be fully consistent with the differential structure of relativity, and to derive relativistic QM directly from the resulting constrained (compact) relativistic mechanics. Our analysis of the ECs geometrodynamics  (reported in more detail in previous papers)  demonstrates that ECs physics do not break relativistic symmetries and invariances. Indeed  PBCs on compact spacetime manifolds fulfil the variational principle of relativistic (bosonic) field and string theories.  

We want to point out that, actually, ECs theory must be regarded a novel class of String Theory, \cite{Dolce:cycles,Dolce:tune,Dolce:ADSCFT}. Ordinary String Theory is characterised by a compact world-sheet (2D) with PBCs or other kinds of BCs (consistently with the variational principle) of the ends of the compact world-sheet dimension, corresponding respectively to closed or open strings. As well-known, the assumption of bidimensional (2D) world-sheet  implies, for reasons of self-consistency of String Theory, the (problematic) extra-dimensions on the target spacetime of ordinary String Theory. 

Similarly ECs theory is characterised by a compact  (one dimensional) world-line  encoding the Compton clock --- the Compton clock is a periodicity on the world-line of a particle, \ie on the proper time --- with PBCs or other types (consistently to the variational principle) of BCs depending on the particle dynamics (bosonic or fermionic) that we want to describe \cite{Dolce:cycles}. In this way the target spacetime of the theory is consistently purely 4D as for ordinary relativity (without necessarily involving extra-dimensions), though it turns out to be an intrinsically compact (cyclic) spacetime. %The compactification length of the world-line parameter is fixed by the mass of the particle that we want to represent, according to the Compton relation $\lambda_C = T_C = 2 \pi / m$. 

The possibility of a consistent description of \emph{quantum} particle physics by means of compact world-lines  is provided by the fact that QM (in particular the wave-particle duality) implies that a massive particle has a word-line recurrence of Compton length $\lambda_C$, \ie a recurrence in the proper-time of duration $\lambda_C = T_C = 2 \pi / m$,  sec.(\ref{EC:MASS}).  This simple consideration (probably neglected in the initial development rush of String Theory in favour of an excess of formalism and abstractness) shows that actually it is possible to define a String Theory in a (1D) compact world-line, instead of a compact (2D) world-sheet, avoiding the non-compact world-sheet parameter of ordinary String Theory. The resulting theory, \ie ECs theory, is able to reproduce quantum and relativistic mechanics in a unified way, and inherits fundamental aspects of String Theory.   Indeed ECs theory confirms important historical motivations of String Theory and justifies most of its mathematical beauty \cite{Dolce:ADSCFT,Dolce:tune,Dolce:cycles,Dolce:2009ce}. 

On one hand the emphatically compact description of spacetime in ECs theory (resulting from the assumption of PBCs on a  compact world-line encoding the Compton clock) preserves relativistic dynamics and invariances. On the other hand they imply some kind of new relativistic phenomena in addition to the purely relativistic ones. According to our results, the effective statistical description of the resulting cyclic (undulatory) relativistic dynamics are equivalent to relativistic QM in all its fundamental aspects. The relativistic cyclic dynamics of an EC are identical, at a statistical level, to the quantum dynamics of the corresponding relativistic particle. 

The PBCs are the quantisation conditions of the ECs theory (similarly to the BCs of a particle in a ``time box''). The spacetime periodicities  are fixed by  the content of energy-momentum in a given spacetime point, according to de Broglie undulatory mechanics. We know from Einstein that relativity is the manifestation of the differential structure of spacetime.  Similarly, we can state that \emph{ QM, in all its phenomenology, is the manifestation of the compact (cyclic) nature of relativistic spacetime} --- \ie of the compact (cyclic) nature of the particles world-lines.   

All in all the assumption of intrinsic periodicity is a Coulomb egg. Paraphrasing Einstein, ECs theory is ``simple but not simpler'', in the sense that it exactly yields \emph{de facto} a unified description of relativistic and quantum theories in an unexpected simple and elegant way (it is not the first example of this kind in the history of science) but, at the same time, it is groundbreaking as it revolutionises  the concept of relativistic spacetime so much that  it may erroneously appear absurd from an orthodox point of view. The undeniable fact is that the ECs theory really yields \emph{de facto} a unified description of quantum and relativistic physics. 

Physics of composite systems (\ie systems of interacting elementary particles) can be consistently described as composition (similarly to 't Hooft ``beables'') of ECs dynamics. The universe is composed by elementary particles and every elementary particle can be represented as an EC. 
Hence the universe can be regarded as the ensemble (``beable'') of  all the ECs associated to all the elementary particles contained in it. Notice that, even though ECs are characterised by cyclic dynamics, the evolution of a system of many ECs interacting each other is in general chaotic --- see the considerations about the thermal noise in sec.(\ref{therm:QM}). Interactions, due to the exchange of four-momentum,  modulate locally and contravariantly the ECs spacetime periodicities.  Thus our description does not necessarily imply a cyclic universe. ECs theory is able to reproduce the extreme complexity of ordinary physical system despite its basic constituents, if isolated, are intrinsically cyclic, exactly as Newton mechanics is able to describe the complexity of classical systems (thermodynamics) despite its basic constituents, if isolated, have uniform rectilinear motion.  

Roughly  speaking, in such a description \emph{every elementary particle composing our universe can be regarded as an independent ultra-fast cyclic universe itself}. The combination and interaction of the cyclic dynamics associated to each elementary particle reproduces the chaotic behaviour of our universe. Notice that, even though many phenomena on every day life appears to be non-periodic or even irreversible due to the statistical laws, the elementary quantum constituents of our universe (elementary particles) are perfectly cyclic  if isolated (free ECs). Indeed our results clearly show that the quantum dynamics of elementary particles are the manifestation of fundamental relativistic cyclic dynamics, \ie elementary spacetime cycles \cite{Dolce:cycles,Dolce:ADSCFT}, see also \cite{Penrose:cycles}. Recall that, in ECs theory \cite{Dolce:cycles,Dolce:tune,Dolce:ADSCFT,Dolce:2009ce,Dolce:licata,Dolce:SuperC,Dolce:EPJP,Dolce:Dice2014,Dolce:TM2012,Dolce:Dice2012,Dolce:cycle,Dolce:ICHEP2012,Dolce:FQXi,Dolce:Dice,Dolce:2010ij,Dolce:2010zz,Dolce:2009cev4}, particles can  also be equivalently described as vibrations of intrinsically compact cyclic spacetime dimensions. Such a description is a fascinating novel, unprecedented  interpretation of relativistic spacetime. 

In short, the undeniable validity of the ordinary axiomatic description of QM represents the undeniable  experimental proof  that allows us to claim \emph{the discovery of the cyclic nature of relativistic spacetime}. 

In ECs theory, the local nature of spacetime is enforced \emph{w.r.t.} the ordinary description of relativistic spacetime. Indeed undulatory mechanics (wave-particle duality)  is encoded directly into relativistic geometrodynamics \cite{Dolce:cycles}. 
In this view the time flow of the universe has a relational interpretation, with some analogies to Rovelli's  \cite{Rovelli:2009ee} and Penrose ideas \cite{Penrose:cycles}, as pointed out in previous publications, \emph{e.g.} \cite{Dolce:TM2012,Dolce:2009ce,Dolce:ADSCFT,Dolce:FQXi}. The irreversibility of the time flow is a statistical consequence of the fact that ordinary physical systems are the combinations of many interacting elementary cycles, similarly to the irreversibility in classical thermodynamics which follows from the fact that systems are composed by many classical particles interacting chaotically (thermal noise).

 The ordinary, emphatically non-compact, relativistic spacetime is inferred as an emerging collective phenomenon from ECs cyclic dynamics. This aspect of ECs theory has been described  in several papers, with particular emphasis on the emerging relativistic time flow in \cite{Dolce:TM2012,Dolce:Dice2012,Dolce:cycle,Dolce:2009ce,Dolce:FQXi}. We can imagine that each particle of our universe defines an independent relativistic time coordinate. Indeed in ECs theory every particle can be regarded as a relativistic clock with its own time. For instance we may assume that the particle $j$-th of our universe defines a relativistic (``ontic'') time $t_j$ coordinate of periodicity $T_j$. Now it is sufficient to chose one of these clocks,Êfor instance the particle $k$-th  --- or another periodicity phenomenon such as that associated to the Cesium atomic transition --- as reference clock to have a relation, emergent description of ordinary relativistic time $t_k = t$.   Each relativistic time $t_j$ now can be expressed in terms of the time of the reference clock, $t_j(t)$, and its period referred to the reference period as well as.

 In this description it is important to bear in mind that these ``clocks'' can interact, for instance, by exchanging photons, and that photons have ``frozen'' internal clocks, \ie infinite Compton periodicity. As also noticed by R. Penrose ``any stable massive particle behaves as a very precise quantum clock, which ticks away with [Compton periodicity]'' and  a ''photon would take until eternity [infinite Compton periodicity] before its internal clock gets the first tick! To put this another way, it would appear that rest-mass is necessary ingredient for the building of a clock'' \cite{Penrose:cycles}. The mass is thus essential to differentiate the role of relativistic time with respect to that of relativistic space in physics (in relativity they are linked by Lorentz transformations). Similarly, without energy and momentum is not possible to define relativistic spacetime. 

ECs interpretation of spacetime can be pushed even further. Actually we can say that every elementary particle defines its own spacetime since every elementary particle, interpreted as EC, is a reference clock and ruler. Thus in ECs theory it would be more appropriate to speak about many vibrating spacetimes, one for each particle. Ordinary spacetime is a collective description of these many spacetimes, similarly to the collective description of time given above. As an EC has zero entropy whereas a composite system of ECs has high content of entropy for statistical reasons, we actually see that the arrow of time emerges from the EC spacetimes in a statistical way, contrarily to the ordinary interpretation of relativistic spacetime in which the arrow of time cannot be justified with similar arguments.   

The Feynman conjecture that antiparticles are particles travelling backward in time is absolutely consistent with EC physics due to the fact that the relativistic time has an enforced local character ---whereas it is not admissible in ordinary interpretation of ordinary spacetime. In ECs theory, since every particle is a clock defining its own time, it is possible to invert the arrow of time in a single particle without affecting the arrow of time of the other particles. That is, we can imagine to invert the rotation of a clock from clockwise to anti-clockwise, but this doesn't mean that as a result all the other clocks turn out to be inverted, nor that the flow of time of the whole universe result to be inverted. This is because in ECs theory every particle has its own relativistic time, whose ticks are determined by the particle mass and \emph{vice versa}.  

The result of this inversion is to transform the particle to the corresponding antiparticle. This is in agreement with our interpretation of the Hamiltonian operator of the theory, whose ``negative'' eigenstates describe antiparticles. Indeed ECs offer an elegant solution of the problem of the arrow of time by enforcing the local nature of time: the fundamental postulate of ECs theory, from which all the results of the theory can be derived, can be equivalently stated in the following way: \emph{every particle is a clock!} 

 This paper is exclusively focused on the equivalence between ECs dynamics and QM. However, many further important applications of ECs to modern physics has been developed in \cite{Dolce:cycles,Dolce:tune,Dolce:ADSCFT,Dolce:2009ce,Dolce:licata,Dolce:SuperC,Dolce:EPJP,Dolce:Dice2014,Dolce:TM2012,Dolce:Dice2012,Dolce:cycle,Dolce:ICHEP2012,Dolce:FQXi,Dolce:Dice,Dolce:2010ij,Dolce:2010zz,Dolce:2009cev4}. Applications in condensed matter, such as a novel description of superconductivity and graphene physics has been reported in \cite{Dolce:SuperC,Dolce:EPJP,Dolce:Dice2014}, and summarised in sec.{\ref{therm:QM}}.
 
Last but not least,  ECs dynamics also share fundamental mathematical and phenomenological properties with extra-dimensional theories \cite{Dolce:ADSCFT}. Remarkably, this advanced aspect of ECs --- see the concept of ``virtual extra dimension'' in ECs theory --- provides an unprecedented, intuitive, yet rigorous demonstration of the central relation of the AdS/CFT (Anti de Sitter / Conformal Field Theory) correspondence. That is, in Witten's words, in the AdS/CFT correspondence ``quantum phenomena [...] are encoded in classical [extra dimensional] geometries''. The detailed demonstration of the AdS/CFT correspondence from ECs physics is given in \cite{Dolce:cycles,Dolce:ADSCFT,Dolce:ICHEP2012,Dolce:2009ce}. Actually we have seen that in ECs physics quantum phenomena are encoded in classical (compact spacetime) geometrodynamics. 
  Another spectacular aspect is that, in the duality of ECs theory to extra-dimensional theory, the geometrodynamical description of gauge interactions, already reported in this paper and analogous to gravitational one, turns out to be nothing but ``Kaluza's miracle'' of unification of gravitational and electromagnetic interactions \cite{Dolce:cycles,Dolce:ADSCFT,Dolce:tune}.   
  
  All these results are abundant, clear, rigorous, mathematical proofs of the absolute validity and consistency of the ECs theory, as well as of its potentiality to face open problems of theoretical physics. In addition to this, ECs theory also inherits some fascinating aspects of 't Hooft CA \cite{Hooft:2014kka,hooft-2009-0,'tHooft:2007xi,'tHooft:2006sy,'tHooft:2001ct,'tHooft:2001fb,'tHooft:2001ar,'tHooft:1998fa}. In particular CA  brings interesting new insights on foundations of quantum gravity. 
  
  The description, on one hand, of QED from spacetime geometrodynamics and, on the other hand, the correspondence with gravitational geometrodynamics, actually opens an unexplored scenario to approach the problem of the quantisation of gravity (does it make sense to quantise the boundary which quantises, through BCs,  elementary particles dynamics?), Black-Hole physics (a Black-Hole can be regarded as the macroscopic $T$-dual of an elementary particle due to the intrinsic Minkowskian time periodicity $8 \pi G M_\odot$ coming from its metric?) or cosmology (has the rate of the elementary clocks of the universe been always the same or it is accelerating/decellarating? How this contribute to the Big Bang Theory).

\section{Conclusions} 
In this paper we have presented the unified description of quantum and relativistic mechanics allowed by the Elementary Cycles (ECs) theory \cite{Dolce:cycles,Dolce:tune,Dolce:ADSCFT,Dolce:2009ce,Dolce:licata,Dolce:SuperC,Dolce:EPJP,Dolce:Dice2014,Dolce:TM2012,Dolce:Dice2012,Dolce:cycle,Dolce:ICHEP2012,Dolce:FQXi,Dolce:Dice,Dolce:2010ij,Dolce:2010zz,Dolce:2009cev4}. Every elementary quantum particle is described as the manifestation of corresponding ultra-fast cyclic spacetime dynamics, classical in the essence. We have found that the effective description of the ECs relativistic dynamics is indistinguishable from ordinary relativistic QM.

 Remarkably, ECs theory clearly proofs that a  derivation of QM from pure relativistic dynamics is definitively possible. The idea to  derive QM by constraining relativistic equations of motion was after all originally proposed also by Einstein, and many other fathers of modern physics \cite{Dolce:licata,Pais:einstein,Einstein:1910}. Actually ECs physics do not involves hidden variables of any sort and it can violate Bell's inequalities as much as ordinary QM. ECs theory really represents a viable solution to the riddles of quantum physics. ECs approach to physics also shares some fundamental aspects with Cellular Automata models proposed by 't Hooft \cite{Hooft:2014kka,hooft-2009-0,'tHooft:2007xi,'tHooft:2006sy,'tHooft:2001ct,'tHooft:2001fb,'tHooft:2001ar,'tHooft:1998fa} --- though ECs is a completely independent theory, based on a continuum of spacetimes.

ECs are characterised by classical-relativistic dynamics and the constraint of intrinsic (covariant) periodicities (not to be confused with the problematic theory of CTC) whose time scales are uniquely fixed by the Compton times of the corresponding elementary particles. In simple words, the basic postulate of EC theory can be stated in the following way: \emph{every elementary particle is an elementary relativistic reference clock}. That is (paraphrasing Newton's principles and de Broglie's hypothesis), \emph{every free elementary quantum particle of (persistent) energy $\omega$, observed from an inertial reference frame, is an elementary  relativistic cyclic system, classical in the essence, of persistent time periodicity $T = 2 \pi / \omega$}. 

The resulting ECs theory has been investigate in detail in several peer-reviewed papers \cite{Dolce:cycles,Dolce:SuperC,Dolce:TM2012,Dolce:Dice2012,Dolce:cycle,Dolce:ICHEP2012,Dolce:tune,Dolce:ADSCFT,Dolce:FQXi,Dolce:Dice,Dolce:2010ij,Dolce:2010zz,Dolce:2009ce}. Indeed the statistical description of ECs classical dynamics leads to  a manifest, exact mathematical equivalence to ordinary relativistic QM. The correspondence has been rigorously proven for all the fundamental aspects of QM. 

From ECs dynamics we have exactly derived, for instance: all the axioms of QM; the Dirac quantisation rule based on commutation relations (second quantisation of quantum fields); the Heisenberg uncertainty relation; the semi-classical methods of QM; the quantisation of thermal systems. Furthermore we have proven (independently \emph{w.r.t} the results just mentioned) that \emph{the ECs classical-relativistic evolutions are exactly described by the ordinary Feynman path integrals of elementary relativistic particles}. We have described how to build the second quantised field at the base of ordinary QFT from ECs. 
All these equivalences (and the many other not reported in this paper because objectively too much to be contained here) have been first proven for the free case, and then they have been fully generalised to interactions (as learnt from Newton's description of classical mechanics which starts with the ideal case of free elementary systems and then generalising to interactions). 

Remarkably, we have also proven that gauge interactions can be derived, without postulating gauge invariance, directly from a particular subclass of ECs spacetime geometrodynamics, in perfect correspondence with gravitational interaction. In this way we have derived QED directly from the resulting ECs classical-relativistic geometrodynamics. Indeed, ECs theory not only yields \emph{de facto} a unified description of quantum and relativistic mechanics, it also yields a unified description of gauge and gravitational interactions \cite{Dolce:tune}, including a mathematical demonstration from first principles of the central relation of the AdS/CFT correspondence, also known as gauge/gravity duality \cite{Dolce:ADSCFT}. 

Such a long list of remarkable, rigorously proven equivalencies resulting from ECs physics cannot be the fruit of mere mathematical coincidences --- coincidences don't exist in mathematical demonstrations. ECs theory must be therefore regarded as a new formulation of QM, equivalent to the axiomatic formulation of  QM, to the Feynman formulation of QM, to the Dirac quantisation rule, and so on. Furthermore, the details of such  exact equivalences to QM are extremely fascinating and clearly indicate a possible way out to long standing problems of physics, as well as possible new physics beyond QM. 

ECs theory clearly indicates that QM emerges from ultra-fast cyclic spacetime dynamics associated to elementary particles and that the ordinary interpretation of relativistic spacetime itself, based on emphatically non-compact dimensions, must be reconsidered. \emph{The price to pay for a unified description of quantum and relativistic mechanics is to give up with the ordinary emphatically non-compact description of relativistic spacetime}. QM is the manifestation of intrinsic boundaries of relativistic spacetime.  

In his recent review paper, t' Hooft has raised the --- venerable --- question: is  ``a [classical] view on the quantum nature of our universe, compulsory or impossible?'' \cite{Hooft:2014kka}.  The clear answer resulting from our analysis, obtained through rigourous mathematical proofs, all certified by peer-reviews and published on scientific journals, is that a view of QM based on simple relativistic physical systems, classical in the essence,  is definitively \emph{compulsory}, rather than impossible.  The real question now is: are we physicists ready to consider a new description of relativistic reality beyond QM? Pretending to not see scientific facts is against the very essence of science:

\vspace{1cm}
\begin{small}
%\null\vspace{\stretch{1}}
\begin{flushright}

\textit{``I wish, my dear Kepler, that we could have a good laugh \\ 
together at the extraordinary mediocrity of the mob.  \\ 
 What do you think of the foremost philosophers of\\ 
this University, to whom I have offered  a  thousand \\ 
times of my own accord to show my  studies, \\ 
but who, with the lazy obstinacy of a  serpent \\ 
who has eaten his fill, have never consented\\ 
 to look  at planets, nor  moon, nor through my glass? \\
 Verily, just as serpents close their ears, so do these \\ 
 men close their eyes to the light of truth. These are \\
 great matters; yet they do not occasion any surprise. \\
$[ \dots ]$  In questions of science, the authority of a thousand \\
is not worth the humble reasoning of a single individual.'' \\}

\vspace{0.5em}

{From the second (1610) of the two Galileo's letters to Kepler.}

%\end{quotation}
\end{flushright}
%\vspace{\stretch{2}}\null
\end{small}

 \section*{Acknowledgements} 
We acknowledge, in chronological order, G. 't Hooft, L. Chiatti and I. Licata for the precious comments.

%
%
%\bibliographystyle{utphys}
%%\bibliographystyle{apsrev4-1}
%\bibliography{comp3+1}
%%\bibliography{cycles}

\bibliography{}

\begin{thebibliography}{10}

\bibitem{2013arXiv1301.1069S}
M.~{Schlosshauer}, J.~{Kofler}, and A.~{Zeilinger}, ``{A Snapshot of
  Foundational Attitudes Toward Quantum Mechanics},'' {\em ArXiv e-prints}
  (Jan., 2013) \href{http://www.arXiv.org/abs/1301.1069}{{\tt 1301.1069}}.

\bibitem{Hooft:2014kka}
G.~t. Hooft, ``{The Cellular Automaton Interpretation of Quantum Mechanics. A
  View on the Quantum Nature of our Universe, Compulsory or Impossible?},''
\href{http://www.arXiv.org/abs/1405.1548}{{\tt 1405.1548}}.
%%CITATION = ARXIV:1405.1548;%%.

\bibitem{Wilczek:2012jt}
F.~Wilczek, ``{Quantum Time Crystals},'' {\em Phys.Rev.Lett.} {\bf 109} (2012)
  160401,
\href{http://www.arXiv.org/abs/1202.2539}{{\tt 1202.2539}}.
%%CITATION = ARXIV:1202.2539;%%.

\bibitem{Weinberg:2014ewa}
S.~Weinberg, ``{Quantum Mechanics Without State Vectors},''
\href{http://www.arXiv.org/abs/1405.3483}{{\tt 1405.3483}}.
%%CITATION = ARXIV:1405.3483;%%.

\bibitem{'tHooft:2005ym}
G.~'t~Hooft, L.~Susskind, and E.~Witten, ``{A theory of everything?},'' {\em
  Nature} {\bf 433} (2005), no.~PRESSCUT-H-2005-091, 257--259.

\bibitem{Dolce:cycles}
D.~Dolce, ``{Elementary spacetime cycles},'' {\em Europhys. Lett. ,} {\bf 102}
  (2013) 31002,
\href{http://www.arXiv.org/abs/1305.2802}{{\tt 1305.2802}}.
%%CITATION = ARXIV:1305.2802;%%.

\bibitem{Dolce:tune}
D.~Dolce, ``{Gauge Interaction as Periodicity Modulation},'' {\em Annals Phys.}
  {\bf 327} (2012) 1562--1592,
\href{http://www.arXiv.org/abs/1110.0315}{{\tt 1110.0315}}.
%%CITATION = ARXIV:1110.0315;%%.

\bibitem{Dolce:ADSCFT}
D.~Dolce, ``{Classical geometry to quantum behaviour correspondence in a Virtual
  Extra Dimension},'' {\em Annals Phys.} {\bf 327} (2012) 2354--2387,
\href{http://www.arXiv.org/abs/1110.0316}{{\tt 1110.0316}}.
%%CITATION = ARXIV:1110.0316;%%.

\bibitem{Dolce:2009ce}
D.~Dolce, ``{Compact Time and Determinism for Bosons: foundations},'' {\em
  Found. Phys.} {\bf 41} (2011) 178--203,
\href{http://www.arXiv.org/abs/0903.3680v5}{{\tt 0903.3680v5}}.
%%CITATION = 0903.3680;%%.

\bibitem{Dolce:licata}
D.~Dolce, ``Introduction to the Quantum Theory of Elementary Cycles,'' Chapter 4, {\em
  Beyond Peaceful Coexistence; The Emergence of Space, Time and Quantum},
  I.~Licata, ed.
\newblock World Scientific, May 2016, 740pp, ISBN: 978-1-78326-831-3 

\bibitem{Dolce:SuperC}
D.~{Dolce} and A.~{Perali}, ``{The role of quantum recurrence in
  superconductivity, carbon nanotubes and related gauge symmetry breaking},''
  {\em Found.Phys.} {\bf 44} (Sept., 2014) 905--922,
  \href{http://www.arXiv.org/abs/1307.5062}{{\tt 1307.5062}}.

\bibitem{Dolce:EPJP}
D.~{Dolce} and A.~{Perali}, ``{On the Compton clock and the undulatory nature
  of particle mass in graphene systems},'' {\em EPJ Plus} {\bf 41} (Jan., 2015) 130, \href{http://www.arXiv.org/abs/1403.7037}{{\tt 1403.7037}}.

\bibitem{Dolce:Dice2014}
D.~Dolce and A.~Perali, ``Testing Elementary Cycles formulation of quantum
  mechanics in carbon nanotubes and superconductivity,'' {\em Journal of
  Physics: Conference Series} {\bf 626} (2015), no.~1, 012062.

\bibitem{Dolce:TM2012}
D.~Dolce, ``{Elementary cycles of time},'' {\em EPJ Web Conf.} {\bf 58} (2013)
  01018,
\href{http://www.arXiv.org/abs/1306.0579}{{\tt 1306.0579}}.
%%CITATION = ARXIV:1306.0579;%%.

\bibitem{Dolce:Dice2012}
D.~Dolce, ``{Intrinsic periodicity: the forgotten lesson of quantum
  mechanics},'' {\em J. Phys.: Conf. Ser.} {\bf 442} (2013) 012048,
\href{http://www.arXiv.org/abs/1304.4167}{{\tt 1304.4167}}.
%%CITATION = ARXIV:1304.4167;%%.

\bibitem{Dolce:cycle}
D.~Dolce, ``{On the intrinsically cyclic nature of space-time in elementary
  particles},'' {\em J. Phys.: Conf. Ser.} {\bf 343} (2012) 012031,
\href{http://www.arXiv.org/abs/1206.1140}{{\tt 1206.1140}}.
%%CITATION = ARXIV:1206.1140;%%.

\bibitem{Dolce:ICHEP2012}
D.~Dolce, ``{AdS/CFT as classical to quantum correspondence in a Virtual Extra
  Dimension},'' {\em PoS} {\bf ICHEP2012} (2013) 478,
\href{http://www.arXiv.org/abs/1309.2646}{{\tt 1309.2646}}.
%%CITATION = ARXIV:1309.2646;%%.

\bibitem{Dolce:FQXi}
D.~Dolce, ``{Clockwork Quantum Universe},''. IV Prize, QFXi contest (2011).

\bibitem{Dolce:Dice}
D.~Dolce, ``{de Broglie Deterministic Dice and Emerging Relativistic Quantum
  Mechanics},'' {\em J. Phys.: Conf. Ser.} {\bf 306} (2011) 10.

\bibitem{Dolce:2010ij}
D.~Dolce, ``{Deterministic Quantization by Dynamical Boundary Conditions},''
  {\em AIP Conf. Proc.} {\bf 1246} (2010) 178--181,
\href{http://www.arXiv.org/abs/1006.5648}{{\tt 1006.5648}}.
%%CITATION = 1006.5648;%%.

\bibitem{Dolce:2010zz}
D.~Dolce, ``{Quantum Mechanics from Periodic Dynamics: the bosonic case},''
  {\em AIP Conf. Proc.} {\bf 1232} (2010) 222--227,
\href{http://www.arXiv.org/abs/1001.2718}{{\tt 1001.2718}}.
%%CITATION = 1001.2718;%%.

\bibitem{Dolce:2009cev4}
D.~Dolce, ``{Compact Time and Determinism for Bosons (version 4)},''
  \href{http://www.arXiv.org/abs/0903.3680v4}{{\tt 0903.3680v4}}. The
  foundational part of this paper (i.e. par.2, par.3 and par.4.2) has been
  published [1].

\bibitem{Einstein:1923}
A.~Einstein, ``{Bietet die Feldtheorie M\"oglishKeiten f\"ur L\"osung des
  Quantenproblems?},'' {\em S.B. Press. Aked. Wiss.} {\bf 33} (1923).

\bibitem{Pais:einstein}
A.~Pais, {\em {Subtle is the Lord: The science and the life of Albert
  Einstein}}.
\newblock Oxford University Press, 1982.

\bibitem{hooft-2009-0}
G.~'t~Hooft, ``Entangled quantum states in a local deterministic theory,''
  2009.

\bibitem{'tHooft:2007xi}
G.~'t~Hooft, ``{Emergent Quantum Mechanics and Emergent Symmetries},'' {\em AIP
  Conf. Proc.} {\bf 957} (2007) 154--163,
\href{http://www.arXiv.org/abs/0707.4568}{{\tt 0707.4568}}.
%%CITATION = ARXIV:0707.4568;%%.

\bibitem{'tHooft:2006sy}
G.~'t~Hooft, ``{The Mathematical basis for deterministic quantum mechanics},''
  {\em J. Phys. Conf. Ser.} {\bf 67} (2007) 012015,
\href{http://www.arXiv.org/abs/quant-ph/0604008}{{\tt quant-ph/0604008}}.
%%CITATION = QUANT-PH/0604008;%%.

\bibitem{'tHooft:2001ct}
G.~'t~Hooft, ``{Quantum mechanics and determinism},'' in {\em {8th
  International Symposium on Particles Strings and Cosmology (PASCOS 2001)
  Chapel Hill, North Carolina, April 10-15, 2001}}.
\newblock 2001.
\newblock
\href{http://www.arXiv.org/abs/hep-th/0105105}{{\tt hep-th/0105105}}.
\newblock
%%CITATION = HEP-TH/0105105;%%.

\bibitem{'tHooft:2001fb}
G.~'t~Hooft, ``{How does God play dice? (Pre)determinism at the Planck
  scale},''
\href{http://www.arXiv.org/abs/hep-th/0104219}{{\tt hep-th/0104219}}.
%%CITATION = HEP-TH/0104219;%%.

\bibitem{'tHooft:2001ar}
G.~'t~Hooft, ``{Determinism in free bosons},'' {\em Int. J. Theor. Phys.} {\bf
  42} (2003) 355--361,
\href{http://www.arXiv.org/abs/hep-th/0104080}{{\tt hep-th/0104080}}.
%%CITATION = HEP-TH/0104080;%%.

\bibitem{'tHooft:1998fa}
G.~'t~Hooft, ``{TransPlanckian particles and the quantization of time},'' {\em
  Class. Quant. Grav.} {\bf 16} (1999) 395--405,
\href{http://www.arXiv.org/abs/gr-qc/9805079}{{\tt gr-qc/9805079}}.
%%CITATION = GR-QC/9805079;%%.

\bibitem{Broglie:1924}
L.~d. Broglie {\em Phil. Mag.} {\bf 47} (1924) 446.

\bibitem{2008FoPh...38..659C}
P.~{Catillon}, N.~{Cue}, M.~J. {Gaillard}, R.~{Genre}, M.~{Gouan{\`e}re}, R.~G.
  {Kirsch}, J.-C. {Poizat}, J.~{Remillieux}, L.~{Roussel}, and M.~{Spighel},
  ``{A Search for the de Broglie Particle Internal Clock by Means of Electron
  Channeling},'' {\em Foundations of Physics} {\bf 38} (July, 2008) 659--664.

\bibitem{Lan01022013}
S.-Y. Lan, P.-C. Kuan, B.~Estey, D.~English, J.~M. Brown, M.~A. Hohensee, and
  H.~M\"uller, ``A clock directly linking time to a particle's mass,'' {\em
  Science} {\bf 339} (2013), no.~6119, 554--557.

\bibitem{1996FoPhL}
R.~{Ferber}, ``{A missing link: What is behind de Broglie's ``periodic
  phenomenon''?},'' {\em Foundations of Physics Letters, Volume 9, Issue 6,
  pp.575-586} {\bf 9} (Dec., 1996) 575--586.

\bibitem{Penrose:cycles}
R.~Penrose, {\em {Cycles of Time. An Extraordinary View of The Universe}},
  ch.~2.3.
\newblock Knopf, New York, 2011.

\bibitem{thooft:heavy}
M.~B. {van der Mark} and G.~W. {'t Hooft}, ``{Light is Heavy},'' {\em ArXiv
  e-prints} (Aug., 2015) \href{http://www.arXiv.org/abs/1508.06478}{{\tt
  1508.06478}}.

\bibitem{Elze:2005gv}
H.-T. Elze, ``A quantum field theory as emergent description of constrained
  supersymmetric classical dynamics,'' {\em Brazilian Journal of Physics} {\bf
  35 no 2A+2B} (2005) 205--529,
\href{http://www.arXiv.org/abs/hep-th/0508095}{{\tt hep-th/0508095}}.
%%CITATION = HEP-TH 0508095;%%.

\bibitem{Nikolic:2008sn}
H.~Nikolic, ``{Probability in relativistic Bohmian mechanics of particles and
  strings},'' {\em Found. Phys.} {\bf 38} (2008) 869--881,
\href{http://www.arXiv.org/abs/0804.4564}{{\tt 0804.4564}}.
%%CITATION = 0804.4564;%%.

\bibitem{Feynman:1942us}
R.~P. Feynman, ``{The principle of least action in quantum mechanics},'' {\em
  {PhD thesis}}.

\bibitem{Feynman:1948art}
R.~P. Feynman, ``Space-time approach to non-relativistic quantum mechanics,''
  {\em Rev. Mod. Phys.} {\bf 20} (1948) 367--387.

\bibitem{Nielsen:2006vc}
H.~B. Nielsen and M.~Ninomiya, ``{Intrinsic periodicity of time and non-maximal
  entropy of universe},'' {\em Int. J. Mod. Phys.} {\bf A21} (2006) 5151--5162,
\href{http://www.arXiv.org/abs/hep-th/0601021}{{\tt hep-th/0601021}}.
%%CITATION = HEP-TH/0601021;%%.

\bibitem{feynman1990q}
R.~Feynman, {\em Q E D:}.
\newblock Penguin Books. Penguin, 1990.

\bibitem{peskin1995introduction}
M.~Peskin and D.~Schroeder, {\em An Introduction to Quantum Field Theory}.
\newblock Advanced book classics. Addison-Wesley Publishing Company, 1995.

\bibitem{Birrell:1982ix}
N.~D. Birrell and P.~C.~W. Davies, {\em {Quantum fields in curved space}}.
\newblock Cambridge, Uk: Univ. Pr., 1982.

\bibitem{Stephens:1993an}
C.~R. Stephens, G.~'t~Hooft, and B.~F. Whiting, ``{Black hole evaporation
  without information loss},'' {\em Class.Quant.Grav.} {\bf 11} (1994)
  621--648,
\href{http://www.arXiv.org/abs/gr-qc/9310006}{{\tt gr-qc/9310006}}.
%%CITATION = GR-QC/9310006;%%.

\bibitem{ArkaniHamed:2000ds}
N.~Arkani-Hamed, M.~Porrati, and L.~Randall, ``{Holography and
  phenomenology},'' {\em JHEP} {\bf 08} (2001) 017,
\href{http://www.arXiv.org/abs/hep-th/0012148}{{\tt hep-th/0012148}}.
%%CITATION = HEP-TH/0012148;%%.

\bibitem{Casalbuoni:2007xn}
R.~Casalbuoni, S.~De~Curtis, D.~Dominici, and D.~Dolce, ``{Holographic approach
  to a minimal Higgsless model},'' {\em JHEP} {\bf 08} (2007) 053,
\href{http://www.arXiv.org/abs/0705.2510}{{\tt 0705.2510}}.
%%CITATION = 0705.2510;%%.

\bibitem{Ohanian:1995uu}
H.~Ohanian and R.~Ruffini, {\em {Gravitation and space-time}}.
\newblock New York, USA: Norton, 1994.

\bibitem{Hestenes:zbw:1990}
D.~Hestenes, ``The zitterbewegung interpretation of quantum mechanics,'' {\em
  Foundations of Physics} {\bf 20} (1990) 1213--1232.

\bibitem{Weinberg:1996kr}
S.~Weinberg, ``The quantum theory of fields. vol. 2: Modern applications,''.
  Cambridge, UK: Univ. Pr. (1996) 489 p.

%\bibitem{2014arXiv1403.7037D}
%D.~{Dolce} and A.~{Perali}, ``{Probing the geometric nature of particles mass
%  in graphene systems},'' {\em ArXiv e-prints} (Mar., 2014)
%  \href{http://www.arXiv.org/abs/1403.7037v1}{{\tt 1403.7037}}.

\bibitem{deCordoba:2013cda}
P.~Fern\'andez De~C\'ordoba, J.~M. Isidro, and M.~H. Perea, ``{Emergent quantum
  mechanics as a thermal ensemble},'' {\em Int. J. Geom. Meth. Mod. Phys.} {\bf
  11} (2014), no.~08, 1450068,
\href{http://www.arXiv.org/abs/1304.6295}{{\tt 1304.6295}}.
%%CITATION = ARXIV:1304.6295;%%.

\bibitem{licata:2015}
I.~Licata and L.~Chiatti, ``{Timeless Approach to Quantum Jumps},'' {\em Int.
  J. Quant. Found.} {\bf 4} (2015), no.~01, 10--26.

\bibitem{Zinn-Justin:2002ru}
J.~Zinn-Justin, ``Quantum field theory and critical phenomena,'' {\em Int. Ser.
  Monogr. Phys.} {\bf 113} (2002)
1--1054.
%%CITATION = IMPHA,113,1;%%.

\bibitem{Kapusta:1989tk}
J.~I. Kapusta, ``Finite temperature field theory,''.

\bibitem{Zinn-Justin:2000dr}
J.~Zinn-Justin, ``Quantum field theory at finite temperature: An
  introduction,''
\href{http://www.arXiv.org/abs/hep-ph/0005272}{{\tt hep-ph/0005272}}.
%%CITATION = HEP-PH 0005272;%%.

\bibitem{Cooperstock2005}
Cooperstock, F. I. and Tieu, S, ``Closed Timelike Curves and Time Travel: Dispelling the Myth'', 
{\em Foundations of Physics}, {\bf 35} (2005) 497--1509


\bibitem{Jaffe:2005vp}
{Jaffe, R. L.}, ``The Casimir effect and the quantum vacuum,'' {\em Phys. Rev. D} {\bf 72} (2005), 021301.


\bibitem{refId0}
{Solov\'{}ev, E.A.}, ``Classical approach in atomic physics,'' {\em Eur. Phys.
  J. D} {\bf 65} (2011), no.~3, 331--351.

\bibitem{Margolis:2014}
H.~{Margolis}, ``{Timekeepers of the future},'' {\em Nature Physics} {\bf 10}
  (Jan., 2014) 82--83.

\bibitem{York1986}
J.~York, JamesW., ``Boundary terms in the action principles of general
  relativity,'' {\em Foundations of Physics} {\bf 16} (1986), no.~3, 249--257.

\bibitem{Rovelli:2009ee}
C.~Rovelli, ``{'Forget time'},'' {\em Found.Phys.} {\bf 41} (2011) 1475--1490,
\href{http://www.arXiv.org/abs/0903.3832}{{\tt 0903.3832}}.
%%CITATION = ARXIV:0903.3832;%%.

\bibitem{Einstein:1910}
A.~Einstein, ``Principe de relativit\'e,'' {\em Arch. Sci. Phys. Natur.} {\bf
  29} (1910).

\end{thebibliography}

\providecommand{\href}[2]{#2}\begingroup\raggedright\endgroup

\end{document}